\documentclass[final,5p,times,authoryear,twocolumn]{elsarticle}
\usepackage{graphicx}
\usepackage{amsmath}
\usepackage{amsthm}

\usepackage{enumitem}

\usepackage{amsfonts}
\usepackage{amssymb}
\usepackage{dsfont}
\usepackage{mathrsfs}
\usepackage{stfloats}
\usepackage{subcaption}
\usepackage{caption}
\usepackage{bm}
\usepackage{comment}
\usepackage{xcolor}
\usepackage[table]{xcolor}
\usepackage{tabularx}
\usepackage{adjustbox}
\usepackage{makecell}
\usepackage{algorithm} 
\usepackage{algpseudocode}
\usepackage{multirow}
\usepackage{booktabs}
\usepackage{placeins}
\usepackage[final]{microtype}
\usepackage{longtable}
\usepackage{array}

\usepackage[nolist,printonlyused]{acronym}

\newacro{e2e}[E2E]{End-to-End}
\newacro{qos}[QoS]{Quality of Service}
\newacro{ttl}[TTL]{Time to Live}
\newacro{cnf}[CNF]{Cloud Network Flow}
\newacro{rti}[RTI]{Real-Time Interactive}

\newacro{rl}[RL]{Reinforcement Learning}
\newacro{drl}[DRL]{Deep Reinforcement Learning}
\newacro{madrl}[MADRL]{Multi-Agent Deep Reinforcement Learning}
\newacro{ddpg}[DDPG]{Deep Deterministic Policy Gradient}
\newacro{bc}[BC]{Behavioral Cloning}
\newacro{mdp}[MDP]{Markov Decision Process}
\newacro{rlfd}[RLfD]{Reinforcement Learning from Demonstrations}
\newacro{bc}[BC]{Behavioral Cloning}
\newacro{gpr}[GPR]{General Policy Reward}
\newacro{mgarl}[MGA-RL]{Model-Guided Annealed Reinforcement Learning}
\newacro{mse}[MSE]{Mean Squared Error}
\newacro{dqfd}[DQfD]{Deep Q-Learning from Demonstrations}

\newacro{bp}[BP]{Backpressure}
\newacro{ldp}[LDP]{Lyapunov Drift-Plus-Penalty}
\newacro{umw}[UMW]{Universal Max-Weight}
\newacro{dcnc}[DCNC]{dynamic cloud network control}
\newacro{ucnc}[UCNC]{Universal Cloud Network Control}
\newacro{lelf}[LELF]{Lower Effective Lifetime First}
\newacro{mwp}[MWP]{Minimum Weight Path}
\newacro{mwprc}[MWP RC]{Minimum Weight Path \ac{rc}}
\newacro{upg}[UPG]{Uniform Path Grouping}
\newacro{upgecp}[UPG EC \protect\ensuremath{(p)}]{Uniform Path Grouping \ac{ec} \protect\ensuremath{(p)}}
\newacro{upgecps}[UPG EC \protect{\ensuremath{(p^*)}}]{Uniform Path Grouping \ac{ec} \protect\ensuremath{(p^*)}}
\newacro{mwpecp}[MWP EC (\protect\ensuremath{p})]{Minimum Weight Path \ac{ec} \protect\ensuremath{(p)}}
\newacro{mwpecps}[MWP EC \protect{\ensuremath{(p^*)}}]{Minimum Weight Path \ac{ec} \protect\ensuremath{(p^*)}}
\newacro{rcnc}[RCNC]{Reliable Cloud Network Control}
\newacro{marlrc}[\ac{madrl} RC]{\ac{madrl} Regular Congestion}
\newacro{marllac}[\ac{madrl} LAC]{\ac{madrl} Lifetime Aware Congestion}
\newacro{marlecps}[\acs{madrl} \acs{ec} \protect{\ensuremath{(p^*)}}]{Multi-Agent Deep Reinforcement Learning Effective Congestion \protect{\ensuremath{(p^*)}}}

\newacro{dcmt}[DCMT]{Delay-Constrained Maximum-Throughput}
\newacro{o2o}[O{\footnotesize{2}}O]{Offline-to-Online}

\newacro{ood}[OOD]{Out-of-Distribution}
\newacro{el}[EL]{Effective Lifetime}
\newacro{ec}[EC]{Effective Congestion}
\newacro{rc}[RC]{Regular Congestion}
\newacro{lac}[LAC]{Lifetime Aware Congestion}
\newacro{ecp}[EC \protect\ensuremath{p}]{Effective Congestion \protect\ensuremath{(p)}}
\newacro{ecps}[EC \protect\ensuremath{p^*}]{Effective Congestion \protect\ensuremath{(p^*)}}

\usepackage{tikz}
\usetikzlibrary{matrix, positioning, calc, backgrounds, arrows.meta}
\pgfdeclarelayer{background}
\pgfsetlayers{background,main}

\newcommand{\bigO}{\mathcal{O}}

\newcommand{\probP}{\text{I\kern-0.15em P}}

\theoremstyle{remark}

\journal{Journal of Network and Computer Applications}

\begin{document}
	
	\begin{frontmatter}
		
		\title{From Prior-Guided Heuristics to Deployable Agents: Accelerating Demonstration-Driven Reinforcement Learning for Deadline-Constrained Network Control}
		
		\author[a]{Vincenzo Norman Vitale\corref{cor1}}
		\ead{vincenzonorman.vitale@unina.it}
		\cortext[cor1]{Corresponding author}
		\author[a]{Mohammad Solki}
		\ead{mohammad.solki@unina.it}
		\author[a,e]{Antonia Maria Tulino}
		\ead{antoniamaria.tulino@unina.it}
		\author[b]{Andreas F. Molisch}
		\ead{molisch@usc.edu}
		\author[c,d,e]{Jaime Llorca}
		\ead{jllorca@ctts.es}

		\affiliation[a]{organization={DIETI, University of Naples Federico II},
			addressline={Via Claudio 21}, 
			city={Naples},
			postcode={80125}, 
			state={Italy}
		}
		
		\affiliation[b]{organization={University of Southern California},
			addressline={3740 McClintock Ave.}, 
			city={Los Angeles},
			postcode={90089}, 
			state={California},
			country={USA}
		}
		\affiliation[c]{organization={University of Trento},
			addressline={Via Sommarive, 9}, 
			city={Trento},
			postcode={38123}, 
			state={Italy}
		}
		\affiliation[d]{organization={Centre Tecnològic de Telecomunicacions de Catalunya (CTTC)},
			addressline={Avinguda Carl Friedrich Gauss, 7},
			city={Castelldefels,Barcelona},
			postcode={08860},
			state={Spain}
		}
		\affiliation[e]{organization={New York University},
			addressline={6 MetroTech Center},
			city={Brooklyn, New York},
			postcode={11201},
			state={USA}}
		
		\begin{abstract}
			Timely delivery of delay-sensitive information over dynamic, heterogeneous networks is essential for NextG interactive applications, yet providing strict \acf{e2e} peak latency guarantees remains an open challenge. Two obstacles limit the adoption of learning-based network control in this setting: traditional volume-based routing metrics, while highly effective for general traffic management, are not designed to capture traffic urgency; and \acf{drl} controllers trained from scratch suffer from sample inefficiency, long training times, and early-stage exploration volatility. This paper introduces a deployment-focused network control framework that addresses both obstacles. First, we present \textit{Effective Congestion (EC)}, a deadline-aware metric family that quantifies interface congestion by packet urgency and proactively filters non-viable traffic, coupled with a Uniform Path Grouping (UPG) distribution heuristic promoting robust load-balancing; the resulting policies are embedded into \textit{\acf{marlecps}}, a hybrid architecture combining a distributed scheduler with a centralized RL-based router. 
			Second, we introduce a unified training objective that generalizes existing policy-learning paradigms---behavioral cloning, offline \ac{rl}, online \ac{rl}, and offline-to-online schemes---as special cases, combining a live-reward term, a pre-collected-reward term, and a policy-imitation term. From this objective, we derive the \textit{\acf{mgarl}} protocol, instantiated on a \acf{ddpg} backbone: a deployment-oriented, demonstration-driven training approach that generalizes conventional \ac{o2o} schemes, in which trajectories from a lightweight prior-guided heuristic drive the offline pre-training of the RL-based router before online fine-tuning.
			Because the reference policy is analytical and queryable at any state, its imitation signal extends beyond the states seen offline to those visited online, while a pre-collected reward term keeps the Critic aligned with the Actor throughout; together, these jointly prevent the extrapolation error and Actor--Critic misalignment that otherwise arise across the offline-to-online transition, with the imitation weight decaying automatically as a function of the effective amount of live experience collected, rather than on a fixed or manually tuned schedule. Simulations across three structurally diverse topologies show that EC-based policies improve delivery reliability by up to 40\% over traditional volume-based routing, and that the proposed \ac{mgarl} protocol preserves early-stage stability while reducing online interaction cost by a factor of seven relative to from-scratch training, with the Vectorial congestion state representation striking the most favorable reliability-to-online-cost balance among the learning-based controllers.
		\end{abstract}
		
		
	\end{frontmatter}
	
	\section{Introduction}
	The rapid rise of \acf{rti} applications---such as connected autonomous vehicles, smart factories, and remote robotic control---has imposed unprecedented requirements on modern communication networks. Consequently, 5G and emerging 6G systems have evolved to meet not only high throughput but, especially, strict \ac{e2e} latency guarantees~\citep{cai2022compute}. In these environments, information loses its value if it is not delivered within application-imposed deadlines: the traditional paradigm of throughput maximization is no longer sufficient, and managing urgent traffic becomes a paramount objective for modern networked applications.
	
	To satisfy these strict requirements, NextG networks rely on \ac{e2e} service orchestration, jointly optimizing service placement, flow routing, and resource provisioning under application-specific \ac{qos} constraints---a process prominently abstracted by the \ac{cnf} framework~\citep{barcelo2016iot}, for which several polynomial-time centralized strategies have been proposed~\citep{Poularakis2020Mobihoc,mauro2024end}. Actualizing \ac{e2e} service delivery, however, requires pairing these macroscopic, long-term strategies with responsive dynamic control policies that fine-tune routing and scheduling at a faster timescale to cope with instantaneous network variations~\citep{pagliuca2024dual}.
	
	In the context of dynamic network control, the literature has progressively moved from throughput-optimal but loop-prone control (BP, LDP), to acyclic, delay-aware routing (UMW, UCNC), and finally to explicit per-packet deadline guarantees (RCNC). Foundational algorithms like \ac{bp}~\citep{tassiulas1990stability} and \ac{ldp}~\citep{neely2022stochastic} provide fully distributed, throughput-optimal routing and scheduling, with \ac{dcnc}~\citep{feng2018optimal} extending the \ac{ldp} approach to jointly optimize flow processing and routing across distributed cloud networks. However, their purely queue-driven nature often leads to routing loops and high average delays. Hybrid architectures combining centralized routing with distributed scheduling, such as \ac{umw}~\citep{sinha2017optimal} and \ac{ucnc}~\citep{zhang2021optimal}, enforce acyclic path selection, eliminating routing loops and reducing the expected \ac{e2e} delay---yet they optimize only for average metrics (e.g., delay, throughput). As a pivotal attempt to move beyond average performance, \ac{rcnc}~\citep{cai2022ultra} integrated packet lifetime constraints into the \ac{ldp} framework, tracking deadlines on a per-packet basis to improve reliability. Nevertheless, per-packet deadline constraints do not admit the time-average structure that \ac{ldp} optimization relies upon, rendering \ac{ldp}-based solutions computationally expensive and endowed with weaker performance guarantees in this setting.
	
	On the other hand, the \ac{dcmt} problem~\citep{vitale2025flexible} provides a mathematical representation of the \ac{rti} application requirements, that can be formulated as a \ac{mdp}, making reinforcement learning approaches a highly viable alternative. In~\citep{vitale2025flexible}, a \ac{madrl} framework combining a distributed scheduler with a centralized routing agent---the same hybrid architectural principle adopted throughout this work---demonstrated the strong potential of this paradigm, outperforming baselines like \ac{umw} in deadline-constrained scenarios. Their results, achieved with the introduction of a new family of deadline-aware \emph{scheduling} agents, combined with computationally efficient \emph{routing} agents relying on traditional queue-occupancy metrics, paved the way for the definition of novel deadline-aware occupancy metrics aimed at improving performance with the lowest impact on computational efficiency.
	
	Despite the potentially high performances, \ac{madrl} algorithms that learn entirely \textit{from scratch} via online trial-and-error exploration (whether simulated or real) usually exhibit high computational complexity and low sample efficiency~\citep{levine2020offline}: the potentially massive volume of interactions required to discover an effective policy, which depends on the specific problem instance, translates into unacceptable training times and poor initial decisions that hinder practical deployment~\citep{prudencio2024survey}, calling for a reshaping of standard training pipelines~\citep{ball2023efficient}. Leveraging stable prior-guided policies to pre-train and accelerate the convergence of \ac{madrl} agents reduces the initial exploration phase and favors reliable performance from the very first interaction~\citep{nair2020awac}. This approach falls under the broader paradigm of \emph{\acf{rlfd}}~\citep{schaal1997learning}, where the policy search of the learning agent is guided by externally generated trajectories. \ac{dqfd}~\citep{hester2018deep} provides an influential instantiation of \ac{rlfd}, pre-training a deep Q-learning agent on demonstrations that remain in the replay buffer throughout online learning. While the application of \ac{rl} to Traffic Engineering has been extensively surveyed~\citep{riosguiral2025leveraging}, the use of \ac{rlfd} principles in deadline-constrained network control remains, to the best of our knowledge, largely unexplored.
	
	This work contributes in both directions by introducing novel deadline-aware congestion metrics and extending the \ac{madrl} framework of~\citep{vitale2025flexible} with a deployment-focused training methodology. Concretely, the proposed system results from the composition of three largely independent design choices---a congestion metric, a control architecture, and a training protocol---whose contributions are summarized as follows:
	
	\begin{enumerate}
		\item \textbf{Congestion metric.} We introduce \acf{ec}, a family of deadline-aware congestion metrics tailored for \ac{rti} services. Unlike volume-based congestion, which counts all queued packets indiscriminately, \ac{ec} restricts an interface's congestion to only those packets that will actually compete with a given packet under \ac{lelf} scheduling---filtering out packets that would have already expired by the time of arrival, as well as packets with lower priority---yielding a congestion estimate that is both path- and urgency-aware. To control the acquisition cost of this path-dependent metric, we further introduce an approximate \emph{$p^*$-model}, which anchors the evaluation of each interface to a single reference path rather than to every path traversing it, reducing the observation dimensionality from one value per (interface, path) pair to one value per interface, with a negligible loss in accuracy. We leverage both the exact and approximate \ac{ec} formulations, in scalar and vectorial form, to design new urgency-grounded, prior-guided routing strategies.
		\item \textbf{Prior-guided routing policies.} Building on the proposed \ac{ec} metrics, we design a family of five model-based routing policies obtained by combining two orthogonal design choices: the congestion type (regular, EC~$p$, or EC~$(p^*)$) and the path-assignment strategy (greedy, or \acf{upg}, which balances traffic across paths of comparable weight rather than funnelling it onto a single least-congested path). We show that \ac{upg} consistently prevents the localized hotspot formation that affects greedy assignment under tight deadlines or near saturation, and that its combination with the computationally lightweight EC~$(p^*)$ metric, \ac{upgecps}, achieves near-optimal reliability at a fraction of the state-space complexity of the exact formulation---making it, in turn, the natural reference policy for \ac{mgarl}.
		\item \textbf{Control architecture.} We extend the \ac{madrl} framework of~\citet{vitale2025flexible} into \acf{marlecps}: a hybrid architecture combining a distributed \ac{lelf} scheduler---which forwards packets at each interface in strict order of increasing \ac{el}---with a centralized routing agent, whose observation space incorporates the proposed \ac{ec} metrics, turning the urgency awareness that~\citet{vitale2025flexible} encode in the queueing dynamics into an explicit, spatio-temporally filtered congestion observation for the routing agent. The agent returns, for each commodity, a continuous split of traffic across its feasible paths, enforced to satisfy per-commodity flow conservation by construction; its observation space can be instantiated in either a compact scalar form or a higher-dimensional vectorial form that retains the full per-lifetime urgency profile at each interface, allowing us to trade observation expressiveness for policy complexity.
		\item \textbf{Training protocol.} We introduce the \acf{gpr}, a single training objective that unifies existing policy-learning paradigms---behavioral cloning, offline \ac{rl}, online \ac{rl}, and offline-to-online schemes---as special cases governed by three coefficients weighting live reward, pre-collected reward, and policy imitation, respectively. From the \ac{gpr}, we derive \acf{mgarl}, a deployment-oriented protocol that pre-trains the routing agent of \ac{marlecps} on an initially sub-optimal, computationally light prior-guided policy before transitioning to online fine-tuning. Unlike prior offline-to-online methods, which fix the imitation weight or release it on a manually tuned schedule, \ac{mgarl} decays it automatically as a function of the effective amount of live experience collected, avoiding both an abrupt phase boundary and the need for manual tuning of a training-time schedule. Because the reference policy is analytical and therefore queryable at any state, the imitation signal is not confined to the states seen offline, but extends to those the agent visits online as well; combined with the pre-collected reward term, which keeps the Critic aligned with the Actor before and throughout online interaction, this jointly prevents the extrapolation error and Actor--Critic misalignment that otherwise arise when transitioning from offline pre-training to online fine-tuning.
		
		\item \textbf{Empirical validation.} Through simulations spanning three complementary structural regimes---hierarchical concentration, backbone asymmetry, and symmetric path diversity---we show that the proposed \ac{ec} metrics yield reliability gains of up to 40\% over traditional volume-based routing, and that \ac{mgarl} preserves competitive reliability while reducing online interaction cost by a factor of seven with respect to from-scratch training. Among the two congestion-state representations enabled by \ac{ec}, the Vectorial form---which retains the full per-lifetime urgency profile at each interface---consistently matches or exceeds the reliability of fully online training, whereas the more compact Scalar form trades a modest reliability margin for faster convergence and lower state-space complexity; \ac{mgarl} narrows this gap further, achieving comparable reliability under both representations while cutting the online interaction budget by the same factor of seven.
	\end{enumerate}
	
	The remainder of the paper is organized as follows: Section~\ref{sec:rl_background} reviews the fundamentals of Reinforcement Learning, and Section~\ref{sec:actor_critic_framework} develops the actor--critic backbone and the empirical reward estimators used throughout this work. Section~\ref{sec:system_model} introduces the system model and the deadline-aware queueing dynamics. Section~\ref{ssec:model_data_driven_rl_dcs} presents the key concepts and the novel \ac{ec} metrics, and Section~\ref{sec:model_based_policies} details the proposed prior-guided routing policies. Section~\ref{sec:policy_learning_paradigms} introduces the \ac{gpr} and the \ac{mgarl} framework, which Section~\ref{sec:mga_rl_instantiation} instantiates for deadline-constrained routing as \ac{marlecps}. The experimental setting and numerical results are discussed in Sections~\ref{sec:experimental_setting} and~\ref{sec:numerical_results}, respectively. Finally, Section~\ref{sec:conclusions} concludes the paper and outlines future directions. To preserve the narrative, an extended analysis of the simulation results is reported in~\ref{sec:appendinx_spatial_drop}, while a table of notations is provided in~\ref{sec:appendix_table}.

	
	\section{Reinforcement Learning: Background}
	\label{sec:rl_background}
	
	Since the \ac{dcmt} problem is formulated as a \acf{mdp}, 
	we begin by reviewing the fundamental concepts of RL that underlie the multi-agent extension developed in Section \ref{sec:mga_rl_instantiation}.
	Specifically, we first outline the core elements of \ac{rl}, including value functions and policy optimization, and then introduce the actor--critic backbone adopted in this work.
	
	\subsection{Reinforcement Learning and Value Functions}
	\label{sec:rl_value_functions}
	
	\ac{rl} provides a powerful framework for modelling sequential decision-making problems in complex and uncertain environments. The interaction between a learning agent and its environment is commonly formalised as an \ac{mdp}~\citep{bellman1957markovian}. An agent observes the current state $s(t)\in\mathcal{S}$ and selects an action $a(t)\in\mathcal{A}$ according to a stochastic policy $\pi(a(t)\mid s(t))$. Upon executing the action, the agent receives a scalar reward $r(t)$, which is a function of the current state and action, reflecting the immediate utility of the decision. Concurrently, the environment transitions to a new state $s(t+1)$, according to the probability distribution $P(s(t+1)\mid s(t),a(t))$. Under a stationary \ac{mdp}, the probability of each possible pair of next state and reward, $s'$ and $r$, is denoted as
	\begin{equation}
		p(s',r\mid s,a) =
		\Pr\bigl\{s(t+1)=s',\;r(t)=r\mid s(t)=s,\;a(t)=a\bigr\}.
	\end{equation}
	
	The overarching objective in \ac{rl} is to discover a policy $\pi^*$ that maximises the expected cumulative reward, formalised as the expected discounted return \begin{equation}
		\label{eq:expectedreturn}
		J=\mathbb{E}[R(\xi)],
	\end{equation}  
	with
	\begin{equation}
		\label{eq:return}
		R(\xi) = \sum_{t=0}^{T} \gamma^t\,r(t),
	\end{equation}
	$\gamma$ denoting the discount factor and $\xi=\{s(t),a(t)\}_{t=0}^{T}$ the state--action trajectory sampled under $\pi$.
	
	The state-value and action-value (Q) functions under policy $\pi$ are:
	\begin{align}
		V^\pi(s) &= \mathbb{E}\bigl[R(\xi)\mid s_0=s\bigr], \label{eq:vfunction}\\
		Q^\pi(s,a) &= \mathbb{E}\bigl[R(\xi)\mid s_0=s,\;a_0=a\bigr]. \label{eq:qfunction}
	\end{align}
	Given an optimal policy $\pi^*$, the associated optimal action-value function $Q^*(s,a)$ 
	satisfies the \emph{Bellman optimality equation}:
	\begin{equation}
		\label{eq:bellman}
		Q^*(s,a) = \mathbb{E}\!\left[r + \gamma\max_{a'}Q^*(s',a')\right].
	\end{equation}
	For finite stationary \acp{mdp}, an optimal policy always exists in deterministic form and is obtained as:
	\begin{equation}
		\label{eq:optimal_policy}
		\pi^*(s) = \arg\max_{a\in\mathcal{A}}\,Q^*(s,a).
	\end{equation}
	Therefore, the optimal policy can be derived directly from $Q^*$ without requiring an explicit policy representation.
	
	\subsection{Practical Limitations and the Actor--Critic Backbone}
	\label{sec:actor_critic_intro}
	Two obstacles prevent the direct application of Eq.~\eqref{eq:bellman} in real-world network control. First, the transition dynamics $P(s(t+1)\mid s(t),a(t))$ are unknown in general: the Bellman backup cannot be evaluated in closed form, and the agent must learn from sampled tuples $(s(t),a(t),r(t),s(t+1))$ collected by interacting with the environment. This is the \emph{model-free} setting adopted throughout this work. Second, for the continuous, high-dimensional state--action spaces of routing and scheduling tasks, tabular representations of $Q^*(s,a)$ are infeasible, and solving $\pi^*(s)=\arg\max_{a\in\mathcal{A}}Q^*(s,a)$ at each step is computationally intractable for continuous $\mathcal{A}$.
	
	These obstacles motivate \emph{actor--critic architectures}~\citep{sutton1999policy,silver2014deterministic}: a parametric actor $\mu_\theta$ directly outputs actions and a parametric critic $Q_\phi$ estimates their value, bypassing the explicit maximization over $\mathcal{A}$. This work adopts \acf{ddpg}~\citep{lillicrap2015continuous} as the backbone; the full actor--critic derivation, the DDPG policy gradient, and the empirical estimation of all reward signals used in training are developed in Section~\ref{sec:actor_critic_framework}.
	While the formulation above considers a single decision-making agent, Section ~\ref{sec:mga_rl_instantiation} extends this actor–critic backbone to the multi-agent setting, coordinating a centralized routing agent with distributed per-interface scheduling agents within the \ac{madrl} architecture of \citep{vitale2025flexible}.

	\section{Reward Estimation and the Actor--Critic Framework}
	\label{sec:actor_critic_framework}
	
	Section~\ref{sec:actor_critic_intro} identified two obstacles to a direct application of Eq.~\eqref{eq:bellman}: the transition kernel $P(s(t+1)\mid s(t),a(t))$ is unknown, and the state--action space is continuous and high-dimensional. The first obstacle is not specific to Eq.~\eqref{eq:bellman}: since $P$ is unknown, it equally prevents the closed-form evaluation of the individual expected reward terms in Eq.~\eqref{eq:return}. This section addresses both obstacles in turn, following the approach usually taken in the literature~\citep{sutton1998reinforcement}. Specifically, Section~\ref{sec:reward_estimation} shows how every such expectation — whether the reward terms of Eq.~\eqref{eq:return} or the Bellman backup of Eq.~\eqref{eq:bellman} — is replaced by a sample average computed over a finite ensemble of observed transitions, a single estimation principle applied consistently throughout this work, independently of any specific learning architecture. Section~\ref{sec:actor_critic_backbone} then introduces the actor--critic architecture that resolves the second obstacle.
	
	\subsection{Reward Estimation}
	\label{sec:reward_estimation}
	
	As previously noted, $P(s(t+1)\mid s(t),a(t))$ is unknown in general — as is the case in this work — and consequently the expectations required by Eq.~\eqref{eq:bellman} cannot be evaluated in closed form. The standard approach in the literature is to approximate any such expectation by a \emph{sample average} (a Monte Carlo estimate) computed over a finite ensemble of transition tuples $(s,a,r,s')$, collected either through direct interaction with the environment or gathered offline~\citep{sutton1998reinforcement,mnih2015human}. This subsection reviews the data structures that supply these ensembles and defines the resulting empirical estimators used throughout this work.
	
	\subsubsection{Replay Buffer and Pre-collected Dataset}
	Following~\citep{lillicrap2015continuous,mnih2015human}, transitions are stored in a circular replay buffer of fixed capacity $N_{\text{buf}}$: each new tuple $(s,a,r,s')$ is appended and, once full, the oldest entry is overwritten. Let $t\in\mathbb{N}$ count the total number of transitions collected. The buffer at step $t$ contains:
	\begin{equation}
		\label{eq:replay_buffer}
		\mathcal{D}(t) = \left\{ (s(\tau),a(\tau),r(\tau),s(\tau+1)) : \max(0,t-N_{\text{buf}}) \le \tau < t \right\}.
	\end{equation}
	
	For $t < N_{\text{buf}}$ the buffer is partially filled; for $t \ge N_{\text{buf}}$ it holds the $N_{\text{buf}}$ most recent transitions. At each gradient step, a mini-batch $\mathcal{B}(t) \subset \mathcal{D}(t)$ of size $N$ is drawn uniformly at random~\citep{mnih2015human}.
	
	Independently, a pre-collected dataset $\mathcal{D}_{\text{off}}$ of $N_{\text{off}}$ fixed transitions may be available, gathered prior to any live environment interaction by executing a reference policy $\mu^{\mathrm{MB}}$ --- sometimes referred to in the literature as a \emph{demonstrator}.
	
	At each gradient step, a mini-batch $\mathcal{B}_{\text{off}} \subset \mathcal{D}_{\text{off}}$ of size $M$ is drawn uniformly. Since $\mathcal{D}_{\text{off}}$ is fixed, its sampling distribution is stationary, in contrast to the non-stationary, policy-dependent distribution underlying $\mathcal{D}(t)$. Throughout this work, $N$ and $M$ denote the sizes of $\mathcal{B}(t)$ and $\mathcal{B}_{\text{off}}$, respectively; unlike the replay buffer capacity $N_{\text{buf}}$ and the fixed dataset size $N_{\text{off}}$, these mini-batch sizes are training hyperparameters (see Section~\ref{sec:mga_rl_instantiation}).
	
	\subsubsection{Empirical Reward and Policy-Deviation Estimates}
	Recalling Eq.~\eqref{eq:return}, the expected discounted return can be written, by linearity of expectation, as a sum of individual expected reward terms:
	\begin{equation}
		\label{eq:return_expectation}
		J = \mathbb{E}[R(\xi)] = \mathbb{E}\left[\sum_{t=0}^{T} \gamma^t r(t)\right] = \sum_{t=0}^{T} \gamma^t\, \mathbb{E}[r(t)].
	\end{equation}
	Since $P(s(t+1)\mid s(t),a(t))$ is unknown, each term $\mathbb{E}[r(t)]$ in Eq.~\eqref{eq:return_expectation} cannot be evaluated in closed form, and is instead approximated by a sample average computed over a mini-batch of observed transitions. Substituting the sample-average estimator for $\mathbb{E}[r(t)]$ at each $t$ yields the empirical estimate of $J$:
	\begin{equation}
		\label{eq:return_estimate}
		\hat{J} = \sum_{t=0}^{T} \gamma^t\, \hat{r}(t),
	\end{equation}
	where $\hat{r}(t)$ denotes the sample-average estimator of $\mathbb{E}[r(t)]$ at step $t$, computed over the mini-batch available at that step. Depending on the provenance of that mini-batch, $\hat{r}(t)$ takes one of two forms: $\hat{r}_{\text{on}}(t)$, when the mini-batch is drawn from the live replay buffer $\mathcal{D}(t)$, and $\hat{r}_{\text{off}}(t)$, when it is drawn from the stationary pre-collected dataset $\mathcal{D}_{\text{off}}$. We define each in turn.
	\begin{itemize}
		\item\textbf{Reward from Interaction.} When the mini-batch is drawn from $\mathcal{D}(t)$, the estimator of $\mathbb{E}[r(t)]$ takes the form
		\begin{equation}
			\label{eq:r_on}
			\hat{r}_{\text{on}}(t) = \frac{1}{N} \sum_{(s,a,r,s') \in \mathcal{B}(t)} r.
		\end{equation}
		As $t$ grows, $\mathcal{D}(t)$ (Eq.~\eqref{eq:replay_buffer}) covers a broader and more recent distribution of states under the current $\mu_\theta$, so $\hat{r}_{\text{on}}(t)$ tracks reward performance of the evolving policy.
		\item\textbf{Reward from Pre-collected Data.} When the mini-batch is instead drawn from the stationary dataset $\mathcal{D}_{\text{off}}$, the estimator of $\mathbb{E}[r(t)]$ takes the form
		\begin{equation}
			\label{eq:r_off}
			\hat{r}_{\text{off}}(t) = \frac{1}{M} \sum_{(s,a,r,s') \in \mathcal{B}_{\mathrm{off}}} r.
		\end{equation}
		Although $\mathcal{D}_{\text{off}}$ is fixed and its sampling distribution stationary, $\hat{r}_{\text{off}}(t)$ retains the index $t$ for notational consistency with $\hat{r}_{\text{on}}(t)$, since both are combined at each training step $t$ in the unified objective of Section~\ref{sec:mga_rl_instantiation}.
	\end{itemize}
	The two estimators above both target $\mathbb{E}[r(t)]$, and therefore both serve reward maximisation. Rather than relying solely on a reward signal to define the optimal policy, an alternative and widely adopted paradigm in the literature instead trains the agent to imitate a given prior-guided policy, minimising the discrepancy between the learned policy $\mu_\theta$ and a reference policy $\mu^{\text{MB}}$ rather than maximising an accumulated reward. This second paradigm requires an estimator of a fundamentally different quantity: not a reward term, but a policy discrepancy.
	
	\textbf{Policy Deviation.} 
	Given the reference policy $\mu^{\text{MB}}$, the sample-average discrepancy between $\mu_\theta$ and $\mu^{\text{MB}}$ over the pre-collected dataset is
	\begin{equation}
		\label{eq:mse_estimate}
		\hat{\mathcal{M}}_{\text{MSE}} = \frac{1}{N_{\mathrm{off}}} \sum_{(s,a,r,s')\in\mathcal{D}_{\mathrm{off}}} \left\lVert \mu_\theta(s) - \mu^{\text{MB}}(s) \right\rVert^2,
	\end{equation}
	where $\mu^{\text{MB}}$ is an analytical reference policy queryable at any $s\in\mathcal{S}$ without additional environment interaction. Unlike $\hat{r}_{\text{on}}(t)$ and $\hat{r}_{\text{off}}(t)$, which estimate a reward term of Eq.~\eqref{eq:return_expectation}, $\hat{\mathcal{M}}_{\text{MSE}}$ measures a discrepancy between $\mu_\theta$ and $\mu^{\text{MB}}$, evaluated over the entirety of the fixed, fully available dataset $\mathcal{D}_{\text{off}}$ rather than a resampled mini-batch. consistently with its role as a stable anchor to the reference policy.
	
	Having introduced both paradigms --- reward maximisation, driven by the sample-average estimators $\hat{r}_{\text{on}}(t)$ and $\hat{r}_{\text{off}}(t)$, and policy imitation, driven by the sample-average estimator $\hat{\mathcal{M}}_{\text{MSE}}$ --- Sections~\ref{sec:policy_learning_paradigms} and~\ref{sec:mga_rl_instantiation} unify them into a single training objective (the \ac{mgarl} training objective), in which all three sample-average quantities are combined.
	
	\subsection{Actor--Critic Framework}
	\label{sec:actor_critic_backbone}
	Section~\ref{sec:reward_estimation} addressed the first obstacle of Section~\ref{sec:actor_critic_intro} --- $P(s(t+1)\mid s(t),a(t))$ unknown --- which prevents $Q^{\mu_\theta}(s,a)$ from being evaluated in closed form and necessitates estimating it from sampled transitions instead. The second obstacle --- the continuous, high-dimensional state--action space of routing and scheduling tasks --- further rules out tabular representations of $Q^*(s,a)$ and makes $\arg\max_{a\in\mathcal{A}}Q^*(s,a)$ intractable at every step.
	
	The actor $\mu_\theta: \mathcal{S}\to\mathcal{A}$ and critic $Q_\phi: \mathcal{S}\times\mathcal{A}\to\mathbb{R}$ are parametric functions, implemented as deep neural networks, that jointly resolve both obstacles: $Q_\phi$ is a learned estimate of $Q^{\mu_\theta}(s,a)$, updated via temporal-difference bootstrapping from sampled tuples rather than computed exactly, addressing the first obstacle; its parametrisation as a neural network, rather than a lookup table, together with the actor $\mu_\theta$ directly outputting actions and bypassing the explicit maximisation over $\mathcal{A}$, addresses the second. Both networks are updated from a mini-batch of transition tuples, without requiring knowledge of $P(s'\mid s,a)$. Depending on data availability, this mini-batch is drawn from $\mathcal{B}(t)$, from $\mathcal{B}_{\text{off}}$ --- both defined in Section~\ref{sec:reward_estimation} --- or, as in this work, from a mixture of both, whose composition is specified in Section~\ref{sec:mga_rl_instantiation}. In what follows, we denote the mini-batch used at a given update generically as $\mathcal{B}$, leaving its specific instantiation --- drawn from $\mathcal{B}(t)$, from $\mathcal{B}_{\text{off}}$, or from both --- to context.
	
	We stress that the Actor and Critic updates presented in this subsection belong entirely to the reward-maximisation paradigm of Section~\ref{sec:reward_estimation}: both rely on the reward $r$ observed in each sampled tuple, and neither makes use of the reference policy $\mu^{\text{MB}}$ or the policy-deviation estimator $\hat{\mathcal{M}}_{\text{MSE}}$. How the two paradigms are combined into a single set of parameter updates is deferred to Section~\ref{sec:mga_rl_instantiation}.

	\textbf{Actor Update.} Recalling the objective $J=\mathbb{E}[R(\xi)]$ of Eq.~\eqref{eq:return}, the actor-parameterised form is $J(\theta) = \mathbb{E}_{\mathcal{B}}[Q_\phi(s,\mu_\theta(s))]$, where the expectation is taken with respect to the empirical distribution of states induced by the sampled mini-batch $\mathcal{B}$. By the Deterministic Policy Gradient (DPG) theorem~\citep{silver2014deterministic}, its gradient with respect to $\theta$ is:
	\begin{equation}
		\label{eq:dpg_gradient}
		\nabla_\theta J(\theta) = \mathbb{E}_{\mathcal{B}}\Bigl[\nabla_\theta \mu_\theta(s)\cdot\nabla_a Q_\phi(s,a)\big|_{a=\mu_\theta(s)}\Bigr].
	\end{equation}
	In practice, Eq.~\eqref{eq:dpg_gradient} is estimated via stochastic gradient ascent over the mini-batch $\mathcal{B}$~\citep{lillicrap2015continuous}:
	\begin{equation}
		\label{eq:dpg_gradient_sample}
		\nabla_\theta J(\theta) \approx \frac{1}{|\mathcal{B}|}\sum_{s\in\mathcal{B}} \nabla_\theta \mu_\theta(s)\cdot\nabla_a Q_\phi(s,a)\big|_{a=\mu_\theta(s)},
	\end{equation}
	with the actor parameters updated by gradient ascent, i.e.
	\begin{equation}
		\label{eq:actor_ascent}
		\theta \leftarrow \theta + \eta_\theta \nabla_\theta J(\theta),
	\end{equation}
	where $\eta_\theta$ is the actor learning rate.
	
	\textbf{Critic Update.} The critic is updated by minimising the mean squared temporal-difference (TD) error~\citep{sutton1998reinforcement} over the mini-batch $\mathcal{B}$. For each sampled tuple $(s,a,r,s')\in\mathcal{B}$, define the TD target
	\begin{equation}
		\label{eq:td_target}
		y(r,s') = r + \gamma\, Q_{\phi'}\bigl(s',\mu_{\theta'}(s')\bigr),
	\end{equation}
	where $Q_{\phi'}, \mu_{\theta'}$ are target networks — exponentially smoothed copies of the critic and actor that stabilise the bootstrap target~\citep{lillicrap2015continuous}. The critic loss is
	\begin{equation}
		\label{eq:critic_loss0}
		L(\phi) = \frac{1}{|\mathcal{B}|}\sum_{(s,a,r,s')\in\mathcal{B}} \bigl(y(r,s') - Q_\phi(s,a)\bigr)^2,
	\end{equation}
	with the critic parameters updated by gradient descent,
	\begin{equation}
		\label{eq:critic_update0}
		\phi \leftarrow \phi - \eta_\phi\, \nabla_\phi L(\phi),
	\end{equation}
	where $\eta_\phi$ is the critic learning rate. The TD target of Eq.~\eqref{eq:td_target} depends only on the observed tuple $(s,a,r,s')$: no model of $P$ is required.
	
	\emph{Remark.} The Actor and Critic updates of Eqs.~\eqref{eq:dpg_gradient}--\eqref{eq:critic_update0} address only reward maximisation. Neither update makes use of the policy-deviation estimator $\hat{\mathcal{M}}_{\text{MSE}}$ of Section~\ref{sec:reward_estimation}, which is unrelated to reward maximisation. How all three estimators of Section~\ref{sec:reward_estimation} are instead combined into a unified training objective is the subject of Section~\ref{sec:gpr_to_actor_critic} and Section~\ref{sec:mga_rl_instantiation}.

	\section{System Model}
	\label{sec:system_model}
	Given the nature of modern time-critical applications, it is necessary to define network, service, and queuing models capable of expressing their stringent requirements. Grounded in a network control architecture that combines centralized routing with distributed scheduling \citep{vitale2025flexible}, we describe (i)  the network and service models, and associated parameters, (ii) the network control variables, and (iii) the queuing model that captures the dynamics of deadline-aware queues at each communication interface. Finally, we introduce two concepts that are crucial to the design of effective deadline-aware control policies: \textit{\ac{el}} and \textit{\ac{lelf}} scheduling.
	
	\subsection{Network and service parameters}
	\label{sec:network_and_service_parameters}
	
	We consider a communications network described by a directed graph \( \mathcal{G} = (\mathcal{V}, \mathcal{E}) \), where \( \mathcal{V} \) and \( \mathcal{E} \) denote the set of nodes and links, respectively.
	We use $\rho_{i}^{+} \subset \mathcal{V}$ and $\rho_{i}^{-} \subset \mathcal{V}$ to denote the sets of outgoing and incoming neighbors of node $i \in \mathcal{V}$, respectively.
	
	The system operates in time-slotted fashion, with equally-sized slots, indexed by \( t \in \{0, 1, \ldots \} \). For each link \( (i, j) \in \mathcal{E} \), the link capacity \( C_{ij}(t) \) specifies the maximum number of packets that can be transmitted over link \( (i, j) \) during time slot \( t \).
	
	The network supports latency-sensitive services that require the timely delivery of packets across multiple source--destination pairs. \ac{qos} requirements are imposed by associating a maximum \emph{lifetime}, or \ac{ttl}, with each service packet. A packet with a positive lifetime is considered effective and continues to flow through the network. Conversely, a packet is considered outdated once its lifetime reaches zero, and is immediately dropped from the network.
	
	We identify latency-sensitive services as a set of commodities, where each commodity \( c \in \mathcal{C} \) is associated with:
	\begin{itemize}
		\item  a source node \( s^{c} \in \mathcal{V} \),
		\item a destination node \( d^{c} \in \mathcal{V} \), and
		\item an initial lifetime \( L^c \in \mathcal{L} = \{1, \dots, L_{max}\} \).
	\end{itemize}
	Note that a lower initial lifetime $L^{c}$ indicates a more latency-sensitive service. 
	The stochastic number of commodity-$c$ packets arriving at source node $s^{c}$  at time $t$ is denoted by $b^{c}(t)$, with $\bar{b}^{c} = \mathbb{E} \big[ b^{c}(t) \big]$ denoting its mean arrival rate. We use $\bm{b}(t) \triangleq \big\{b^{c}(t), \forall c \in \mathcal{C}\big\}$ to denote the packet arrival vector at time $t$.
	
	\subsection{Network control variables}
	We adopt a network control architecture characterized by a hybrid \emph{centralized routing} and \emph{distributed scheduling} paradigm \citep{vitale2025flexible}. In this architecture, a centralized routing agent dynamically assigns feasible paths to newly arrived packets based on global congestion state information.
	Simultaneously, distributed scheduling agents located at each network interface independently prioritize packet transmissions based on local congestion states and packet lifetimes.
	
	Consistent with the \emph{centralized routing} nature of the architecture, we define the set of candidate paths for each commodity $c \in \mathcal{C}$ as $\mathcal{P}^c$. This set consists of all feasible\footnote{A path is feasible for commodity~$c$ if its length does not exceed the initial lifetime of commodity~$c$. Formally, the set of feasible paths for commodity~$c$ consists of all source--destination paths whose traversal time, in the absence of queuing delays, is within the initial lifetime of commodity~$c$. This set depends only on the network topology and the commodity's attributes (initial lifetime and source--destination pairs) and is computed offline.} paths from the source to the destination of commodity $c$. The set of all commodity paths is denoted by $\mathcal{P}= \bigcup_{c \in \mathcal{C}} \mathcal{P}^c$. Furthermore, we define $\mathcal{P}_{ij} = \{ p \in \mathcal{P} \mid (i, j) \in p \}$ as the subset of commodity paths that traverse link $(i, j)$, and $\mathcal{P}_{ij}^{c} = \{ p \in \mathcal{P}^c \mid (i, j) \in p \}$ as the subset of paths for commodity $c$ that traverse link $(i, j)$.
	We then denote by $b^c_{ij}(t)$ the number of commodity-$c$ packets that exogenously arrive at node $i$ and are assigned to a path in $\mathcal{P}_{ij}^{c}$ at time $t$.
	
	The routing and scheduling decisions are encoded into flow variables \( f_{ij}^{(c, l)}(t) \). Here, \( f_{ij}^{(c, l)}(t) \) denotes the number of packets of commodity \( c \in \mathcal{C} \) with lifetime \( l \) transmitted over link \( (i, j) \in \mathcal{E} \) at time \( t \). Furthermore, $\mathbf{f}(t) \triangleq \big\{ f_{ij}^{(c,l )}(t), \forall (i,j) \in \mathcal{E}, \forall c\in\mathcal{C}, l \in \mathcal{L} \big\}$ denotes the network-wide set of flow variables at time \( t \).
	
	To account for scheduling policies that actively drop packets from the network before their lifetime expires, we introduce the packet-dropping variables \( g_{ij}^{(c, l)}(t) \). These variables represent the number of commodity-$c$ packets with lifetime \( l \) that are intentionally dropped at interface $(i,j)$ during time slot \( t \).
	The complete set of packet-dropping variables across the network at time \( t \) is defined as:
	\[
	\mathbf{g}(t) \triangleq \big\{ g_{ij}^{(c, l)}(t) \mid (i,j) \in \mathcal{E}, c \in \mathcal{C}, l \in \mathcal{L} \big\}.
	\]
	
	\subsection{Deadline-Aware Queueing Model}
	
	Note that given the centralized routing and distributed scheduling architecture, when a packet arrives at a network node, it immediately gets queued at the outgoing interface corresponding to the assigned path. Accordingly, we denote with $q_{ij}^{(c,l)}(t)$ the number of packets of commodity $c$ with lifetime $l$ queued at interface $(i,j)$ (the outgoing interface of node $i$ to communicate with node $j$), and with $\mathbf{q}_{ij}(t) \triangleq \{ q_{ij}^{(c,l)}(t), c\in \mathcal{C}, l \in 1 \ldots L^{c} \}$ the queuing state of interface $(i,j)\in\mathcal E$.

	\noindent We can then characterize the deadline-aware queue dynamics as: 
	\begin{align}
		& q_{ij}^{(c, l)}(t) =  q_{ij}^{(c,l + 1)}(t-1) \! - \! f_{ij}^{(c,l + 1)}(t-1) - g_{ij}^{(c,l+1)}(t-1) \notag\\
		& +  f_{\rightarrow ij}^{(c,l + 1)}(t-1) + b_{ij}^{(c,l)}(t), 
		\quad \forall (i,j) \in \mathcal{E}, \forall l \in \mathcal{L},\forall c \in \mathcal{C}, \forall t \label{eq:lifetime-queue}
	\end{align}
	with $b_{ij}^{(c,l)}(t)$ denoting the total number of packets exogenously arriving at node $i$ that are assigned a path in $\mathcal{P}_{ij}^{c}$, and $f_{\rightarrow ij}^{(c, l)}(t)$ denoting the total number of packets arriving at node $i$ from the set of incoming neighbors $\rho_{i}^{-}$ that are assigned a path in $\mathcal{P}_{ij}^{c}$.
	
	We assume that packets with zero lifetime, namely {\em expired packets}, are immediately dropped from the queue backlog:
	\begin{align} \label{eq:expired-packets}
		q_{ij}^{(c,0)}(t) &= 0, &&\forall (i,j) \in \mathcal{E}, \forall c \in \mathcal{C} , \forall t
	\end{align}
	while packets reaching the destination are immediately consumed:
	\begin{align}
		q_{i \, d_{c}}^{(c,l)}(t) &= 0, && d_{c} \in \mathcal{V}, 
		\forall i \in  \rho_{d_{c}}^{-}, \forall l \in \mathcal{L}, \forall c \in \mathcal{C}, \forall t
		\label{eq:packets-destination}
	\end{align}

	\subsection{Effective Lifetime}
	\label{ssec:effective_lifetime}
	
	In deadline-aware systems, the lifetime of a packet, i.e., its TTL, plays a crucial role in determining its urgency for transmission. To effectively capture such urgency, \citep{vitale2025flexible} introduced the concept of \emph{\ac{el}}. The \ac{el} denotes the maximum number of time slots a packet can wait in a network queue before having no chance to reach its destination over its assigned path.
	It is computed as the difference between the packet's lifetime and the minimum time required to reach its destination over its assigned path. 
	Mathematically, the \ac{el} of a packet assigned to path $p$, currently at node $i$, with \emph{absolute} lifetime $l$, denoted as $EL(p,l,i)$, is computed as:  
	\begin{equation}
		\label{eq:effective_lifetime}
		EL(p,l,i)= l - dist(i,p) + 1
		\,\, \forall p \in \mathcal{P}, \forall i \in p , 
	\end{equation}
	where $dist(i,p)$ denotes the distance between node $i$ and the destination of path $p$,  which, under the assumption of one time slot per network hop, corresponds to the minimum time required to reach the destination following the assigned path. 
	
	Compared to the {\em absolute} lifetime $l$, which decreases every time slot irrespective of network actions, the {\em effective} lifetime $\ell$:
	\begin{itemize}
		\item Decreases only when packets wait in a network queue, and remains unchanged when packets traverse a network link. 
		\item Is always smaller or equal than the absolute lifetime, $\ell \leq l$.
	\end{itemize}
	
\begin{figure}[htb]
		\centering
		\begin{subfigure}{0.48\textwidth}
			\centering
			\includegraphics[width=0.8\linewidth]{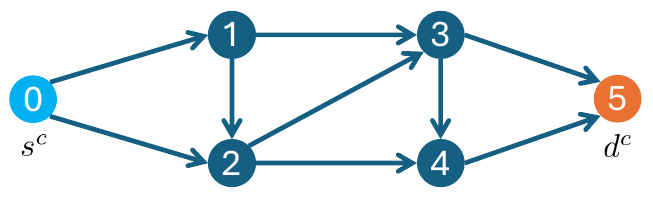}
			\caption{6-node network. Single commodity with source $s^c = 0$, destination $d^c = 5$, and initial lifetime $L^c = 6$, which leads to 8 feasible paths: $p_1 = [0, 1, 3, 5], p_2 = [0, 2, 4, 5], p_3 = [0, 2, 3, 5], p_4 = [0, 1, 2, 3, 5], p_5 = [0, 1, 2, 4, 5], p_6 = [0, 1, 3, 4, 5], p_7 = [0, 2, 3, 4, 5]$, and $p_8 = [0, 1, 2, 3, 4, 5]$.}
			\label{fig:network_0_toy_example}
		\end{subfigure}
		
		\begin{subfigure}{0.48\textwidth}
			\centering
			\includegraphics[width=0.8\linewidth]{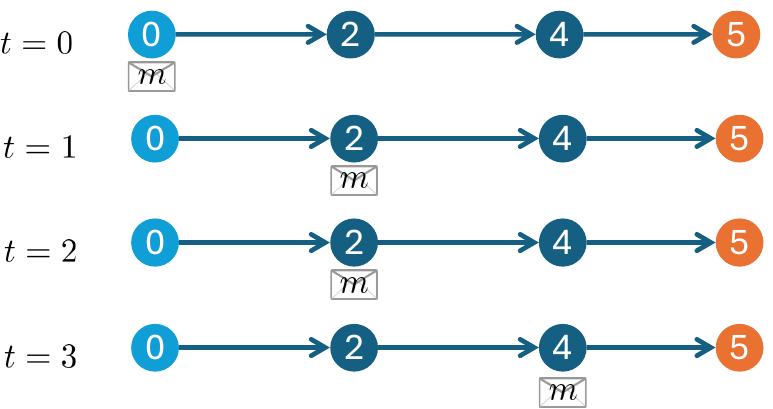}
			\caption{Evolution of packet $m$ traveling along path $p_2 = [0, 2, 4, 5]$ over time slots $t = 0,\dots,3$.
			}
			\label{fig:effective_lifetime_example}
		\end{subfigure}
		\hfill
		\begin{subfigure}{0.48\textwidth}
			\centering
			\setlength{\tabcolsep}{6pt}
			\begin{tabular*}{\linewidth}{@{\extracolsep{\fill}}ccccc}
				\toprule
				$t$ & $(i,j)$ & $l$ & $\ell$ & $a_{\text{scheduler}(i,j)}$ \\
				\midrule
				$t=0$ & (0,2) & 6 & 4 & \text{Send} \\
				$t=1$ & (2,4) & 5 & 4 & \text{Keep} \\
				$t=2$ & (2,4) & 4 & 3 & \text{Send} \\
				$t=3$ & (4,5) & 3 & 3 & \text{Send} \\
				\bottomrule
			\end{tabular*}
			\caption{Evolution of \emph{absolute} lifetime $l$, \emph{effective} lifetime $\ell$, 
				and scheduler action $a_{\text{scheduler}(i,j)}$ 
				along path $p_2$. While absolute lifetime decreases by 1 every time slot, effective lifetime $\ell$ remains unchanged during time slots $t=0,1$ and $t=2,3$ because the packet keeps moving through the network.}
			\label{tab:effective_lifetime_example}
		\end{subfigure}
		\caption{Illustrative running example. (a) network topology.
			(b) packet trajectory over chosen path $p_2$ under scheduling actions shown in (c).}
		\label{fig:effective_lifetime_combo}
	\end{figure}

	In Figure~\ref{fig:effective_lifetime_combo}, we illustrate the concept of effective lifetime through the example network reported in Figure~\ref{fig:network_0_toy_example}. In particular, we follow the evolution of the absolute lifetime \textit{l} and effective lifetime $\ell$ for a packet flowing over path $p_2$ (Figure~\ref{fig:effective_lifetime_example}) according to the sequence of actions reported in Table \ref{tab:effective_lifetime_example}.
	
	Given the strict advantage of using the \ac{el} instead of the absolute lifetime to track packets' urgency and relevance, from now on we will assume an \ac{el}-driven network. That is, a network that tracks and queues packets according to \ac{el}, and naturally drops packets when their \ac{el} reaches zero. {\em An \ac{el}-driven deadline-aware network hence avoids keeping packets congesting network queues and competing for network resources when they have no chance to reach their destination}.

	\subsection{\acf{lelf} Scheduler}
	\label{sssec:lelf_scheduler}

	A key related concept introduced in \citep{vitale2025flexible} is the \textit{\ac{lelf}} scheduling policy. The \ac{lelf} scheduling policy leverages the \ac{el} metric to maximize the total number of packets successfully delivered on time. Specifically, \ac{lelf} gives priority to forwarding packets with the lowest \ac{el}. Recall that a packet's \ac{el} represents the number of time slots it can wait in a network queue before having no chance to reach the destination, and that a packet's \ac{el} does not decrease as long as the packet is in transit through the network. Therefore, the underlying goal of the \ac{lelf} policy is to ensure that packets with low \ac{el} keep moving through the network. By keeping them moving, their \ac{el} values are preserved, which increases the probability of successful timely delivery. In practical terms, the \ac{lelf} policy operates by selecting packets for transmission in strict ascending order of their current \ac{el}.
	
	From an agentic point of view, the states and actions of the \textbf{\ac{lelf} Scheduler} at a given interface $(i,j)$ are characterized as:
	\begin{itemize}
		\item 
		The  \textbf{State} at time $t$ is the vector whose entries specify the queue state of interface $(i,j)$ at time $t$, for each commodity and effective lifetime, i.e., $s_{\mathrm{scheduler},ij}(t) = [q_{ij}^{(c,\ell)}(t) : \forall c \in \mathcal{C}, \forall \ell \in \mathcal{L}]$, with dimension $|\mathcal{C}| \cdot |\mathcal{L}|$.
		\item 
		The \textbf{Action} at time $t$ is the vector indexed by commodity and residual EL, whose entries specify the number of packets to forward over interface $(i,j)$: $a_{\mathrm{scheduler},ij}(t) = [F^{(c,\ell)}(t): \forall c \in C, \forall \ell \in L]$, where $F^{(c,\ell)}(t)$ is the number of packets of commodity $c$ and residual lifetime $\ell$ forwarded over interface $(i,j)$ at time $t$. Recall that packets are forwarded in order of increasing residual lifetime, subject to interface capacity constraints.
	\end{itemize}
	
	As shown in~\citep{vitale2025flexible}, the \ac{lelf} scheduling policy is very effective in terms of timely throughput performance when compared with RL-based alternatives. Furthermore, it significantly reduces overall computational complexity by eliminating the training phase inherent to \ac{rl} approaches.

	\begin{figure*}
		\centering
		\begin{tikzpicture}[
			scale=1.0,
			font=\small,
			vcell/.style={
				draw, 
				minimum width=1cm,   
				minimum height=0.7cm, 
				align=center, 
				anchor=center,
				inner sep=0pt
			},
			excluded/.style={
				vcell, 
				fill=gray!5,        
				draw=gray!40,       
				text=black!30,      
			},
			v_fix_left/.style={
				excluded,
				draw=none, 
				append after command={
					(\tikzlastnode.north west) edge [draw=gray!40, thin] (\tikzlastnode.north east)
					(\tikzlastnode.north east) edge [draw=gray!40, thin] (\tikzlastnode.south east)
					(\tikzlastnode.south east) edge [draw=gray!40, thin] (\tikzlastnode.south west)
				}
			},
			vvec/.style={
				matrix of nodes,
				nodes={vcell},
				row sep=-\pgflinewidth,
				column sep=-\pgflinewidth,
				nodes in empty cells,
				anchor=west
			},
			v_orange/.style={fill=orange!20, dashed, thick},
			v_cyan/.style={fill=cyan!20, dotted, thick},
			param/.style={
				minimum height=0.7cm,
				anchor=center,
				text width=0.9cm,
				align=center
			},
			sumarrow/.style={->, -Latex, thick, gray!80, shorten >=2pt, shorten <=2pt},
			colarrow/.style={->, -Latex, thick, gray!50, shorten >=1pt, shorten <=1pt},
			sepdots/.style={
				font=\large\bfseries, 
				text=gray!80, 
				align=center,
				minimum height=0.4cm
			}
			]
			
			\node[param, font=\bfseries, text width=1.2cm, align=center] (head-lbl) at (0,0) {};
			
			\node[param, right=0pt of head-lbl, font=\bfseries] (head-L) {$L^{c}$};
			\node[param, right=0pt of head-L, font=\bfseries] (head-P) {$p$};
			\node[param, right=0pt of head-P, font=\bfseries] (head-T) {$T_{4,5}^p$};
			\node[param, right=0pt of head-T, font=\bfseries] (head-Lp) {$\text{\sf L}^{p}$};
			
			\coordinate (vec-start) at ($(head-Lp.east)+(1.2cm,0)$);
			
			\node[param, below=0.5cm of head-lbl, font=\bfseries] (lac-lbl) {LAC};
			\node[param] at (lac-lbl -| head-L) {6};
			\node[param] at (lac-lbl -| head-P) {N/A};
			\node[param] at (lac-lbl -| head-T) {N/A};
			\node[param] at (lac-lbl -| head-Lp) {N/A};
			
			\node[anchor=east, font=\small, xshift=-5pt] at (lac-lbl -| vec-start) {$\mathbf{Q}_{4,5}(t)$};
			
			\matrix (vec-lac) [vvec] at (lac-lbl -| vec-start) {
				|[fill=gray!20]| 15 & |[fill=gray!20]| 26 & |[fill=gray!20]| 20 & |[fill=gray!20]| 21 & |[fill=gray!20]| 18 & |[fill=gray!20]| 15 & |[fill=gray!20]| 12 \\
			};
			
			\node[vcell, right=1.5cm of vec-lac, fill=gray!20] (scal-lac) {\textbf{127}};
			\draw[sumarrow] (vec-lac.east) -- node[midway, above, text=black] {$\Sigma$} (scal-lac.west);
			\node[right=3pt] at (scal-lac.east) {$Q_{4,5}(t)$};
			
			\foreach \i/\col in {1/red!50, 2/red!42, 3/red!34, 4/red!26, 5/red!18, 6/red!10, 7/red!5} {
				\node (ell-\i) [vcell, draw=none, fill=\col] 
				at (vec-lac-1-\i.center |- head-lbl.center) {$\ell=\i$};
			}
			
			\node[font=\scriptsize\bfseries, align=center, anchor=east, text=gray!70] 
			at ($(ell-1.west)+(-0.2,0)$) {Priority \\ Higher};
			
			\node[font=\scriptsize\bfseries, align=center, anchor=west, text=gray!70] 
			at ($(ell-7.east)+(0.2,0)$) {Priority \\ Lower};

			\node[param, below=0.6cm of lac-lbl, font=\bfseries] (ec1-lbl) {};
			\node[param] at (ec1-lbl -| head-L) {6};
			\node[param] at (ec1-lbl -| head-P) {$p_2$};
			\node[param] at (ec1-lbl -| head-T) {2};
			\node[param] at (ec1-lbl -| head-Lp) {4}; 
			
			\node[anchor=east, font=\small, xshift=-5pt] at (ec1-lbl -| vec-start) {$\mathbf{\bar{Q}}_{4,5}^{p_2}(t)$};
			
			\matrix (vec-ec1) [vvec] at (ec1-lbl -| vec-start) {
				|[excluded]| 15 & |[excluded]| 26 & |[v_orange]| 20 & |[v_orange]| 21 & |[v_orange]| 18 & |[v_orange]| 15 & |[v_fix_left]| 12 \\
			};
			
			\node[vcell, right=1.5cm of vec-ec1, v_orange] (scal-ec1) {\textbf{74}};
			\draw[sumarrow] (vec-ec1.east) -- node[midway, above, text=black] {$\Sigma$} (scal-ec1.west);
			\node[right=3pt] at (scal-ec1.east) {$\bar{Q}_{4,5}^{p_2}(t)$};
			
			\draw[black] (vec-ec1-1-6.north east) -- (vec-ec1-1-6.south east);
			
			\node[sepdots, below=0.1cm of ec1-lbl] (dots1) {};
			\node[sepdots] at (dots1 -| head-L) {$\vdots$};
			\node[sepdots] at (dots1 -| head-P) {$\vdots$};
			\node[sepdots] at (dots1 -| head-T) {$\vdots$};
			\node[sepdots] at (dots1 -| head-Lp) {$\vdots$};
			\node[sepdots, minimum width=6cm] at (dots1 -| vec-ec1) {$\vdots$};
			\node[sepdots] at (dots1 -| scal-ec1) {$\vdots$};
			
			\node[param, below=0.1cm of dots1, font=\bfseries] (ec2-lbl) {EC}; 
			\node[param] at (ec2-lbl -| head-L) {6};
			\node[param] at (ec2-lbl -| head-P) {$p_5$};
			\node[param] at (ec2-lbl -| head-T) {3};
			\node[param] at (ec2-lbl -| head-Lp) {3}; 
			
			\node[anchor=east, font=\small, xshift=-5pt] at (ec2-lbl -| vec-start) {$\mathbf{\bar{Q}}_{4,5}^{p_5}(t)$};
			
			\matrix (vec-ec2) [vvec] at (ec2-lbl -| vec-start) {
				|[excluded]| 15 & |[excluded]| 26 & |[excluded]| 20 & |[v_orange]| 21 & |[v_orange]| 18 & |[v_orange]| 15 & |[v_fix_left]| 12 \\
			};
			
			\node[vcell, right=1.5cm of vec-ec2, v_orange] (scal-ec2) {\textbf{54}};
			\draw[sumarrow] (vec-ec2.east) -- node[midway, above, text=black] {$\Sigma$} (scal-ec2.west);
			\node[right=3pt] at (scal-ec2.east) {$\bar{Q}_{4,5}^{p_5}(t)$};
			
			\draw[black] (vec-ec2-1-6.north east) -- (vec-ec2-1-6.south east);
			
			\node[sepdots, below=0.1cm of ec2-lbl] (dots2) {};
			\node[sepdots] at (dots2 -| head-L) {$\vdots$};
			\node[sepdots] at (dots2 -| head-P) {$\vdots$};
			\node[sepdots] at (dots2 -| head-T) {$\vdots$};
			\node[sepdots] at (dots2 -| head-Lp) {$\vdots$};
			\node[sepdots, minimum width=6cm] at (dots2 -| vec-ec2) {$\vdots$};
			\node[sepdots] at (dots2 -| scal-ec2) {$\vdots$};
			
			\node[param, below=0.1cm of dots2, font=\bfseries] (ec3-lbl) {};
			\node[param] at (ec3-lbl -| head-L) {6};
			\node[param] at (ec3-lbl -| head-P) {$p_8$};
			\node[param] at (ec3-lbl -| head-T) {4};
			\node[param] at (ec3-lbl -| head-Lp) {2}; 
			
			\node[anchor=east, font=\small, xshift=-5pt] at (ec3-lbl -| vec-start) {$\mathbf{\bar{Q}}_{4,5}^{p_8}(t)$};
			
			\matrix (vec-ec3) [vvec] at (ec3-lbl -| vec-start) {
				|[excluded]| 15 & |[excluded]| 26 & |[excluded]| 20 & |[excluded]| 21 & |[v_orange]| 18 & |[v_orange]| 15 & |[v_fix_left]| 12 \\
			};
			
			\node[vcell, right=1.5cm of vec-ec3, v_orange] (scal-ec3) {\textbf{33}};
			\draw[sumarrow] (vec-ec3.east) -- node[midway, above, text=black] {$\Sigma$} (scal-ec3.west);
			\node[right=3pt] at (scal-ec3.east) {$\bar{Q}_{4,5}^{p_8}(t)$};
			
			\draw[black] (vec-ec3-1-6.north east) -- (vec-ec3-1-6.south east);
			
			\node[param, below=0.5cm of ec3-lbl, font=\bfseries] (ecp-lbl) {EC($p^*$)};
			\node[param] at (ecp-lbl -| head-L) {6};
			\node[param] at (ecp-lbl -| head-P) {$p^*$};
			\node[param] at (ecp-lbl -| head-T) {2};
			\node[param] at (ecp-lbl -| head-Lp) {4}; 
			
			\node[anchor=east, font=\small, xshift=-5pt] at (ecp-lbl -| vec-start) {$\mathbf{\hat{Q}}_{4,5}^{p^*}(t)$};
			
			\matrix (vec-ecp) [vvec] at (ecp-lbl -| vec-start) {
				|[excluded]| 15 & |[excluded]| 26 & |[v_cyan]| 20 & |[v_cyan]| 21 & |[v_cyan]| 18 & |[v_cyan]| 15 & |[v_fix_left]| 12 \\
			};
			
			\node[vcell, right=1.5cm of vec-ecp, v_cyan] (scal-ecp) {\textbf{74}};
			\draw[sumarrow] (vec-ecp.east) -- node[midway, above, text=black] {$\Sigma$} (scal-ecp.west);
			\node[right=3pt] at (scal-ecp.east) {$\hat{Q}_{4,5}^{p^*}(t)$};
			
			\draw[black] (vec-ecp-1-6.north east) -- (vec-ecp-1-6.south east);
			
			\draw[gray!50, thick] (head-L.south west) -- (head-Lp.south east |- head-L.south west);
			
			\draw[gray!50, thick] (ell-1.south west |- head-L.south west) -- (ell-7.south east |- head-L.south west);
			
			\draw[gray!30] ([yshift=-2mm]lac-lbl.south west) -- ([yshift=-2mm]lac-lbl.south west -| head-Lp.east);
			\draw[gray!30] ([yshift=2mm]ecp-lbl.north west) -- ([yshift=2mm]ecp-lbl.north west -| head-Lp.east);
			
			\foreach \i in {1,...,7} {
				\draw[colarrow] (ell-\i.south) -- (vec-lac-1-\i.north);
			}
			
		\end{tikzpicture}
		\caption{Visual breakdown of the congestion metrics introduced in Section~\ref{ssec:model_data_driven_rl_dcs}, applied to communication interface $(4,5)$ consistent with the toy network topology reported in Figure~\ref{fig:network_0_toy_example}. While regular congestion accounts for the entire queue occupancy without considering any path information, \ac{ec} explicitly differentiates the congestion perceived by each path, resulting in a more precise metric that varies based on both the path's effective lifetime $\text{\sf L}^{p}$ and transit time $T_{4,5}^p$. The colored cells indicate packets satisfying the filtering condition $T_{4,5}^p < \ell \leq \text{\sf L}^{p} + T_{4,5}^p$: for $p_2$ ($\text{\sf L}^{p}=4$, $T=2$) packets with $\ell \in \{3,4,5,6\}$ are included (orange); for $p_5$ ($\text{\sf L}^{p}=3$, $T=3$) packets with $\ell \in \{4,5,6\}$; for $p_8$ ($\text{\sf L}^{p}=2$, $T=4$) packets with $\ell \in \{5,6\}$. The excluded cells (grayed out) illustrate the boundary-filtering effect. Finally, the $EC(p^*)$ metric (cyan cells) constitutes a strategic trade-off, reducing complexity by providing a single congestion value for the interface based on a representative path $p^*$. In this example, $p^* = p_2$ is selected as it satisfies Eq.~\eqref{eq:effective_congestion_p_star} by minimizing the distance between the interface and the path source, resulting in the same filtering range $\ell \in \{3,4,5,6\}$.}
		\label{fig:scalar_to_vector_obs}
	\end{figure*}
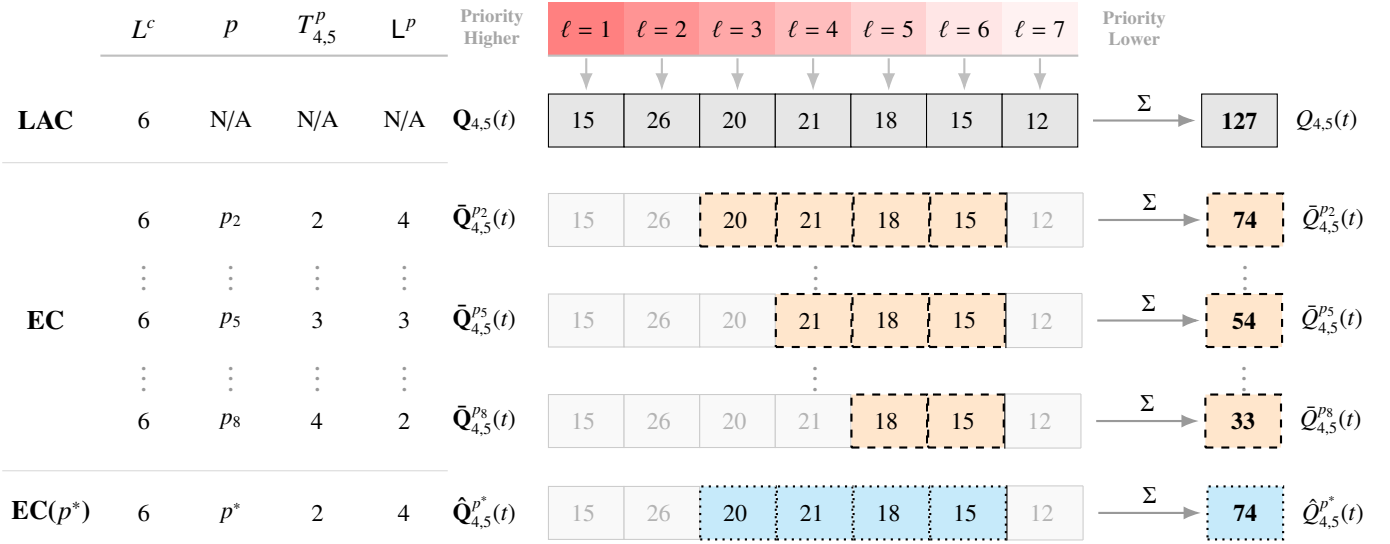
	
	\section{Key Concepts for Effective Routing in Deadline-Constrained Systems}
	\label{ssec:model_data_driven_rl_dcs}
	
	Routing decisions heavily rely on congestion metrics that indicate the extent to which the load on network interfaces causes packets to wait before transmission.
	
	\subsection{Regular Congestion}
	\label{sec:regular_congestion}
	The standard way to compute congestion is to count the number of packets enqueued in the given interface, which in our system would result in a scalar given by:  
	\begin{equation}
		\label{eq:congestion_interface}
		Q_{ij}(t) = \sum_{c \in \mathcal{C}} \sum_{\ell=1}^{L^c} q_{ij}^{(c,\ell)}(t)
	\end{equation}
	
	\subsection{\acf{lac}}
	\label{sec:lifetime_aware_congestion}
	In the considered deadline-constrained system, where packets have a finite time to reach their destination, standard volume-based metrics are unable to fully capture deadline-related aspects. Purely volume-based metrics do not account for packet expiration and therefore overlook the impact of deadline-aware policies (e.g., \ac{lelf}), whose forwarding decisions are intrinsically linked to packet lifetime. In particular, under \ac{lelf} scheduling, packets with low \ac{el} create stronger competition for network resources than packets with high \ac{el}.
	
	Consequently, we define the \acf{lac} of an interface as a vector containing the number of packets for each \ac{el}:
	
	\begin{equation}
		\label{eq:lifetime_aware_congestion_vector_def}
		\mathbf{{Q}}_{ij}(t) = \Big[ v_{1}(t), \, \dots, \, v_{\ell}(t), \, \dots, \, v_{L_{max}}(t) \Big]
	\end{equation}
	where each component $v_{\ell}(t)$ represents the aggregate occupancy of packets with effective lifetime $\ell$:\footnote{Recall that in \ac{el}-driven systems $q_{ij}^{(c,\ell)}(t)=0$ for all $\ell > L^c$.}
	\begin{equation}
		\label{eq:effective_congestion_component}
		v_{\ell}(t) = \sum_{c \in \mathcal{C}} q_{ij}^{(c,\ell)}(t)
	\end{equation}

	\subsection{Effective congestion: a path-dependent metric}
	\label{ssec:effective_congestion_metric}
	
	Furthermore, under centralized routing, where packets are assigned to paths upon arrival at their respective sources, routing decisions can benefit from inferring the congestion a packet may encounter along a given path. However, existing metrics often overestimate this congestion by including traffic that will not actually compete with the considered packet.
	
	To this end, we introduce the \textit{\ac{ec}} metric, a refined measure that characterizes the congestion of an interface with respect to a path, by considering only those packets expected to coexist with the traveling packet (the packet to be routed) at the time of arrival to the interface. In particular, it filters out packets that will have already expired, as well as packets with higher \ac{el} (i.e., lower priority under \ac{lelf} scheduling).
	
	Specifically, the \ac{ec} of an interface $(i,j)$ with respect to a path $p$ is designed to quantify the congestion that a packet currently at the source of $p$ will encounter when reaching interface $(i,j)$ traveling over path $p$. Let $T^{p}_{ij}$ denote the minimum time required for a packet\footnote{Recall that under the assumption of one time slot per one-hop transmission, this is equal to the number of hops between the source of $p$ and node $i$.}, currently at the source of $p$, to reach interface $(i,j)$ along path $p$.
	The \ac{ec} of interface $(i,j)$ with respect to path $p$ restricts its focus to packets that satisfy the following two conditions simultaneously:
	\begin{itemize}
		\item They can survive the minimum transit time of the considered packet, i.e., their \ac{el} satisfies $\ell > T^{p}_{ij}$.
		\item They can maintain an equal or higher priority under \ac{lelf} scheduling (i.e., equal or lower \ac{el}) than the considered packet. That is, their \ac{el} satisfies $\ell \leq \text{\sf L}^{c,p} + T^{p}_{ij}$, where $\text{\sf L}^{c,p}$ denotes the initial \ac{el} of the considered packet, of commodity $c$, when it gets assigned to path $p$, i.e., $\text{\sf L}^{p} = EL( p, L^{c}, s^{c})$. For ease of notation, in the following we focus on settings where a path belongs only to a single commodity, which allows dropping the dependence on $c$, and use $\text{\sf L}^{p}$ in place of $\text{\sf L}^{c,p}$. 
	\end{itemize}
	
	We will refer to this metric as EC $p$-model. We first define the \textit{vectorial form} for the EC $p$-model, denoted as $\mathbf{\bar{Q}}_{ij}^{(p)}(t)$. The length of $\mathbf{\bar{Q}}_{ij}^{(p)}(t)$ is determined by the range of lifetimes that can actually compete with the considered packet  at the time of arrival at interface $(i,j)$, namely $\ell \in [T^{p}_{ij} + 1, \text{\sf L}^{p} + T^{p}_{ij}]$:
	
	\begin{equation}
		\label{eq:effective_congestion_vector_def}
		\mathbf{\bar{Q}}_{ij}^{(p)}(t) = \Big[ v_{ T^{p}_{ij}+1}(t), \, \dots, \, v_{\ell}(t), \, \dots, \, v_{\text{\sf L}^{p} + T^{p}_{ij}}(t) \Big]
	\end{equation}
	
	Each component  $v_{\ell}(t)$ represents the aggregate occupancy of packets with effective lifetime $\ell$, provided they satisfy the filtering conditions.
	Then, the {\em scalar form} of the \ac{ecp} metric is obtained by simply summing the entries of the vectorial \ac{ecp} metric.
	
	\begin{equation}
		\label{eq:effective_congestion_scalar}
		\bar{Q}_{ij}^{(p)}(t) = 
		\sum_{\ell = T^{p}_{ij}+1}^{\text{\sf L}^{p} + T^{p}_{ij}}
		v_{\ell}(t) 
	\end{equation}
	
	An example of this metric is shown in Figure~\ref{fig:scalar_to_vector_obs}, where the orange-filled (dash bordered) cells represent the packets considered by the EC $p$-model congestion metric at interface $(4,5)$ of the considered example network reported in Figure~\ref{fig:network_0_toy_example}. For instance, for path $p_2$ with $L^{c}=6$ and $T_{4,5}^{p_2}=2$, the \ac{ecp} metric includes packets with $\ell \in \{3,4,5,6\}$, reflecting those that will survive transit and maintain competitive priority upon arrival.
	
	As it will be clear from later sections, the scalar version of the EC metric will be used by prior-guided routing agents to characterize and compare the congestion of the different paths that a packet of a given commodity can take upon arrival to the network. To that end, the \ac{ecp} metric  of path $p$ is computed as:
	
	\begin{equation}
		\label{eq:EC_path}
		\bar{Q}^{(p)}(t) = \sum_{(i,j) \in p} \bar{Q}_{ij}^{(p)}(t) 
	\end{equation}

	\subsection{Effective Congestion: $p^*$-model}
	\label{ssec:effective_congestion_p_star_metric}
	As remarked throughout Sec.~\ref{ssec:effective_congestion_metric}, the \textit{\ac{ecp}} defined in Eqs.~(\ref{eq:effective_congestion_vector_def})--(\ref{eq:effective_congestion_scalar}) is a path-dependent metric, i.e., it characterizes the congestion of an interface with respect to a packet expected to travel over a given path $p$. This per-path evaluation provides a fine-grained congestion picture, enabling precise routing decisions; its price lies in the acquisition of the network observation: since the filtering range depends on the reference path, one evaluation is required for every (interface, path) pair, and the acquisition time grows accordingly.
	To reduce this acquisition cost, we introduce the \textit{Effective Congestion $p^*$-model}: rather than computing the metric for every candidate path, this formulation anchors the evaluation of each interface to a single reference path $p^*$, selected from $\mathcal{P}_{ij}$, the subset of all-commodity paths traversing interface $(i,j)$, so that each interface is evaluated once and all paths crossing it share the same congestion measure. For learning-based controllers, using a reference path brings an additional benefit: with one congestion value per interface, the resulting observation has dimensionality $|\mathcal{E}|$, i.e., one component per network interface.
	
	Specifically, we define $p^*$ as the path minimizing the traversal distance to the interface among the considered subset:
	
	\begin{equation}
		\label{eq:effective_congestion_p_star}
		p^*= \arg\min_{p\in \mathcal{P}_{ij}} D(s^p, (i,j))
	\end{equation}
	where $D(s^p, (i,j))$ represents the distance (or minimum travel time) between the source node of path $p$ and the interface $(i,j)$. 
	
	Consequently, letting $T^{p^*}_{ij}$ denote the minimum time required for a packet to reach interface $(i,j)$ when traveling over path $p^*$, the packet filtering conditions change as: they can (i) survive the transit time ($\ell > T^{p^*}_{ij}$), and (ii) maintain a priority level higher than or equal to that of the considered packet ($\ell \leq \text{\sf L}^{p^*} + T^{p^*}_{ij}$).
	
	Based on equations (\ref{eq:effective_congestion_vector_def}) and (\ref{eq:effective_congestion_p_star}), we define the \textit{EC $p^*$-model} of interface $(i,j)$ as the congestion metric computed with respect to the reference path $p^*$. In its vectorial form, denoted as $\mathbf{\hat{Q}}_{ij}^{(\mathcal{P}_{ij})}(t)$, it is expressed as:
	
	\begin{equation}
		\label{eq:effective_congestion_pstar_vector_def}
		\mathbf{\hat{Q}}_{ij}^{(\mathcal{P}_{ij})}(t) = \Big[ v_{ T^{p^*}_{ij}+1}(t), \, \dots, \, v_{\ell}(t), \, \dots, \, v_{\text{\sf L}^{p^*} + T^{p^*}_{ij}}(t) \Big]
	\end{equation}
	
	Accordingly, the Effective Congestion $p^*$ for interface $(i,j)$ in its scalar form is defined as:
	
	\begin{equation}
		\label{eq:effective_congestion_interface_path_ref}
		\hat{Q}_{ij}^{(\mathcal{P}_{ij})}(t) =
		\sum_{\ell = T^{p^*}_{ij}+1}^{\text{\sf L}^{p^*} + T^{p^*}_{ij}} v_{\ell}(t)
	\end{equation}
	
	An example of the EC $p^*-model$ metric is shown in Figure~\ref{fig:scalar_to_vector_obs}, where the cyan-filled (dot-bordered) cells highlight the packets satisfying the filtering conditions. Since $p^* = p_2$ in this example (with $T_{4,5}^{p^*}=2$), the EC $p^*$-model includes the same set of packets as EC($p_2$): those with $\ell \in \{3,4,5,6\}$.
	
	\subsection{Scalar vs. Vectorial representation tradeoff}
	\label{ssec:integrated_model_data_driven_rl_design}
	The ability to accurately observe network congestion states is fundamental for routing decisions in deadline-constrained systems. This holds true for both model-driven and \ac{rl} approaches, as the dimensionality and expressiveness of the available observations dictate the control policy's complexity and performance. While Model-driven approaches directly rely on state representation to evaluate the network via predefined rules or heuristics, RL agents leverage them to learn latent feature representations and derive optimal decision-making strategies.
	A trade-off between observation complexity and expressiveness becomes more evident with the introduction of lifetime-aware metrics which are represented (as shown in Fig. \ref{fig:scalar_to_vector_obs}) in scalar or vectorial forms:
	
	\begin{itemize}
		\item The \emph{Vectorial Representation} --see Eqs.\ref{eq:lifetime_aware_congestion_vector_def},\ref{eq:effective_congestion_vector_def},\ref{eq:effective_congestion_pstar_vector_def}-- preserves the full distributional information of the queue state, mapping packets to their specific lifetimes, allowing the system to distinguish high-urgency congestion. 
		However, employing such a high-dimensional representation 
		increases the state space size and, therefore, the policy complexity. Vectorial congestion metrics are denoted with $\mathbf{Q}$.
		\item The \emph{Scalar Representation} --see Eqs.\ref{eq:congestion_interface},\ref{eq:effective_congestion_scalar}, \ref{eq:effective_congestion_interface_path_ref}-- aggregates the queue state to a single scalar, reducing complexity. This state compression masks granular details, preventing policies from distinguishing between a queue filled with urgent packets and non-urgent ones. Scalar congestion metrics are denoted with $Q$.
	\end{itemize}
	
	To clarify this impact, consider the EC($p^*$) row in Fig.~\ref{fig:scalar_to_vector_obs}, where the maximum packet lifetime is $L_{max}=7$. The \emph{Vectorial} form is given by $[20, 21, 18, 15]$, which explicitly reveals the urgency distribution. Conversely, the \emph{Scalar} form (the sum of the vectorial form) yields a total of $74$ packets, shown at the bottom, hides the internal queue composition, making it impossible to infer the nature of the congestion.
	
	As will become clear in the following sections, our proposed prior-guided policies utilize the scalar representation. Clearly, alternative heuristic policies could leverage the more expressive vectorial form, albeit at the cost of increased computational complexity. 
	Conversely, the RL-based policies we propose exploit both representations, enabling us to study the resulting expressiveness–complexity trade-off, as described in Section~\ref{sec:marl_ec_p_star}.

	\section{Prior-guided Policies}
	\label{sec:model_based_policies}
	This section presents our prior-guided policies. From an agent-oriented perspective, each policy is characterized by three design dimensions: (i) the \textit{congestion} type, which governs how the observed load is filtered and evaluated with respect to packet urgency (\acs{rc}, \acs{ecp}, or \acs{ecps}); (ii) the \textit{assignment} strategy, which indicates how traffic is distributed across paths (greedy or grouped); and (iii) the resulting \textit{state} space dimension, which determines the amount of information the agent observes. All policies share a common \textit{action} space of dimension $|\mathcal{P}|$, where each element $A^p$ specifies the number of newly arrived packets allocated to path $p \in \mathcal{P}$.
	
	The five policies summarized in Table~\ref{tab:model_policy_summary} arise from combining these dimensions with increasing deadline-awareness. The baseline \ac{mwprc} (Sec.~\ref{ssec:MWP}), introduced in~\citep{vitale2025flexible}, pairs the volume-based \ac{rc} metric (Eq.~\eqref{eq:congestion_interface}) with a greedy assignment. The \ac{mwpecp} (Sec.~\ref{ssec:MWP_EC}) replaces the \ac{rc} with the \ac{ecp} (Eq.~\eqref{eq:effective_congestion_scalar}), and the \ac{upgecp} (Sec.~\ref{ssec:UPG}) further extends it with a uniform packet assignment strategy. Finally, the \textit{\ac{mwpecps}} and \textit{\ac{upgecps}} (Secs.~\ref{ssec:MWP_EC_Pstar} and~\ref{ssec:UPG_Pstar}) employ the \emph{\ac{ecps}} metric (Eq.~\eqref{eq:effective_congestion_interface_path_ref}) in place of the \ac{ecp}, reducing the state space size---and therefore complexity---and facilitating a smoother transition to our  proposed RL-based strategies (see Section~\ref{sec:mga_rl_instantiation}). 
	The individual algorithms underlying the policies described above are detailed in the following subsections.
	
	\begin{table}[h!]
		\centering
		\caption{Summary of the proposed prior-guided routing policies. The action space has dimension $|\mathcal{P}|$ for all policies, where each element $A^p\in A$ specifies the number of newly arrived packets allocated to path $p$.}
		\label{tab:model_policy_summary}
		\begin{tabularx}{\columnwidth}{Xccc}
			\toprule
			\textbf{Policy} & \textbf{$|\text{State}|$} 
			& \textbf{Congestion} 
			& \textbf{Assignment} \\
			\midrule
			\ac{mwprc}      & $|\mathcal{P}|$   & \ac{rc}              & Greedy \\
			\ac{mwpecp}    & $|\mathcal{P}|$   & \ac{ec} $p$      & Greedy \\
			\ac{upgecp}    & $|\mathcal{P}|$   & \ac{ec} $p$      & Grouped \\
			\ac{mwpecps}  & $|\mathcal{E}|$   & \ac{ec} $p^*$    & Greedy \\
			\ac{upgecps}   & $|\mathcal{E}|$   & \ac{ec} $p^*$    & Grouped \\
			\bottomrule
		\end{tabularx}
	\end{table}
	
	The choice between the two \ac{ec} formulations has direct implications for computational cost, as already reflected in the state space dimensions reported in Table~\ref{tab:model_policy_summary}. The \ac{ecps} formulation assigns a single weight to each interface $(i,j)$—computed once and shared by all paths traversing it (Algorithms~\ref{alg:MWP_EC_PSTAR}--\ref{alg:UPG_PSTAR})—reducing the total number of \ac{ec} evaluations per routing step from $\mathcal{O}(|\mathcal{P}|)$ to $\mathcal{O}(|\mathcal{E}|)$. The exact \ac{ecp} formulation requires a separate evaluation for each (interface, path) pair, since the filter range depends on the specific reference path. This distinction extends to the \ac{rl} setting introduced in Sec.~\ref{sec:marl_ec_p_star}: EC $p$ would yield an observation of dimension $|\mathcal{P}|$ (Scalar) or $|\mathcal{P}| \cdot L_{\max}$ (Vectorial), while EC ($(p^*)$)reduces this to $|\mathcal{E}|$ or $|\mathcal{E}| \cdot L_{\max}$, respectively. This makes the \ac{ecps} formulation more advantageous, both in terms of computational efficiency and policy scalability, when $|\mathcal{E}| \ll |\mathcal{P}|$.
	
	\subsection{\acf{mwprc}}
	\label{ssec:MWP}
	
	The \ac{mwprc} strategy prioritizes paths with lower congestion, distributing traffic away from heavily utilized network segments as follows:\\
	\textit{(I)} --
	Determine the total weight of each path $p \in \mathcal{P}$ by summing the Regular Congestion (Eq.~\eqref{eq:congestion_interface}) over all links $(i,j)$ traversed by that path, normalized by their respective capacities:
	\begin{equation}
		\label{eq:mwp_path_weight}
		WP^p(t) = \sum_{(i,j)\in p} \frac{1}{C_{ij}(t)} Q_{ij}(t)
	\end{equation}
	\textit{(II)} --
	Assign newly arriving packets of each commodity to the path with the minimum total weight. If the capacity of the best path is reached, the algorithm proceeds to the next-lowest-weight path.

	Algorithm~\ref{alg:MWP_BASE} describes the \ac{mwprc} algorithm in detail.
	\begin{algorithm}[H]
		\caption{\acf{mwp}}\label{alg:MWP_BASE}
		\begin{algorithmic}[1]
			\State $WP=[WP^p=0, \forall p \in \mathcal{P}]$
			\State $A=[A^p=0, \forall p \in \mathcal{P}]$
			\For{$p \in \mathcal{P}$}\Comment{Compute Weight for each Path}
			\For{$(i,j)\in  p$} 
			\State $WP^{p} += \frac{1}{C_{ij}} Q_{ij} $
			\EndFor
			\EndFor
			\For{$c \in \mathcal{C}$} \Comment{Packets Routing}
			\State $res\_pkts \leftarrow b^c(t)$
			\State $P_{sort}^c \leftarrow Sort(\mathcal{P}^c, WP)$
			\While{$res\_pkts > 0$}
			\For{$p \in P_{sort}^c$}
			\State $A^p = min(res\_pkts, Cap(p))$
			\State $res\_pckts -= A^p$
			\EndFor
			\EndWhile
			\EndFor
			\State $\text{\textbf{return} }A$
		\end{algorithmic}
	\end{algorithm}
	
	In Algorithm \ref{alg:MWP_BASE}, $P_{sort}^c$ denotes the set of paths $p\in \mathcal{P}^c$ sorted by weight (from lowest to highest), while $Cap(p)= \min_{(i,j) \in p} C_{ij}(t)$ represents the bottleneck capacity of path $p$ defined by the minimum link capacity along its route.
	
	\subsection{\acf{mwpecp}}
	\label{ssec:MWP_EC}
	The \ac{mwpecp} enhances the \ac{mwprc} approach by replacing the volume-based \ac{rc} with the \textit{\ac{ecp}} metric (Section~\ref{ssec:effective_congestion_metric}). Path weights are thus computed only over the packets expected to compete for network resources along each candidate path---filtered by remaining lifetime and transit time---rather than over the entire queue occupancy. Based on Eq.~\eqref{eq:effective_congestion_scalar}, the weight of each path is then calculated as:
	\begin{equation}
		\label{eq:mwp_ec_path_weight}
		WP^{p}(t) = \sum_{(i,j) \in p} \frac{1}{C_{ij}(t)} \bar{Q}_{ij}^{(p)}(t)
	\end{equation}
	
	\noindent In Algorithm~\ref{alg:MWP_EC}, we detail the \ac{mwpecp} procedure, which replaces the \ac{rc}-based weights used by \ac{mwprc} with \ac{ecp} ones.
	
	\begin{algorithm}[H]
		\caption{\acf{mwpecp}}\label{alg:MWP_EC}
		\begin{algorithmic}[1]
			\State $WP=[WP^p=0, \forall p \in \mathcal{P}]$
			\State $A=[A^p=0, \forall p \in \mathcal{P}]$
			\For{$p \in \mathcal{P}$}\Comment{Compute Weight for each Path}
			\For{$(i,j)\in  p$}
			\State $WP^{p} += \frac{1}{C_{ij}} \bar{Q}_{ij}^{(p)}$
			\EndFor
			\EndFor
			
			\State Same as Algorithm \ref{alg:MWP_BASE}  \Comment{Packets Routing}
			\State $\text{\textbf{return} }A$
		\end{algorithmic}
	\end{algorithm}
	
	\subsection{\acf{upgecp}}
	\label{ssec:UPG}
	While \ac{mwpecp} (Section~\ref{ssec:MWP_EC}) effectively identifies the least congested path, it may lead to over-utilization of a single path, potentially causing rapid congestion shifts. The \emph{\acf{upg}} strategy mitigates this issue by balancing traffic among paths of comparable weight, operating as follows:
	\begin{enumerate}
		\item \textit{Weighting:} compute the weights $WP^p$ for all available paths using the \ac{ecp} metric (Eq.~\eqref{eq:mwp_ec_path_weight});
		\item \textit{Grouping:} partition the paths into equivalence groups of equal weight, ordered by ascending weight;
		\item \textit{Routing:} assigns packets uniformly across the paths of the lowest-weight group, capping each path's allocation at the group's minimum bottleneck capacity;
		\item \textit{Iterating:} if newly arrived packets remain unassigned, repeat the process with the next lowest-weight group, until all packets are allocated.
	\end{enumerate}
	In Algorithm \ref{alg:UPG}, we detail the \ac{upgecp} procedure, which extends the  \ac{mwpecp} by integrating the grouping and distribution mechanisms to achieve a more balanced load distribution.
	
	\begin{algorithm}[H]
		\caption{\acf{upgecp}}\label{alg:UPG}
		\begin{algorithmic}[1]
			\State $WP=[WP^p=0, \forall p \in \mathcal{P}]$
			\State $A=[A^p=0, \forall p \in \mathcal{P}]$
			\State Same as Algorithm \ref{alg:MWP_EC} \Comment{Compute Weight for each Path}
			\For{$c \in \mathcal{C}$} \Comment{Packets Routing}
			\State $res\_pkts = b^c(t)$
			\State $G_{sort}^c \leftarrow Grouping(\mathcal{P}^c, WP)$
			\While{$res\_pkts > 0$}
			\For{$g \in G_{sort}^c$}
			\State $pkts\_per\_path \leftarrow \lfloor res\_pkts / |g| \rfloor $
			\For{$p \in g$}
			\State $assign = \min(pkts\_per\_path, Cap(g))$
			\State $A^p += assign$
			\State $res\_pkts -= assign$
			\If{$res\_pkts = 0$} \textbf{break} \EndIf
			\EndFor
			\EndFor
			\EndWhile
			\EndFor
			\State $\text{\textbf{return} }A$
		\end{algorithmic}
	\end{algorithm}
	
	In Algorithm \ref{alg:UPG}, $G_{sort}^c$ denotes the collection of path groups for commodity $c$ ordered by ascending weight, $|g|$ represents the number of paths in group $g$, and $Cap(g)= \min_{p \in g} \left( \min_{ij \in p} C_{ij}(t) \right)$ indicates the minimum bottleneck capacity of group $g$.By spreading traffic across paths of comparable weight, \ac{upg} prevents the over-utilization of a single optimal path and mitigates the onset of new congestion hotspots.
	
	\subsection{\acf{mwpecps}}\label{ssec:MWP_EC_Pstar}
	The \ac{mwpecps} combines the greedy assignment of \ac{mwprc} with the \ac{ecps} metric of Section~\ref{ssec:effective_congestion_p_star_metric}: the policy first computes the \ac{ecps} weight $\hat{Q}_{ij}^{(\mathcal{P}_{ij})}(t)$ of each interface $(i,j) \in \mathcal{E}$, and then obtains the total weight $WP^p$ of each path by summing the weights of the interfaces it traverses, so that all paths crossing an interface share the same contribution from it. Algorithm~\ref{alg:MWP_EC_PSTAR} illustrates the procedure in detail.
	\begin{algorithm}[H]
		\caption{\acf{mwpecps}}\label{alg:MWP_EC_PSTAR}
		\begin{algorithmic}[1]
			\State $WP=[WP^p=0, \forall p \in \mathcal{P}]$
			\State $W=[W_{ij}=0, \forall (i,j) \in \mathcal{E}]$
			\State $A=[A^p=0, \forall p \in \mathcal{P}]$
			\For{$(i,j)\in  \mathcal{E}$}  \Comment{Compute Weight for each Interface}
			\State $W_{ij}(t) = \frac{1}{C_{ij}(t)} \hat{Q}_{ij}^{(\mathcal{P}_{ij})}(t) $
			\EndFor
			\For{$p \in \mathcal{P}$}\Comment{Compute Weight for each Path}
			\For{$(i,j) \in p$} 
			\State $WP^{p}(t) += W_{ij}(t)$
			\EndFor
			\EndFor
			\State Same as Algorithm \ref{alg:MWP_BASE}  \Comment{Packets Routing}
			\State $\text{\textbf{return} }A$
		\end{algorithmic}
	\end{algorithm}

	\subsection{\acf{upgecps}}\label{ssec:UPG_Pstar}
	The \emph{\ac{upgecps}} combines the load-balancing benefits of \ac{upg} (see Sec. \ref{ssec:UPG}) and the computational scalability of the EC ($(p^*)$)metric. Algorithm \ref{alg:UPG_PSTAR} details how these aspects are combined based on previously introduced algorithms.
	
	\begin{algorithm}[H]
		\caption{\acf{upgecps}}\label{alg:UPG_PSTAR}
		\begin{algorithmic}[1]
			\State $WP=[WP^p=0, \forall p \in \mathcal{P}]$
			\State $W=[W_{ij}=0, \forall (i,j) \in \mathcal{E}]$
			\State $A=[A^p=0, \forall p \in \mathcal{P}]$
			\State Same as Algorithm \ref{alg:MWP_EC_PSTAR} \Comment{Compute Weight for each Interface}
			\State Same as Algorithm \ref{alg:MWP_EC_PSTAR} \Comment{Compute Weight for each Path}
			\State Same as Algorithm \ref{alg:UPG} \Comment{Packets Routing}
			\State \text{\textbf{return} }A
		\end{algorithmic}
	\end{algorithm}
	Among the proposed policies, \ac{upgecps} thus pairs the lowest state-space complexity with the balanced \ac{upg} assignment. As the numerical results of Section~\ref{sec:results_model_based_approaches} will show, while the exact \ac{ecp} formulation retains an edge under the most adverse conditions --- tight deadlines in symmetric grids and structurally asymmetric backbones --- the decoupled $p^*$ approximation recovers as deadlines relax, and on the Grid topology under overload it even surpasses the exact model. This combination of near-optimal reliability and substantially lower state-space complexity is what makes \ac{upgecps} central when selecting the reference-policy/demonstrator for the \ac{mgarl} training objective (Section~\ref{sec:mga_rl_instantiation}).
	
	
	\section{Policy Learning Paradigms and \acs{mgarl} Framework}
	\label{sec:policy_learning_paradigms}
	Section~\ref{sec:model_based_policies} proposed a family of lightweight, computationally efficient model-based policies, characterised by a favourable trade-off between reliability and state-space complexity. Any one of these --- or, more generally, any policy sharing similar properties --- is well suited to serve as a reference policy, or \emph{demonstrator}, for a learning agent. The question then becomes how to transfer such a policy's behaviour to an agent that can then improve upon it. 
	Several established training recipes address this question in different ways --- \ac{bc}, offline \ac{rl}, online \ac{rl}, and hybrid \ac{o2o} schemes among them --- each exploiting a different subset of the available data and demonstrator signals. Rather than treating these as separate techniques, this section frames them all as special cases of a single objective, the \acf{gpr} (Sec.~\ref{sec:gpr}--\ref{sec:gpr_special_cases}), revisiting each recipe in Section~\ref{sec:gpr_special_cases} as a specific instantiation of this unified framework, and derives from it the proposed \acf{mgarl} protocol (Sec.~\ref{sec:mgarl_design_rationale}): a two-stage scheme in which the anchor to the demonstrator is released as live experience accumulates. The \ac{mgarl} instantiation on the specific reference policy and backbone chosen for this work is then presented in Section~\ref{sec:mga_rl_instantiation}.
	
	\subsection{\acf{gpr} and  \ac{mgarl} Objective}
	\label{sec:gpr}
	
	We formalise this unifying objective below. We propose the following \ac{gpr}, which unifies all reward-based policy learning objectives considered in Section~\ref{sec:actor_critic_framework}:
	\begin{equation}
		\label{eq:gpr}
		r_{\mathrm{GPR}}(\alpha,\beta,t) =
		\alpha(t)\cdot\hat{r}_{\mathrm{on}}(t)
		+ \beta(t)\cdot\hat{r}_{\mathrm{off}}(t),
	\end{equation}
	where $\hat{r}_{\mathrm{on}}(t)$ and $\hat{r}_{\mathrm{off}}(t)$ are defined in Eqs.~\eqref{eq:r_on}--\eqref{eq:r_off}, and $\alpha(t),\beta(t)\ge 0$ weight the live and pre-collected reward contributions, respectively, with their time dependence allowing the relative weight of live versus pre-collected experience to be adjusted as training progresses. In~\eqref{eq:gpr}, for notational economy, the argument list on the left-hand side omits the explicit time dependence of $\alpha$ and $\beta$; the time-varying nature of both coefficients remains explicit on the right-hand side of the equation.
	
	The policy-imitation paradigm introduced in Section~\ref{sec:reward_estimation}, driven by $\hat{\mathcal{M}}_{\mathrm{MSE}}$ (Eq.~\eqref{eq:mse_estimate}), is incorporated separately, alongside $r_{\mathrm{GPR}}$, into the unified training objective introduced next (Eq.~\eqref{eq:mga_rl_objective}), weighted by a regularisation coefficient $\lambda\bigl(|\mathcal{D}|\bigr)\ge 0$. This coefficient depends on the effective size of the live replay buffer, $|\mathcal{D}|$ (Eq.~\eqref{eq:replay_buffer}):
	while the live buffer remains small, $\hat{r}_{\text{on}}$ is estimated from few samples and is therefore unreliable, so $\lambda$ keeps the policy anchored to $\mu^{\text{MB}}$; as the buffer grows towards its capacity $N_{\text{buf}}$, live experience becomes sufficiently representative on its own, and the anchor is correspondingly relaxed. 
	The specific choice of $\alpha(t),\beta(t)$ and of the function $\lambda\bigl(|\mathcal{D}|\bigr)$ adopted in this work, together with the practical implications of $|\mathcal{D}|$ growing over time as live transitions are collected, is detailed in Section~\ref{sec:mga_rl_instantiation}.
	
	The \ac{mgarl} training objective is:
	\begin{equation}
		\label{eq:mga_rl_objective}
		\hat{J}^{\mathrm{MGA\text{-}RL}}(\theta) =
		\sum_t \gamma^t\, r_{\mathrm{GPR}}(\alpha,\beta,t)
		\;-\; \lambda\bigl(|\mathcal{D}|\bigr)\cdot\hat{\mathcal{M}}_{\mathrm{MSE}}.
	\end{equation}
	progressively relaxing the anchor to the reference policy $\mu^{\text{MB}}$ as live experience accumulates.
	
	\subsection{Existing Methods as Special Cases of the \ac{mgarl} Training Objective}
	\label{sec:gpr_special_cases}
	As specified in Section~\ref{sec:gpr}, the \ac{gpr} is fully general when all its coefficients are simultaneously active and, in principle, adaptive, i.e., allowed to vary with $t$. In practice, three engineering constraints determine which of the reward and policy-deviation terms in Eqs.~\eqref{eq:gpr} and \eqref{eq:mga_rl_objective} are available, and therefore which of $\alpha,\beta,\lambda$ are strictly positive:
	
	\begin{enumerate}[label=(\roman*)]
		
		\item availability of the pre-collected dataset $\mathcal{D}_{\mathrm{off}}$;
		
		\item feasibility of collecting live transitions into $\mathcal{D}(t)$;
		
		\item availability of an analytical $\mu^{\mathrm{MB}}$ queryable at any
		
		$s\in\mathcal{S}$.
		
	\end{enumerate}
	
	Their interplay naturally recovers established methods as shown in Table~\ref{tab:gpr_cases}. We now discuss each case in turn, highlighting the limitation that motivates the next step towards \ac{mgarl}. In the \ac{mgarl} case, the adaptivity of $\alpha,\beta$ is realised as a piecewise-constant schedule across training stages via batch composition, as detailed in Section~\ref{sec:mga_rl_instantiation}.
	
	\begin{table*}[ht]
		\centering
		\small
		\caption{Instantiations of the unified \ac{mgarl} training objective as a function of available signals and engineering constraints. Each row represents a distinct combination of data availability and environment access constraints, leading naturally to a specific coefficient choice and a known method in the literature.}
		\label{tab:gpr_cases}
		\renewcommand{\arraystretch}{1.25}
		\begin{tabular}{clllc}
			\toprule
			& \textbf{Available signals} & \textbf{Coefficients} & \textbf{Regime}
			& \textbf{Method} \\
			\midrule
			(a) & $\mathcal{D}_{\mathrm{off}}$ only (no reward, no $\mathcal{D}(t)$)
			& $\alpha=0,\beta=0,\lambda>0$ fixed & Offline
			& BC~\citep{pomerleau1991} \\
			(b) & $\mathcal{D}_{\mathrm{off}}$ with reward; no $\mathcal{D}(t)$
			& $\alpha=0,\beta>0,\lambda=0$ & Offline
			& Offline \ac{rl}~\citep{levine2020offline,fujimoto2019off} \\
			(c) & $\mathcal{D}(t)$ only; no $\mathcal{D}_{\mathrm{off}}$
			& $\alpha>0,\beta=0,\lambda=0$ & Online
			& Online \ac{rl}~\citep{sutton1998reinforcement} \\
			(d) & $\mathcal{D}_{\mathrm{off}}+\mathcal{D}(t)$; static $\alpha,\beta,\lambda$
			& $\alpha>0,\beta>0,\lambda>0$ fixed & \ac{o2o}/hybrid
			& AWAC~\citep{nair2020awac}; \ac{dqfd}~\citep{hester2018deep} \\
			\midrule
			(e) & \makecell[l]{$\mathcal{D}_{\mathrm{off}}+\mathcal{D}(t)+\mu^{\mathrm{MB}}$; adaptive $\alpha,\beta$;\\decaying $\lambda(t)$}
			&$\alpha, \beta$ adaptive; $\lambda(t)\downarrow$ & \ac{o2o}+Online
			& \textbf{\ac{mgarl} (This paper)} \\
			\bottomrule
		\end{tabular}
		\vspace{-0.3cm}
	\end{table*}

	\paragraph{(a) \acf{bc} --- $\alpha=0$, $\beta=0$, $\lambda>0$ fixed}
	With neither live transitions ($\alpha=0$) nor the reward recorded in the offline data $\mathcal{D}_{\mathrm{off}}$ ($\beta=0$), Eq.~\eqref{eq:mga_rl_objective} reduces to $\hat{J}^{\mathrm{MGA\text{-}RL}}(\theta) = -\lambda\cdot\hat{\mathcal{M}}_{\mathrm{MSE}}$, coinciding with the \acf{bc} training objective $\hat{J}^{\mathrm{BC}}(\theta)$: supervised regression to replicate $\mu^{\mathrm{MB}}$ on the states in $\mathcal{D}_{\mathrm{off}}$~\citep{pomerleau1991}. \ac{bc} suffers from \emph{covariate shift}~\citep{ross2011reduction}: deviations at deployment compound into unseen states, and performance is fundamentally bounded by the quality of $\mu^{\mathrm{MB}}$.
	
	\paragraph{(b) Offline \ac{rl} --- $\alpha=0$, $\beta>0$, $\lambda=0$}
	With no live transitions ($\alpha=0$) and no policy-deviation term ( $\lambda=0$), $r_{\mathrm{GPR}}(0,\beta,t) = \beta\cdot\hat{r}_{\mathrm{off}}(t)$ (Eq.~\eqref{eq:gpr}), and Eq.~\eqref{eq:mga_rl_objective} reduces to $\hat{J}^{\mathrm{MGA\text{-}RL}}(\theta) = \sum_t \gamma^t\, \hat{r}_{\mathrm{off}}(t)$: the Critic $Q_\phi$ is trained on the reward observed in $\mathcal{D}_{\mathrm{off}}$ alone, with no interaction with the environment. This corresponds to \emph{Offline \ac{rl}}~\citep{levine2020offline, fujimoto2019off}, which is appropriate when collecting live transitions is unsafe or infeasible. Its fundamental limitation is that $Q_\phi$ is prone to \emph{overestimation for \ac{ood} actions}~\citep{fujimoto2019off}, since the Bellman backup may query states absent from $\mathcal{D}_{\mathrm{off}}$. Conservative methods such as CQL~\citep{kumar2020conservative} and IQL~\citep{kostrikov2022offline} mitigate this via pessimistic regularisation, yet cannot discover behaviours absent from $\mathcal{D}_{\mathrm{off}}$.
	
	\paragraph{(c) Online \ac{rl} --- $\alpha>0$, $\beta=0$, $\lambda=0$}
	With no pre-collected data ($\beta=0$) and no policy-deviation term ($\lambda=0$), $r_{\mathrm{GPR}}(\alpha,0,t) = \alpha\cdot\hat{r}_{\mathrm{on}}(t)$ (Eq.~\eqref{eq:gpr}), and Eq.~\eqref{eq:mga_rl_objective} reduces to $\hat{J}^{\mathrm{MGA\text{-}RL}}(\theta) = \sum_t \gamma^t\, \hat{r}_{\mathrm{on}}(t)$: only $\mathcal{D}(t)$ grows, through live interaction alone. This corresponds to standard \emph{Online \ac{rl}}~\citep{sutton1998reinforcement}. It is asymptotically optimal but severely \emph{sample-inefficient}: all prior knowledge encoded in $\mu^{\mathrm{MB}}$ is discarded, resulting in unacceptable training times in real-world deployments~\citep{levine2020offline, prudencio2024survey}.
	
	\paragraph{(d) \ac{o2o} \ac{rl} and \ac{rlfd} --- $\alpha>0$, $\beta>0$, $\lambda>0$ fixed}
	With all three \ac{gpr} terms active but held at fixed, static coefficients, Eq.~\eqref{eq:mga_rl_objective} combines live and offline reward with a constant policy-deviation penalty: $\hat{J}^{\mathrm{MGA\text{-}RL}}(\theta) = \sum_t \gamma^t\bigl[\alpha\cdot\hat{r}_{\mathrm{on}}(t) + \beta\cdot\hat{r}_{\mathrm{off}}(t)\bigr] - \lambda\cdot\hat{\mathcal{M}}_{\mathrm{MSE}}$. This corresponds to \ac{o2o} \ac{rl} and \ac{rlfd} methods such as \emph{AWAC}~\citep{nair2020awac}, which pre-trains $\mu_\theta$ on $\mathcal{D}_{\mathrm{off}}$ then fine-tunes with $\mathcal{D}(t)$, and \emph{\ac{dqfd}}~\citep{hester2018deep}, which permanently replays $\mathcal{D}_{\mathrm{off}}$ alongside $\mathcal{D}(t)$. Both improve over cases~(a)--(c), but the fixed $\lambda$ never adapts: as $t$ grows and $\mu^{\mathrm{MB}}$ becomes redundant, the \ac{mse} penalty still penalises divergence from it. \ac{o2o} methods also require a manual phase boundary, with no principled transition mechanism.
	
	\subsection{\acf{mgarl}: Design Rationale and Training Protocol}
	\label{sec:mgarl_design_rationale}
	
	\ac{mgarl} closes all limitations of Section~\ref{sec:gpr_special_cases} simultaneously by adopting three coupled design choices. First, $\alpha,\beta$ are realised as a piecewise-constant schedule across two training stages, via batch composition (Section~\ref{sec:mga_rl_instantiation}): \emph{Stage~1}, corresponding to $t\le T_1$, uses pre-collected data only, since $\mathcal{D}(t)=\emptyset$; \emph{Stage~2}, corresponding to $t>T_1$, begins once live transitions accumulate. Second, $\lambda\bigl(|\mathcal{D}|\bigr)$ decays as the effective size of the live replay buffer grows, from $\lambda_0$ down to a strictly positive floor $\lambda_{\mathrm{res}}>0$, rather than to zero. Third, the policy-deviation estimator is evaluated not on a fixed support, but progressively over $\mathcal{D}_{\mathrm{off}}\cup\mathcal{D}(t)$. We motivate each choice in turn, and then show how they jointly complement one another to prevent the failure modes of Distribution Shift; we conclude with the resulting two-stage training protocol, formalising Stage~1 and Stage~2 in detail.

	\textbf{Piecewise-constant schedule for $\alpha,\beta$.} In full generality, $\alpha(t)$ and $\beta(t)$ could be treated as optimisable parameters, autonomously balancing live and pre-collected reward signals throughout training. In this work, their adaptivity is instead realised as a piecewise-constant schedule across training stages, via the composition of the training batches (Section~\ref{sec:mga_rl_instantiation}): a batch drawn entirely from $\mathcal{D}_{\mathrm{off}}$ yields $(\alpha,\beta)=(0,1)$, while a batch mixing live and pre-collected transitions yields intermediate values determined by their relative proportion. This choice requires no additional tuning beyond the batch composition itself, at the cost of foregoing continuous, gradient-based adaptation of $\alpha,\beta$, which we leave to future work.
	
	\textbf{Persistence of the floor $\lambda_{\mathrm{res}}$.} Under an idealised, unbounded interaction budget and full coverage of the state--action space, $\lambda$ would asymptotically vanish, since live experience alone would eventually suffice to characterise the optimal policy. In practice, both the interaction budget and the exploration horizon are finite, and coverage of $\mathcal{S}\times\mathcal{A}$ remains necessarily partial; $\lambda_{\mathrm{res}}$ therefore acts as insurance against residual out-of-distribution (OOD) states that the available budget does not permit the agent to visit. Unlike prior offline-to-online formulations~\citep{fujimoto2021minimalist, zhao2022adaptive}, which treat the imitation weight as a fixed constant or remove it via a hard switch between stages, \ac{mgarl} lets $\lambda$ decay gradually within the training objective itself; this is crucial to avoid the destabilising effects of Distribution Shift at the stage boundary~\citep{zhao2022adaptive, ball2023efficient}.
	
	\textbf{Extended support of $\hat{\mathcal{M}}_{\mathrm{MSE}}$.} Unlike classical \ac{bc}, where $\hat{\mathcal{M}}_{\mathrm{MSE}}$ (Eq.~\eqref{eq:mse_estimate}) is necessarily evaluated on $\mathcal{D}_{\mathrm{off}}$ alone --- the only data available to that paradigm, which involves no environment interaction --- \ac{mgarl} also collects live transitions into $\mathcal{D}(t)$. Since $\mu^{\mathrm{MB}}$ is an analytical policy queryable at any $s\in\mathcal{S}$, rather than a policy learned only over $\mathcal{D}_{\mathrm{off}}$, the imitation signal can be extended beyond the fixed support of $\mathcal{D}_{\mathrm{off}}$ to states encountered during live interaction as well. Accordingly, the policy-deviation estimator is evaluated as
	
	\begin{equation}
		\label{eq:mse_estimate_mgarl}
		\hat{\mathcal{M}}_{\mathrm{MSE}}^{\mathrm{MGA\text{-}RL}}(\theta) = \frac{1}{|\mathcal{D}_{\mathrm{off}}\cup\mathcal{D}(t)|} \sum_{(s,a,r,s')\in\mathcal{D}_{\mathrm{off}}\cup\mathcal{D}(t)} \left\lVert \mu_\theta(s) - \mu^{\text{MB}}(s) \right\rVert^2,
	\end{equation}
	
	which coincides with Eq.~\eqref{eq:mse_estimate} while $\mathcal{D}(t)=\emptyset$ (Stage~1), and progressively incorporates the states the agent actually visits as live transitions accumulate (Stage~2), rather than remaining confined to the states demonstrated offline.
	
	\textbf{Complementarity of $\hat{r}_{\mathrm{off}}$ and $\hat{\mathcal{M}}_{\mathrm{MSE}}$.} Under a limited computational and interaction budget, training is exposed to the two failure modes of Distribution Shift discussed in Section~\ref{sec:ddpg_bc_design}: Extrapolation Error, whereby a Critic trained solely on $\mathcal{D}_{\mathrm{off}}$ overestimates OOD actions that an unconstrained Actor inevitably queries; and Actor-Critic Misalignment, whereby a Critic trained without live feedback produces uninformative gradients upon transitioning online, degrading both Actor and Critic. $\hat{r}_{\mathrm{off}}$ and $\hat{\mathcal{M}}_{\mathrm{MSE}}$ jointly prevent both. $\hat{r}_{\mathrm{off}}$ provides the Critic with an initial value estimate from $\mathcal{D}_{\mathrm{off}}$, but an Actor trained to maximise it without constraint would query actions outside the offline support, with no guarantee of converging within budget --- precisely the Extrapolation Error above; $\hat{\mathcal{M}}_{\mathrm{MSE}}$ prevents it by driving Actor convergence towards $\mu^{\mathrm{MB}}$ during pre-training and anchoring the Actor against divergent, out-of-support behaviour during fine-tuning, confining optimisation to the region where $Q_\phi$ is grounded in data. Conversely, $\hat{\mathcal{M}}_{\mathrm{MSE}}$ alone lets the Actor imitate $\mu^{\mathrm{MB}}$ from Stage~1 onward, but provides no mechanism for the Critic to remain aligned with the Actor it must evaluate --- precisely the Actor-Critic Misalignment above; $\hat{r}_{\mathrm{off}}$ prevents it by keeping the Critic aligned with the Actor in Stage~1, and by preventing the Critic --- and, by extension, the Actor --- from forgetting the reference policy's actions in Stage~2, which represent the minimum acceptable behaviour. In this sense, $\hat{\mathcal{M}}_{\mathrm{MSE}}$ and $\hat{r}_{\mathrm{off}}$ each protect the other against the failure mode it would otherwise induce in isolation.
	
	Beyond these three design choices, \ac{mgarl} advances the state of the art along two further dimensions. \textbf{Smooth two-stage training}: Stages~1 and~2 are connected by the continuous decay of $\lambda\bigl(|\mathcal{D}(t)|\bigr)$, avoiding the abrupt phase boundary of \ac{o2o} methods~\citep{nair2020awac}. \textbf{General formulation with adaptive coefficients}: framing $\alpha$ and $\beta$ as optimisable parameters subsumes all prior methods as fixed-coefficient special cases of the \ac{gpr} (Table~\ref{tab:gpr_cases}) and opens a concrete research direction for future work.
	
	We now detail the resulting two-stage training protocol.

	\subsubsection{Stage 1 --- Pre-collected Data Only}
	\label{ssec:stage1}
	
	Before any live transitions are collected, $\mathcal{D}(t)$ is empty, so $\alpha=0$ and $|\mathcal{D}(t)|=0$, giving $\lambda\bigl(|\mathcal{D}(t)|\bigr)=\lambda_0$ at its maximum value. The objective of Eq.~\eqref{eq:mga_rl_objective} accordingly reduces to:
	\begin{equation}
		\label{eq:stage1_objective}
		\hat{J}^{\mathrm{MGA\text{-}RL}}_{\mathrm{S1}}(\theta) = \sum_t \gamma^t\beta\cdot\hat{r}_{\mathrm{off}}(t) - \lambda_0\cdot\hat{\mathcal{M}}_{\mathrm{MSE}}.
	\end{equation}
	$\hat{r}_{\mathrm{off}}(t)$ trains $Q_\phi$ on the value of behavior in $\mathcal{D}_{\mathrm{off}}$; $\hat{\mathcal{M}}_{\mathrm{MSE}}$ anchors $\mu_\theta$ to $\mu^{\mathrm{MB}}$, so the actor entering Stage~2 already approximates a structured, domain-consistent behavior rather than a random initialization. This implements an \emph{\ac{rlfd}-style pretraining}~\citep{hester2018deep} with $\mu^{\mathrm{MB}}$ as a universally queryable demonstrator.
	
	\subsubsection{Stage 2 --- Live Transitions with Decaying Regularisation}
	\label{ssec:stage2}
	
	Once live transitions begin to fill $\mathcal{D}(t)$, both terms of $r_{\mathrm{GPR}}$ activate, and the objective becomes:
	\begin{equation}
		\label{eq:stage2_objective}
		\hat{J}^{\mathrm{MGA\text{-}RL}}_{\mathrm{S2}}(\theta) = \sum_t \gamma^t\alpha\cdot\hat{r}_{\mathrm{on}}(t) + \beta\cdot\hat{r}_{\mathrm{off}}(t) - \lambda\bigl(|\mathcal{D}(t)|\bigr)\cdot\hat{\mathcal{M}}_{\mathrm{MSE}}^{\mathrm{MGA\text{-}RL}}.
	\end{equation}
	As live transitions accumulate, $|\mathcal{D}(t)|$ rises monotonically towards the buffer capacity $N_{\mathrm{buf}}$ (Eq.~\eqref{eq:replay_buffer}), and $\lambda\bigl(|\mathcal{D}(t)|\bigr)$ decays from $\lambda_0$ towards the floor $\lambda_{\mathrm{res}}$ discussed above. Since $|\mathcal{D}(t)|$ is a deterministic count driven solely by the data-collection process, this release is commensurate with the live experience the agent has effectively gathered, rather than with elapsed training steps alone.
	
	Section~\ref{sec:mga_rl_instantiation} instantiates each of these mechanisms concretely: the piecewise-constant schedule for $\alpha(t),\beta(t)$, the specific functional form of $\lambda\bigl(|\mathcal{D}(t)|\bigr)$ and its practical implications as $|\mathcal{D}(t)|$ grows over time, and the union $\mathcal{D}_{\mathrm{off}}\cup\mathcal{D}(t)$ underlying $\hat{\mathcal{M}}_{\mathrm{MSE}}^{\mathrm{MGA\text{-}RL}}$.

	\section{\acs{mgarl} for Deadline-Constrained Routing}
	\label{sec:mga_rl_instantiation}
	
	In this section, we instantiate \ac{mgarl} for deadline-constrained network routing, identifying each component of the \ac{gpr} with a concrete network-control quantity and specifying the algorithmic mechanisms required to stabilise the offline-to-online transition. The analytical reference policy is set to $\mu^{\text{MB}} = $ \ac{upgecps},  the prior-guided policy proposed in Section~\ref{ssec:UPG_Pstar}; the quantitative justification for this choice, based on its reliability--complexity trade-off relative to the other model-based policies, is deferred to the numerical results of Section~\ref{sec:results_model_based_approaches}. 
	The adaptive coefficients of case~(e) in Table~\ref{tab:gpr_cases} are realised concretely as follows: $\alpha,\beta$ are piecewise-constant per stage via batch composition, and $\lambda\bigl(|\mathcal{D}(t)|\bigr)$ decays as the effective size of the live replay buffer grows, as detailed in Sections~\ref{sec:gpr_to_actor_critic}-\ref{sec:ddpg_bc_design}.
	
	The scalar reward $r$ of Section~\ref{sec:reward_estimation} is likewise instantiated throughout this section, following the \ac{dcmt} problem~\citep{vitale2025flexible}, as the \emph{aggregated timely throughput}, i.e.\ the total number of packets delivered on time at each time slot: \begin{equation} \label{eq:reward} r(t) = \sum_{c\in\mathcal{C}} \sum_{\ell\in\mathcal{L}} f_{\rightarrow d^c}^{(c,\ell)}(t), \end{equation} where $f_{\rightarrow d^c}^{(c,\ell)}(t)$ is the aggregate number of commodity-$c$ packets with lifetime $\ell$ reaching destination $d^c$ at time $t$ via paths in $\mathcal{P}^c$ from the incoming neighbour set $\rho^-_{d^c}$.
	
	Section~\ref{sec:gpr_to_actor_critic} derives the Actor and Critic updates for the unified objective $\hat{J}^{\mathrm{MGA\text{-}RL}}(\theta)$, architecture-agnostically. Section~\ref{sec:ddpg_bc_design} specialises these updates to a \ac{ddpg}~\citep{lillicrap2015continuous} backbone augmented with Delayed Policy Update and Target Policy Smoothing~\citep{fujimoto2018addressing}, which serves as our reference Actor-Critic substrate; the proposed methodology, however, remains agnostic to the underlying Actor-Critic architecture. While \ac{ddpg}-\ac{bc} combinations have been previously explored in robotics and continuous control~\citep{vecerik2017leveraging, nair2018overcoming,rajeswaran2018learning, fujimoto2021minimalist}, their integration within an \ac{mgarl} protocol tailored to the deadline-constrained dynamics of \ac{rti} applications represents, to our knowledge, the first application of \ac{rlfd} principles to network traffic engineering. Section~\ref{ssec:o2o_training_protocol} formalises the procedural transition through the Two-Stage Training Protocol. Section~\ref{sec:marl_ec_p_star} introduces the \emph{Multi-Agent Reinforcement Learning EC~$p^*$} strategy, which integrates RL-based routing with Effective Congestion metrics and serves as the target policy in our experimental phase. Finally, Section~\ref{sec:mga_rl_positioning} positions \ac{mgarl} within the unified framework of Section~\ref{sec:policy_learning_paradigms}.

	\subsection{From GPR to Actor--Critic Updates}
	\label{sec:gpr_to_actor_critic}
	
	Section~\ref{sec:actor_critic_backbone} derived the Actor and Critic updates (Eqs.~\eqref{eq:dpg_gradient}--\eqref{eq:critic_update0}) from the single-objective return $J=\mathbb{E}[R(\xi)]$ of Eq.~\eqref{eq:return}. We now derive the analogous updates for the unified objective $\hat{J}^{\mathrm{MGA\text{-}RL}}(\theta)$ of Eq.~\eqref{eq:mga_rl_objective}, which combines $r_{\mathrm{GPR}}$ (Eq.~\eqref{eq:gpr}) with the policy-deviation term $-\lambda\bigl(|\mathcal{D}(t)|\bigr)\cdot\hat{\mathcal{M}}_{\mathrm{MSE}}^{\mathrm{MGA\text{-}RL}}$ (Eq.~\eqref{eq:mse_estimate_mgarl}).
	
	By linearity, the gradient of $\hat{J}^{\mathrm{MGA\text{-}RL}}(\theta)$ splits additively into two structurally distinct terms:
	\begin{equation}
		\label{eq:gradient_split}
		\nabla_\theta \hat{J}^{\mathrm{MGA\text{-}RL}}(\theta) = \nabla_\theta\sum_t\gamma^t r_{\mathrm{GPR}}(\alpha,\beta,t) - \nabla_\theta\Bigl[\lambda\bigl(|\mathcal{D}(t)|\bigr)\cdot\hat{\mathcal{M}}_{\mathrm{MSE}}^{\mathrm{MGA\text{-}RL}}\Bigr].
	\end{equation}
	The first term has exactly the form of $J=\mathbb{E}[R(\xi)]$ (Eq.~\eqref{eq:return}): it depends on the trajectory generated by future interaction with the environment, and is therefore subject to the first obstacle of Section~\ref{sec:actor_critic_intro} --- the transition kernel $P(s'\mid s,a)$ is unknown --- which is precisely what necessitates estimating $Q^{\mu_\theta}(s,a)$ via the Critic and invoking the DPG theorem. The second term, by contrast, is directly differentiable in closed form:
	
	\begin{equation}
		\label{eq:mse_gradient}
		\nabla_\theta\, \hat{\mathcal{M}}_{\mathrm{MSE}}^{\mathrm{MGA\text{-}RL}} = \frac{2}{\scriptstyle |\mathcal{D}_{\mathrm{off}}\cup\mathcal{D}(t)|} 
		\sum_{\substack{ (s,a,r,s')\in \\ \mathcal{D}_{\mathrm{off}}\cup\mathcal{D}(t) }} \bigl(\mu_\theta(s)-\mu^{\mathrm{MB}}(s)\bigr)\cdot\nabla_\theta\mu_\theta(s),
	\end{equation}
	
	requiring neither knowledge of $P$ nor any estimate of future consequences: $\mu^{\mathrm{MB}}$ is queryable at any $s\in\mathcal{S}$ by assumption (Section~\ref{sec:reward_estimation}), so Eq.~\eqref{eq:mse_gradient} is computed exactly, not approximated. The first obstacle therefore does not apply to the second term of Eq.~\eqref{eq:gradient_split}, which consequently bypasses $Q_\phi$ entirely and enters the Actor update directly.
	
	\textbf{Critic Update.} Following Eq.~\eqref{eq:gradient_split}, the Critic is updated using the first term of the gradient split alone, i.e., using $r_{\mathrm{GPR}}$. Generalising Eqs.~\eqref{eq:td_target}--\eqref{eq:critic_update0} to the mini-batch $\mathcal{B}$ introduced in Section~\ref{sec:actor_critic_backbone},
	\begin{equation}
		\label{eq:mixed_batch}
		\mathcal{B} = \mathcal{B}(t) \cup \mathcal{B}_{\mathrm{off}}, \qquad |\mathcal{B}(t)|=N,\ \ |\mathcal{B}_{\mathrm{off}}|=M,
	\end{equation}
	where $N$ and $M$ are the mini-batch sizes already introduced as training hyperparameters in Section~\ref{sec:reward_estimation}, and whose values are set as design parameters of the system, as detailed in Section~\ref{sec:ddpg_bc_design}. 
	
	The TD target of Eq.~\eqref{eq:td_target} uses the raw reward $r(t)$ of Eq.~\eqref{eq:reward}, observed in each sampled tuple $(s,a,r,s')\in\mathcal{B}$. 
	It is only at the level of the aggregated Critic loss (Eq.~\eqref{eq:critic_update0}) that $r_{\mathrm{GPR}}(\alpha,\beta,t)$ emerges. Among the possible ways of realising $r_{\mathrm{GPR}}(\alpha,\beta,t)$ (Eq.~\eqref{eq:gpr}) on this mixed batch, we choose to set $\alpha(t)$ and $\beta(t)$ as
	\begin{equation}
		\label{eq:alpha_beta_from_NM}
		(\alpha(t),\beta(t)) =
		\begin{cases}
			(0,\ 1), & t \le T_1, \\[4pt]
			\left(\dfrac{1}{1+\rho},\ \dfrac{\rho}{1+\rho}\right), & t > T_1,
		\end{cases}
		\qquad \rho := \frac{|\mathcal{B}_{\mathrm{off}}|}{|\mathcal{B}(t)|}.
	\end{equation}
	This choice avoids introducing $\alpha(t)$ and $\beta(t)$ as independently tuned hyperparameters: once $N$ and $M$ are fixed as the design parameters specified in Section~\ref{sec:ddpg_bc_design}, $\alpha(t)$ and $\beta(t)$ follow automatically.

	\textbf{Actor Update.} Following Eqs.~\eqref{eq:gradient_split} and~\eqref{eq:mse_gradient}, the policy-deviation term enters the Actor update as an additive regularisation term, alongside the DPG gradient of Eq.~\eqref{eq:dpg_gradient}:
	
	\begin{multline}
		\label{eq:actor_update_gpr}
		\nabla_\theta \hat{J}^{\mathrm{MGA\text{-}RL}}(\theta) = \mathbb{E}_{\mathcal{B}}\Bigl[\nabla_\theta \mu_\theta(s)\cdot\nabla_a Q_\phi(s,a)\big|_{a=\mu_\theta(s)}\Bigr] \\
		- \lambda\bigl(|\mathcal{D}(t)|\bigr)\, \nabla_\theta \hat{\mathcal{M}}_{\mathrm{MSE}}^{\mathrm{MGA\text{-}RL}}.
	\end{multline}
	
	The first term is identical in form to Eq.~\eqref{eq:dpg_gradient}, evaluated on the mini-batch $\mathcal{B}$; the second term directly penalises deviation of $\mu_\theta$ from $\mu^{\mathrm{MB}}$, with the coefficient $\lambda\bigl(|\mathcal{D}(t)|\bigr)$ --- decaying as the effective size of the live replay buffer grows (Section~\ref{sec:mgarl_design_rationale}) --- governing the strength of this anchor.
	
	\emph{Remark.} When $\lambda\bigl(|\mathcal{D}(t)|\bigr)=0$, Eq.~\eqref{eq:actor_update_gpr} reduces to Eq.~\eqref{eq:dpg_gradient}, and the Critic update is unaffected regardless of $\lambda$, since $\hat{\mathcal{M}}_{\mathrm{MSE}}^{\mathrm{MGA\text{-}RL}}$ never enters the TD target. Section~\ref{sec:ddpg_bc_design} instantiates both updates concretely on the DDPG backbone.
	
	\subsection{DDPG Instantiation of \ac{mgarl}}
	\label{sec:ddpg_bc_design}
	We now specialise the Actor and Critic updates of Section~\ref{sec:gpr_to_actor_critic} to the DDPG backbone, putting into practice the complementary roles of $\hat{r}_{\mathrm{off}}$ and $\hat{\mathcal{M}}_{\mathrm{MSE}}^{\mathrm{MGA\text{-}RL}}$ established in Section~\ref{sec:mgarl_design_rationale}. The resulting design combines four mechanisms, illustrated in detail in what follows: a Q-value normalisation that decouples the imitation weight $\lambda$ from the reward's absolute scale; a Symmetric Logarithmic (SymLog) transformation that compresses the sharp reward discontinuities typical of deadline-thresholded traffic; a buffer-size-dependent schedule that governs how $\lambda$ decays in practice; and a set of frozen input statistics that prevent distribution shift across the offline-to-online transition.
	
	\paragraph{Q-Normalised Actor Update}
	The RL and BC terms of Eq.~\ref{eq:actor_update_gpr} differ substantially in scale: the RL term inherits the magnitude of $Q_\phi(s,a)$, which varies with the reward's absolute scale (Eq.~\eqref{eq:reward} can range from a handful to several dozen packets per slot, depending on topology and load), whereas the BC term is a bounded distance between actions. Without correction, a large $Q_\phi$ would dominate the gradient regardless of $\lambda$, silencing the imitation anchor irrespective of its intended weight. To decouple $\lambda$ from this scale, we normalise the RL term by the batch-wise mean magnitude of $Q_\phi$, $\omega = \frac{1}{N+M}\sum|Q_{\phi}(s,a)|+\epsilon$~\citep{fujimoto2021minimalist}. The resulting gradient is evaluated over the mixed mini-batch $\mathcal{B}=\mathcal{B}(t)\cup\mathcal{B}_{\mathrm{off}}$, instantiating the generic mini-batch of Eq.~\eqref{eq:actor_update_gpr}; we repeat the resulting expression below for clarity, with the normalisation factor made explicit:

	\begin{multline}
		\label{eq:constrained_actor_update}
		\nabla_{\theta} \hat{J}^{\mathrm{MGA\text{-}RL}}(\theta)
		\approx
		\frac{1}{\omega}\,
		\mathbb{E}_{s \sim \mathcal{B}(t) \cup \mathcal{B}_{\mathrm{off}}}
		\Bigl[
		\nabla_{\theta}\mu_{\theta}(s)\cdot
		\nabla_{a} Q_{\phi}(s,a)\big|_{a=\mu_{\theta}(s)}
		\Bigr] \\
		- \lambda\bigl(|\mathcal{D}(t)|\bigr)\, \nabla_{\theta}\hat{\mathcal{M}}_{\mathrm{MSE}}^{\mathrm{MGA\text{-}RL}}
	\end{multline}
	
	With the RL term rendered scale-free, $\lambda$ acts as an interpretable relative weight between reward maximisation and imitation, regardless of the reward's absolute magnitude. The ascent step of Eq.~\eqref{eq:actor_ascent} applies unchanged. Unlike prior formulations~\citep{fujimoto2021minimalist, zhao2022adaptive}, which treat the imitation weight as a hard switch, integrating $\lambda$ directly into the BC term preserves the gradual, stage-boundary-free transition motivated in Section~\ref{sec:mgarl_design_rationale}.
	
	\paragraph{Critic Update with SymLog-Stabilised Targets}
	Since $r(t)$ (Eq.~\eqref{eq:reward}) is an integer-valued, deadline-thresholded quantity, clusters of packets can expire within the same slot near saturation, producing sharp step-to-step jumps and high-variance value targets.
	To compress these discontinuities and prevent \emph{Gradient Shock}, we apply a Symmetric Logarithmic (SymLog)~\citep{hafner2023mastering} transformation to the TD target of Eq.~\eqref{eq:td_target}:
	\begin{equation}
		\label{eq:symlog_target}
		y(r,s') = \mathrm{SymLog}(r) + \gamma\, Q_{\phi'}(s', \mu_{\theta'}(s')),
	\end{equation}
	where $\mathrm{SymLog}(x)=\mathrm{sign}(x)\ln(|x|+1)$ is treated as a fixed regression label through which no gradient is propagated; the Critic thus estimates the discounted return of $\mathrm{SymLog}(r(t))$, and we interpret Eq.~\eqref{eq:symlog_target} as a \emph{stabilised surrogate objective}. 
	Generalising Eq.~\eqref{eq:critic_loss0} to the mixed mini-batch, the resulting Critic loss is:
	\begin{multline}
		\label{eq:critic_loss_mgarl}
		L(\phi) =
		\alpha(t)\,
		\underbrace{
			\frac{1}{|\mathcal{B}(t)|}\!\!\sum_{(s,a,r,s') \in \mathcal{B}(t)}\!\!
			\bigl(y(r,s') - Q_{\phi}(s,a)\bigr)^{2}
		}_{\text{mean squared TD error on } \mathcal{B}(t)}
		+ \\
		\beta(t)\,
		\underbrace{
			\frac{1}{|\mathcal{B}_{\mathrm{off}}|}\!\!\sum_{(s,a,r,s') \in \mathcal{B}_{\mathrm{off}}}\!\!
			\bigl(y(r,s') - Q_{\phi}(s,a)\bigr)^{2}
		}_{\text{mean squared TD error on } \mathcal{B}_{\mathrm{off}}}.
	\end{multline}
	
	The descent step of Eq.~\eqref{eq:critic_update0} applies unchanged.
	
	\paragraph{Buffer-Dependent Anchor Release}
	
	We realise the dependence of $\lambda$ on $|\mathcal{D}(t)|$ established in Section~\ref{sec:mgarl_design_rationale} as
	
	\begin{equation}
		\label{eq:lambda_decay}
		\lambda\bigl(|\mathcal{D}(t)|\bigr) = \lambda_{0} \left( \frac{\lambda_{\mathrm{res}}}{\lambda_{0}}\right)^{\min\left(1,\, |\mathcal{D}(t)|/D_{\mathrm{dec}} \right)},
	\end{equation}
	
	where $D_{\mathrm{dec}}\le N_{\mathrm{buf}}$ sets the buffer size at which $\lambda$ reaches its floor $\lambda_{\mathrm{res}}$.
	
	\paragraph{Anchored Transfer Stabilisation Protocol}
	
	Z-score normalisation is applied to state inputs using fixed statistics (mean $\mu_\mathcal{D}$, standard deviation $\sigma_\mathcal{D}$) extracted exclusively from $\mathcal{D}_{\mathrm{off}}$, avoiding the input distribution shift that adaptive updates would introduce during Stage~2. We refer to the combination of the residual floor $\lambda_{\mathrm{res}}>0$~\citep{fujimoto2021minimalist}, the SymLog surrogate of Eq.~\eqref{eq:symlog_target}, and the frozen normalisation as the \emph{Anchored Transfer Stabilisation Protocol}.

	\subsection{Two-Stage Training Protocol}
	\label{ssec:o2o_training_protocol}
	
	The specialised Actor and Critic updates of Section~\ref{sec:ddpg_bc_design} are executed through the two-stage procedure formalised in Algorithm~\ref{alg:training_procedure}, which implements the \ac{mgarl} training protocol of Section~\ref{sec:mgarl_design_rationale}: Stage~1 corresponds to Eq.~\eqref{eq:stage1_objective}, Stage~2 to Eq.~\eqref{eq:stage2_objective}, and the transition between the two follows the buffer-dependent schedule of Eq.~\eqref{eq:lambda_decay} and Eq.~\eqref{eq:alpha_beta_from_NM}.

	\begin{algorithm}[h!]
		\small
		\caption{\ac{mgarl}: BC, Warm-up \& Accelerated Decay}
		\label{alg:training_procedure}
		\begin{algorithmic}[1]
			\State \textbf{Params:} $K$ offline Epochs, $E$ online Episodes, $W$ warm-up.
			\State \textbf{Init:} $\mathcal{D}_{\mathrm{off}}$
			\State \textbf{Stage 1: Pre-collected data only} ($\lambda \leftarrow \lambda_{0}$)
			\For{$\mathrm{epoch} = 1 \dots K$}
			\State Sample batch $\mathcal{B}_{\mathrm{off}} \sim \mathcal{D}_{\mathrm{off}}$.
			\State Compute $\omega$ on $\mathcal{B}_{\mathrm{off}}$.
			\State Update Actor, Critic (Eqs.~\eqref{eq:constrained_actor_update}, \eqref{eq:critic_loss_mgarl}) and Targets.
			\State Validate \& Save Best Model (raw reward).
			\State \textit{Evaluate Early Stopping Criterion.} \Comment{Early Stopping}
			\EndFor
			\State \textbf{Stage 2: Live transitions + decaying regularisation}
			\While{$ep < E$}
			\State $\mathcal{D}(t) \leftarrow$ Online Data.
			\If{$ep\ \%\ \mathrm{learn\_eps} == 0$}
			\State Sample Batches $\mathcal{B}(t)\sim \mathcal{D}(t)$ and $\mathcal{B}_{\mathrm{off}} \sim \mathcal{D}_{\mathrm{off}}$
			\State Compute $\omega$ on $\mathcal{B}_{\mathrm{off}}$ and $\mathcal{B}(t)$.
			\State Update Critic (Eq.~\eqref{eq:critic_loss_mgarl}).
			\If{$ep > W$} \Comment{End of Warm-up}
			\State Update Actor (Eq.~\eqref{eq:constrained_actor_update}).
			\State Update $\lambda$ (Eq.~\eqref{eq:lambda_decay}).
			\State Validate \& Update Best Online Model.
			\EndIf
			\State Update Targets.
			\EndIf
			\EndWhile
		\end{algorithmic}
	\end{algorithm}
	
	\paragraph{Stage 1: Pre-collected Data Only}
	
	The agent is pre-trained exclusively on $\mathcal{D}_{\mathrm{off}}$. Periodic validation is performed every $V_{\mathrm{freq}}$ epochs with exploration noise disabled; Early Stopping halts training if no improvement in validation reward is observed over a patience window $P$, determining the transition point $T_1$ in practice (Section~\ref{sec:mgarl_design_rationale}).
	
	\paragraph{Stage 2: Live Transitions with Decaying Regularisation}
	
	The agent is initialised with the best Stage-1 checkpoint and fine-tuned with live environment interactions. To prevent Catastrophic Forgetting, a \emph{Seeded Experience Replay} buffer~\citep{vecerik2017leveraging} is preloaded with $\mathcal{D}_{\mathrm{off}}$ and continuously updated with new online transitions. A \emph{Critic Warm-up} of $W$ episodes temporarily freezes the Actor, allowing the Critic to align rapidly with the true online rewards before Actor updates resume.
	
	\subsection{\acs{marlecps} Trained via \ac{mgarl}}
	\label{sec:marl_ec_p_star}
	
	Having specialised how \ac{mgarl} trains an Actor-Critic pair (Sections~\ref{sec:ddpg_bc_design}--\ref{ssec:o2o_training_protocol}), we now define the strategy in which this trained Actor-Critic pair operates: a hybrid approach that combines a distributed \ac{lelf} scheduler with a centralized RL-based router, following the architectural principles of \emph{\ac{madrl}~\ac{el}~\ac{lelf}}~\citep{vitale2025flexible}. While the distributed schedulers operate using the fixed \ac{lelf} policy (Section~\ref{sssec:lelf_scheduler}), the centralized routing agent relies entirely on Reinforcement Learning to dynamically route newly arrived packets, leveraging the EC~$p^*$ congestion metric. 
	We refer to this combination of \ac{lelf} scheduling and centralized RL-based routing as \ac{marlecps} --- omitting the \ac{lelf} scheduling component from the name for brevity, though it remains an integral part of the architecture. 
	
	The routing agent's interaction with the environment is defined by the following spaces:
	\begin{itemize}
		\item \emph{Observation Space:} The routing agent receives a global view of the network status, built by concatenating the exogenous arrival vector $\mathbf{b}(t)$ with the network congestion state in one of two forms:
		\begin{itemize}
			\item \emph{Vectorial:} Concatenation of EC~$p^*$ vectors $\hat{\mathbf{Q}}_{ij}^{(\mathcal{P}_{ij})}(t)$ for all interfaces, yielding a space of size $|\mathcal{C}|+\sum_{(i,j)\in\mathcal{E}}\max\!\left(0,L_{\max}-2T^{p^*}_{ij}\right)$.
			\item \emph{Scalar:} Concatenation of scalar EC~$p^*$ values $\hat{Q}_{ij}^{(\mathcal{P}_{ij})}(t)$, reducing the size to $|\mathcal{C}|+|\mathcal{E}|$.
		\end{itemize}
		\item \emph{Action Space:} The routing agent returns a route assignment for all commodities. For each commodity $c$, sub-action $a^c\in\mathbb{R}^{|\mathcal{P}^c|}$ represents traffic split ratios with $\sum_{k\in\mathcal{P}^c}a^c_k=1$. The global action $a$ is the concatenation of these sub-vectors, of total size $|\mathcal{P}|$.
	\end{itemize}
	
	The Vectorial form retains complete urgency profiles at the cost of a high-dimensional state space $\mathcal{O}(|\mathcal{C}|+|\mathcal{E}|\cdot L_{\max})$; the Scalar form compresses congestion to a single value per interface, reducing dimensionality to $\mathcal{O}(|\mathcal{C}|+|\mathcal{E}|)$ and facilitating faster convergence, at the expense of fine-grained queue visibility. Regardless of the chosen granularity, \ac{marlecps} is strictly agnostic to the training paradigm: it can be trained either fully online or through \ac{mgarl}, leveraging $\mu^{\mathrm{MB}}=$ UPG~EC~$(p^*)$ as the bootstrap heuristic.
	
	The complete system proposed in this work results from the composition of three independent design choices: the EC~$p^*$ congestion metric (Section~\ref{ssec:effective_congestion_p_star_metric}), in its Vectorial or Scalar form, as observation feature; \ac{marlecps} (this section) --- centralized RL router with distributed \ac{lelf} scheduling --- as control structure; and \ac{mgarl} (Sections~\ref{sec:gpr}--\ref{sec:ddpg_bc_design}) as training protocol. Recall that the centralized RL router is the sole component trained via \ac{mgarl}. 
	Precisely because \ac{marlecps} is agnostic to the training paradigm, we refer to this specific composition as \emph{\ac{marlecps} Vectorial, trained via \ac{mgarl}} or \emph{\ac{marlecps} Scalar, trained via \ac{mgarl}}, depending on the observation form, to distinguish it from the same architecture trained fully online
	
	\subsection{Relation to the \ac{gpr} Taxonomy and \ac{rlfd} Literature}
	\label{sec:mga_rl_positioning}
	
	Table~\ref{tab:gpr_cases} summarises how \ac{mgarl} --- case~(e) --- relates to the fixed-coefficient special cases (a)--(d) discussed in Section~\ref{sec:gpr_special_cases}. Beyond this positioning within the \ac{gpr} framework, three further aspects distinguish \ac{mgarl} from existing \ac{rlfd} work~\citep{hester2018deep, vecerik2017leveraging, nair2018overcoming, rajeswaran2018learning}, which is largely concentrated in robotics. \emph{First}, the demonstrator is a deterministic, computationally light prior-guided heuristic (UPG~EC~$(p^*)$), programmatically generated, removing the data-acquisition cost typical of robotic settings and aligning with the RLfD-from-imperfect-demonstrations line~\citep{gao2018reinforcement, wu2020rlfd}. \emph{Second}, the \ac{bc} regulariser is integrated into a \ac{ddpg} backbone subject to per-commodity flow-conservation constraints, enforced architecturally through a grouped-Softmax output layer (Section~\ref{sec:neural_network_architecture}) rather than via soft penalties. \emph{Third}, the entire pipeline is tailored to deadline-constrained traffic engineering, a domain in which, despite extensive \ac{rl} surveys~\citep{riosguiral2025leveraging}, \ac{rlfd} has not been previously applied.

	\section{Experimental Setting}
	\label{sec:experimental_setting}

	\begin{table*}[htb]
		\centering
		\caption{Structural characteristics comparison of the three network topologies.}
		\label{tab:networks_structures_comparison}
		
		\begin{tabular*}{\textwidth}{@{\extracolsep{\fill}}lccc@{}}
			\toprule
			\textbf{Characteristic}              & \textbf{Hierarchical} & \textbf{Abilene}    & \textbf{Grid}      \\
			\midrule
			Topology Type                  & Tree-like             & Real-world backbone & Regular mesh       \\
			Number of Nodes                & 7                     & 11                  & 9                  \\
			Number of Edges/Interfaces               & 22                    & 28                  & 24                 \\
			Graph Diameter                 & 3                     & 5                   & 4                  \\
			Avg. Node Degree               & 6.29                  & 5.09                & 5.33               \\
			Average Clustering Coefficient & 0.3333                & 0.1515              & 0.0000             \\
			\bottomrule
		\end{tabular*}
	\end{table*}

	\subsection{Considered Network Topologies}
	
\begin{figure}[h]
		\centering
		\includegraphics[width=0.3\textwidth]{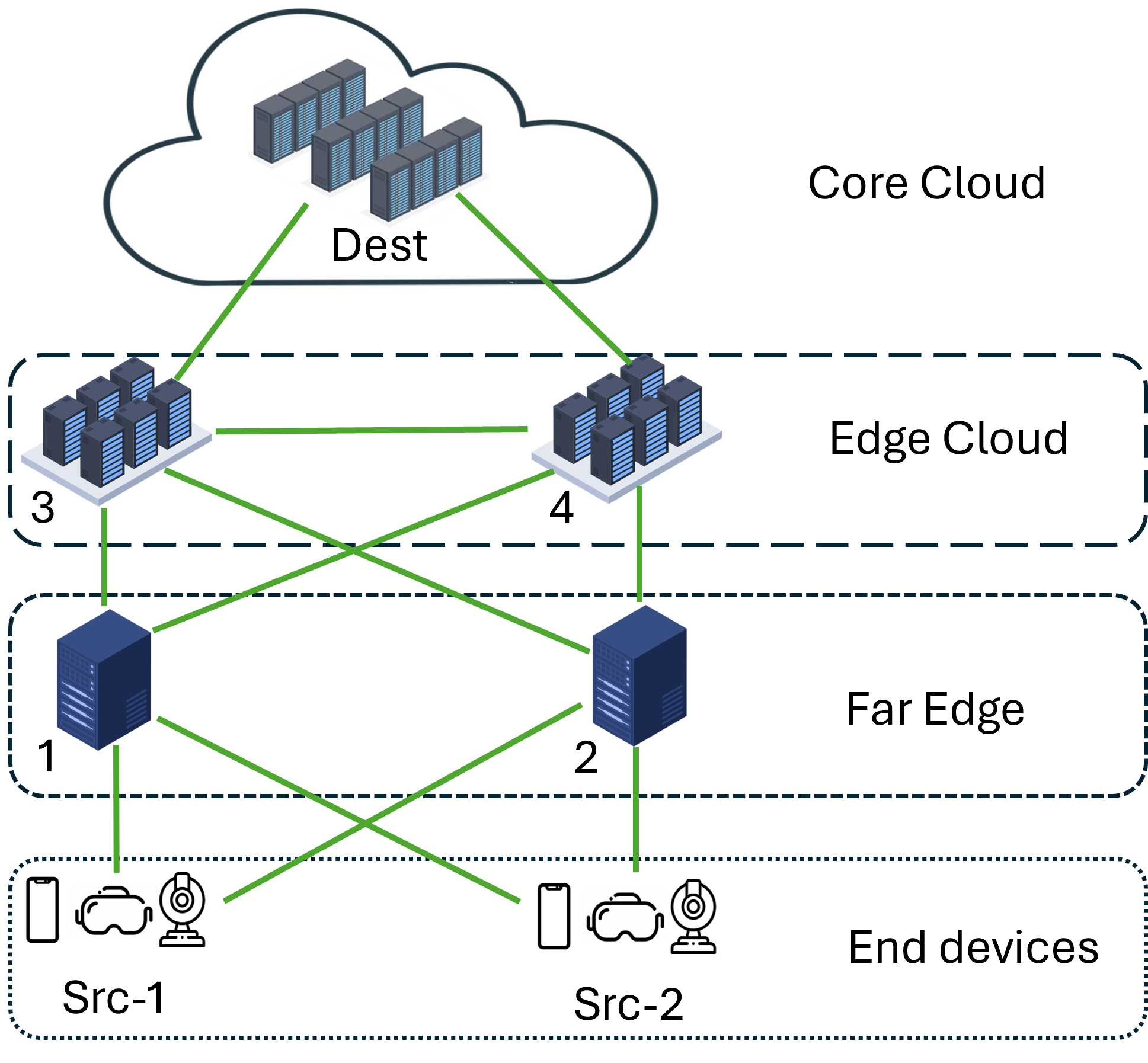}
		\caption{\small{The \textbf{Hierarchical} network topology.}}
		\label{fig:network_1_hierarchical}
		\vspace{-0.4cm}
	\end{figure}

\begin{figure}[h]
		\centering
		\includegraphics[width=0.5\textwidth]{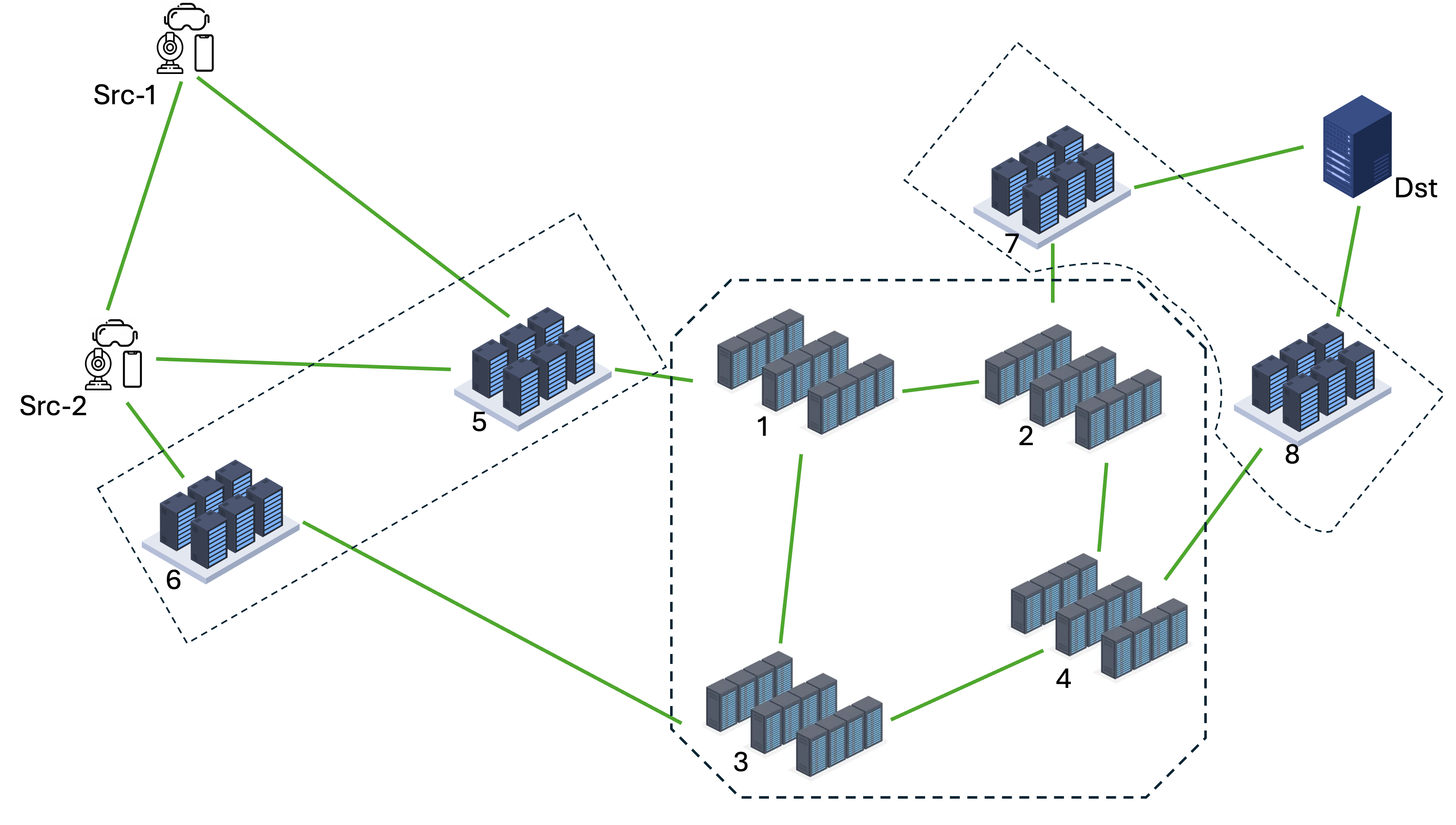}
		\caption{\small{The \textbf{Abilene} network topology.}}
		\label{fig:network_3_abilene}
		\vspace{-0.4cm}
	\end{figure}

\begin{figure}[h]
		\centering
		\includegraphics[width=0.4\textwidth]{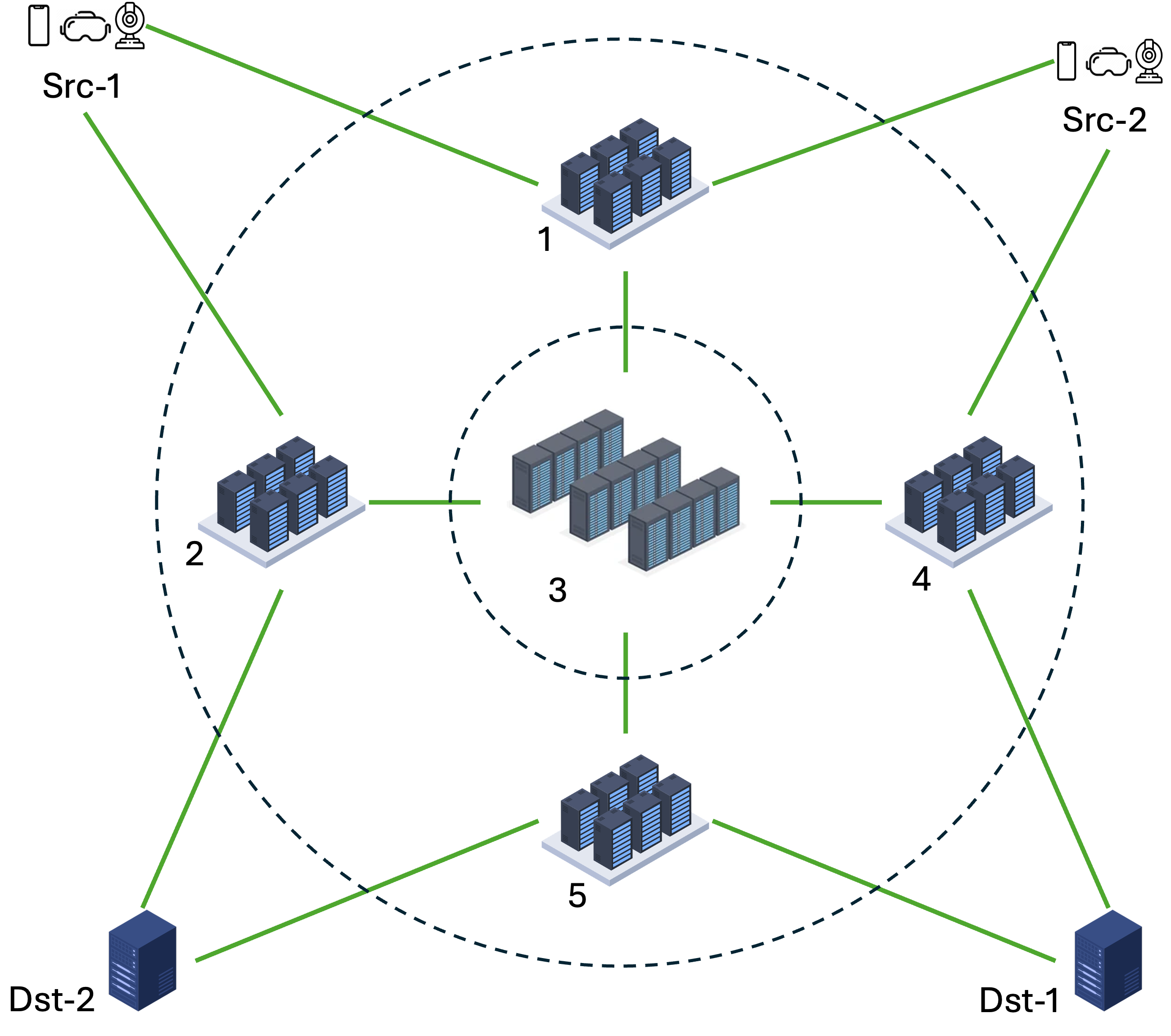}
		\caption{\small{The \textbf{Grid 3×3} network topology.}}
		\label{fig:network_2_grid}
		\vspace{-0.4cm}
	\end{figure}

	To comprehensively assess our proposed methodologies, we selected three network topologies that progressively increase in complexity: Hierarchical, Abilene, and Grid $3\times3$ (Figures~\ref{fig:network_1_hierarchical}, \ref{fig:network_3_abilene}, and \ref{fig:network_2_grid}).
	This selection enables a systematic evaluation across increasing levels of complexity, naturally illustrating the evolution of routing algorithms.

	Table~\ref{tab:networks_structures_comparison} provides a detailed structural comparison of these topologies, highlighting their key differences in connectivity patterns, node degrees, and clustering properties. In these bidirectional networks, the number of edges and interfaces are identical. The graph diameter represents the longest shortest path between any two nodes, while the average node degree indicates the average number of connections per node. All three topologies exhibit strong connectivity, ensuring reachability between all node pairs.
	
	\paragraph{Hierarchical Topology: Foundation for Edge Computing}
	The \emph{Hierarchical} topology (Fig.~\ref{fig:network_1_hierarchical}) represents a paradigmatic edge-computing infrastructure that mirrors real-world deployment scenarios. It features a clear hierarchy with a \emph{Core Cloud data center} serving as the destination, connected to Edge Cloud and Far Edge nodes. At the network periphery, two \emph{source nodes} represent IoT devices generating traffic requiring timely processing at the core.
	
	This topology exhibits a tree-like structure with high node centrality and strong hierarchical organization, making it less prone to routing conflicts. The clear routing preferences inherent to the hierarchy reduce decision complexity while maintaining strong connectivity via efficient paths (diameter of 3 hops). This makes the topology well-suited for evaluating prior-guided routing strategies, as the structured paths yield intuitive decisions effectively captured by our baseline \ac{mwprc} and \ac{upg} algorithms.
	
	\paragraph{Abilene Topology: Real-World Complexity and Asymmetry}
	
	The \emph{Abilene} topology (Fig.~\ref{fig:network_3_abilene}), based on the real-world Internet2 Abilene backbone, introduces asymmetric characteristics that reflect the complexity of practical network deployments. Traffic originates from two source nodes on the left and must reach a shared destination on the right, creating an inherently unbalanced distribution that challenges the efficiency of flow management.
	
	This topology presents a challenging environment with heterogeneous node-centrality levels, asymmetric path distributions (16 paths for one commodity versus 12 for the other), and the largest graph diameter (5 hops). The moderate clustering coefficient (0.15) indicates local patterns typical of real networks, creating bottlenecks and congestion points that are difficult to anticipate with simple prior-guided approaches. Consequently, Abilene serves as a good test case for adaptive learning algorithms, which can exploit the network's asymmetric structure to improve overall performance.
	
	\paragraph{Grid Topology: Symmetric Complexity and Path Diversity}
	The \emph{Grid $\mathbf{3\times3}$} topology (Fig.~\ref{fig:network_2_grid}) represents a regular mesh structure that can model distributed sensor networks, vehicular communication grids, or urban IoT deployments. It features a cross-diagonal traffic pattern that maximizes spatial separation while forcing flows to traverse the grid's overlapping central regions.
	Despite its symmetric structure, the Grid topology presents significant challenges due to high path diversity (12 paths per commodity) and multiple equivalent alternatives. The zero clustering coefficient indicates a pure grid structure without local clustering, yielding a homogeneous yet complex environment in which no single path is obviously superior. This makes the Grid topology the ultimate proving ground for our RL approach, where efficient exploration of the equivalent routing options during offline pre-training demonstrates the potential of hybrid methodologies.
	
	\paragraph{Experimental Configuration}
	The experimental parameters detailed in Table~\ref{table:networks-parameters-comparison} are calibrated to stress-test our approaches under varying load conditions (30\%--120\% of the Min-Cut Capacity). In the table, ``Interface Subset'' denotes the number of directed network interfaces considered, while ``Scale'' shows the ratio of nodes to interfaces. The ``\# of paths per commodity'' specifies the number of routing paths ($|\mathcal{P}^{1}|$ and $|\mathcal{P}^{2}|$), and ``Paths with length \textit{X}'' represents the total count of paths across all commodities. The ``Min-Cut Capacity'' denotes the maximum aggregate throughput the network can sustain under the considered commodities arrangment. Suggested arrival rates specify the parameter ranges used in performance evaluation, where packets arrive according to Poisson distributions. Finally, ``Suggested Lifetimes'' indicates the initial packet lifetime values tested for each topology.
	
	\begin{table*}[htbp]
		\centering
		\caption{Experimental parameters comparison for the three test network topologies. The Link Capacity is 10 packets per time slot for each link. The suggested commodities are reported in each topology image and are indicated with [Src-1,Dst-1], [Src-2,Dst-2].}
		\label{table:networks-parameters-comparison}
		
		\begin{tabular*}{\textwidth}{@{\extracolsep{\fill}}lccc@{}}
			\toprule
			\textbf{Parameter}            & \textbf{Hierarchical}                    & \textbf{Abilene}                           & \textbf{Grid}                              \\
			\midrule
			
			Interface Subset              & 11                                       & 18                                         & 20                                         \\
			Scale (Node/Interface Subset) & 0.63                                     & 0.61                                       & 0.45                                       \\
			
			\# of paths per commodity     & \(|\mathcal{P}^{1}|=6, |\mathcal{P}^{2}|=6\) & \(|\mathcal{P}^{1}|=16, |\mathcal{P}^{2}|=12\) & \(|\mathcal{P}^{1}|=12, |\mathcal{P}^{2}|=12\) \\
			\hline
			\textit{Paths with length 3}  & 8                                        & --                                         & --                                         \\
			
			\textit{Paths with length 4}  & 4                                        & --                                         & 12                                         \\
			
			\textit{Paths with length 5}  & --                                       & 3                                          & --                                         \\
			
			\textit{Paths with length 6}  & --                                       & 9                                          & 8                                          \\
			
			\textit{Paths with length 7}  & --                                       & 10                                         & --                                         \\
			
			\textit{Paths with length 8}  & --                                       & 5                                          & 4                                          \\
			
			\textit{Paths with length 9}  & --                                       & 1                                          & --                                         \\
			\hline
			
			Min-Cut Capacity              & 20                                       & 20                                         & 30                                         \\
			
			Suggested Arrival Rates       & 6, 12, 18, 20, 24                        & 6, 12, 18, 20, 24                          & 9, 18, 27, 30, 36                          \\
			
			Suggested Lifetimes           & 3, 5, 7                                  & 5, 8, 11                                   & 4, 7, 10                                   \\
			\bottomrule
		\end{tabular*}
		\vspace{-0.5cm}
	\end{table*}

	\subsection{RL Approaches Settings}
	This section describes the parameters used for \ac{rl}-based approaches. Note that, all training runs presented in this work were executed on identical hardware, and wall-clock measurements are normalized against the longest run per topology. This relative formulation isolates the impact of the training paradigm from absolute hardware performance, ensuring that the reported reductions reflect the methodological contribution rather than implementation-specific factors.
	
	\subsubsection{Routing Agent's Neural Network Architecture}
	\label{sec:neural_network_architecture}
	The Actor network is implemented as a Multi-Layer Perceptron (MLP) designed to map the high-dimensional observation state $s$ to the continuous action space of routing probabilities. The architecture consists of an input layer matched to the observation dimension, followed by two hidden layers with $n_1=128$ and $n_2=64$ neurons, respectively.
	\begin{figure}[h!]
		\center
		\includegraphics[width=0.6\linewidth]{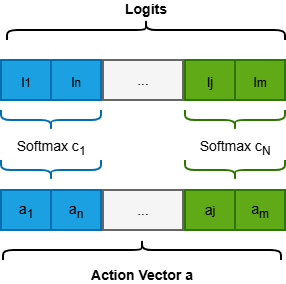}
		\caption{Visual representation of the Grouped Softmax activation mechanism. The raw logits are logically partitioned into sub-vectors corresponding to individual commodities ($c_1, \dots, c_N$). A Softmax function is then applied independently to each partition. This ensures local flow conservation, meaning the routing probabilities sum to 1 exclusively within each commodity group.
			The normalized segments are finally concatenated to form the global Action Vector $a$.}
		\label{fig:nn_scheme_2_grouped_softmax}
		\vspace{-0.2cm}
	\end{figure}
	A distinguishing feature of our architecture is the structure of the output layer, which must satisfy the flow conservation constraints for multiple commodities simultaneously. The raw output of the final linear layer (logits) is partitioned into logical groups, where each group corresponds to the set of feasible paths $\mathcal{P}^c$ associated with a specific commodity $c$. To ensure valid routing configurations, we apply a \emph{grouped} \textit{Softmax} \citep{bridle1990probabilistic} activation: the Softmax function is applied independently to each sub-vector of logits belonging to a commodity $c$. Formally, for every commodity $c \in \mathcal{C}$ and path $p \in \mathcal{P}^c$, the corresponding action $a_{p}^c$ is computed as:
	\[ a_{p}^c = \frac{\exp(z_{p})}{\sum_{k \in \mathcal{P}^c} \exp(z_{k})} \]
	where $z$ represents the raw logits. This mechanism guarantees that $\sum_{p \in \mathcal{P}^c} a_{p}^c = 1$ for all $c$, ensuring that 100\% of the traffic for each commodity is distributed among its feasible paths. The final action vector $a$ is the concatenation of these sub-vectors.
	
	\subsubsection{Choice of the Actor-Critic Backbone} We instantiate the proposed \ac{mgarl} methodology on \ac{ddpg}, the rationale is twofold: First, the routing agent action must satisfy per-commodity flow conservation ($\sum_{p \in \mathcal{P}^c} a^c_p = 1$, $a^c_p \in [0,1]$), which we enforce architecturally through the grouped-Softmax layer at the actor's Neural Network output (Sec .~\ref {sec:neural_network_architecture}), where the network directly outputs the exact continuous action. In contrast, employing stochastic policy algorithms (e.g., PPO or SAC) would require the network to output distribution parameters, introducing significant computational overhead to sample actions that strictly adhere to the flow conservation constraint. Second, the stabilization mechanisms employed operate at the training-objective level: the BC term regularizes the Actor, while the SymLog transformation defines the compressed reward surrogate used in the Critic target. Since these modifications are structurally agnostic to the underlying architecture, the \ac{ddpg} acts as the minimal substrate to support the deterministic, constrained traffic splitting required by our network model. On this substrate, we further adopt target policy smoothing and delayed policy and target updates~\citep{fujimoto2018addressing} to temper the target-value variance induced by stochastic traffic.
	
	\subsubsection{Online Training Hyperparameters}
	For the fully online training setting, the learning process comprises $14{,}000$ episodes of 50 time steps each, organized in two phases: an initial training phase of $10{,}000$ episodes, followed by an improvement phase of $4{,}000$ episodes in which the replay buffer is flushed of all previously collected transitions and training restarts from the best-performing policy of the first phase, discarding the low-quality experience accumulated during early exploration. Both Actor and Critic networks are optimized using the Adam optimizer \citep{kingma2014adam} with a learning rate of $1 \cdot 10^{-3}$, following the update rules of Eqs.~\eqref{eq:dpg_gradient_sample}--\eqref{eq:critic_update0}, with a batch size of $4096$ transitions. To balance exploration and exploitation, with probability $\epsilon$ the Actor output is perturbed with additive Gaussian noise and each per-commodity sub-vector is re-normalized to preserve the flow-conservation constraint $\sum_{p \in \mathcal{P}^c} a^c_p = 1$; $\epsilon$ is initialized to $1.0$ and has a decay rate of $0.95$ after each episode.
	
	\subsubsection{\acs{mgarl} Training Hyperparameters}
	\label{sec:hyperparameters}
	\noindent
	With respect to the fully online training setting, the \acs{mgarl} introduces additional parameters governing the transfer from the deterministic demonstrator. Their values depend on both the network topology and the demonstrator's standalone performance. The rationales behind the choices of the most critical choices are detailed below.
	
	\paragraph{Training Durations} 
	The \acs{mgarl} offline pre-training phase runs for up to 200 epochs. Subsequently, the online fine-tuning phase is strictly limited to $2{,}000$ episodes\footnote{The training budget consumption begins once $\mathcal{D}(t)$ has accumulated enough live transitions to sample a full mini-batch $\mathcal{B}(t)$ of size $N$.}, with a final test phase of $500$ episodes. The seeded replay augments each live batch with pre-collected samples amounting to a fraction $\rho=0.25$ of the live batch size.

	\paragraph{Learning Rates \& Regularization} To promote stable behavior cloning during pre-training, we use a conservative learning rate for both Actor and Critic ($\eta_{act}^{off} = \eta_{crit}^{off} = 1 \cdot 10^{-4}$). Upon online deployment, the Critic's learning rate increases to $\eta_{crit}^{on} = 1 \cdot 10^{-3}$ for rapid adaptation to the dynamic reward landscape, while the Actor's remains at $1 \cdot 10^{-4}$ to mitigate the risk of policy collapse. To mitigate the risk of overfitting on the static dataset, we apply $L_2$ regularization (weight decay $10^{-5}$) exclusively to the Critic, leaving the Actor unregularized to preserve its structural logit magnitude.
	
	\paragraph{Target Update ($\tau$) \& Warm-up ($W$)} We utilize a conservative soft update parameter $\tau = 0.005$ across all phases (see Algorithm~\ref{alg:training_procedure}, lines~7 and~23). This acts as a low-pass filter against the high-frequency variance of Poisson traffic arrivals, mitigating gradient shock. To smooth the initial transition to live-interaction dynamics, we enforce a Critic Warm-up period of $W=50$ episodes (line~18).
	
	\paragraph{BC Factor ($\lambda$)} 
	The initial imitation factor is set to $\lambda_{0} = 1.6$. In the normalized-$Q$ form of Eq.~\eqref{eq:constrained_actor_update}, the balance recommended by \citet{fujimoto2021minimalist} ($\alpha = 2.5$, equivalent to a relative BC weight of $1/\alpha = 0.4$) refers to actions bounded in $[-1,1]$; since the grouped Softmax confines our actions to $[0,1]$, the per-component BC deviation shrinks by up to a factor of $4$, and $\lambda_{0} = 4 \cdot 0.4 = 1.6$ restores the recommended imitation--reinforcement balance. 
	During the transition to online fine-tuning, $\lambda$ decays according to Eq.~\eqref{eq:lambda_decay}, reaching the residual floor $\lambda_{\mathrm{res}}$ once the live replay buffer accumulates $D_{\mathrm{dec}}$ transitions. We set $D_{\mathrm{dec}}$ so that this occurs after a fraction $E_{\mathrm{decay}}$ of the $2{,}000$-episode online fine-tuning budget (50 time steps per episode): $D_{\mathrm{dec}} = E_{\mathrm{decay}} \times 2{,}000 \times 50$.
	For near-optimal demonstrators (e.g., the Hierarchical topology Sec.~\ref{sec:results_model_based_approaches}), we enforce a strong, persistent constraint ($\lambda_{res}=0.8, E_{decay}=40\%$). Conversely, for sub-optimal prior-guided policies (e.g., Abilene, Grid), we employ a rapid decay to a loose constraint ($\lambda_{res}=0.2, E_{decay}=15\%$). The floor $\lambda_{res}$ is kept strictly positive since abruptly removing the \ac{bc} constraint could drive the Critic to diverge as soon as the agent explores \ac{ood} states\citep{ball2023efficient}.
	
	\paragraph{Offline Validation and the Stage-1/Stage-2 Transition} Offline validation is performed every $V_{\mathrm{freq}}=1$ epoch, with exploration noise disabled, and determines the Stage-1/Stage-2 transition $T_1$ (Section~\ref{sec:mgarl_design_rationale}): pre-training halts, and Stage~2 begins, as soon as validation reward shows no improvement over a patience window $P$. $T_1$ is therefore not a fixed epoch count, but adapts to how quickly the Actor and Critic converge on $\mathcal{D}_{\mathrm{off}}$ for each topology and demonstrator quality. The patience parameter $P$ is architecture-dependent: $P=40$ for Vectorial architectures, which tend to oscillate more during learning, and $P=20$ for Scalar architectures, which exhibit more compact convergence.
	
	\paragraph{Online Validation and Model Selection} Once in Stage~2, validation is relaxed to $V_{\mathrm{freq}}=20$ episodes to properly assimilate the dynamic environmental variance.
	Validation begins only once the Critic Warm-up of $W=50$ episodes has elapsed and Actor updates resume, consistently with Algorithm~\ref{alg:training_procedure};
	unlike its offline counterpart, it triggers no further stage transition, and only selects the best checkpoint over the fixed $2{,}000$-episode fine-tuning budget. In both stages, model selection is based on validation reward rather than training loss: losses measure adherence to the static dataset $\mathcal{D}_{\mathrm{off}}$ rather than sequential control performance~\citep{ross2011reduction}, rewarding overfitting if used for selection, whereas evaluation rollouts probe the policy on the states it actually induces, sidestepping the offline model-selection problem~\citep{paine2020hyperparameter} and favouring generalisation to unseen conditions.
	
	\paragraph{Dimensionality-Aware Tuning} To stabilize gradients under near-saturation traffic regimes, the Batch Size $N$ is selectively scaled. Specifically, Hierarchical topologies maintain standard batches ($N=4096$), while Abilene and Grid topologies scale to macro-batches of $N=8192$ when the arrival rate approaches their respective Min-Cut capacities $20.0$ and $30.0$ (Table \ref{table:networks-parameters-comparison}). Furthermore, the number of gradient updates ($U$) per step is modulated based on observation granularity. Compressed Scalar architectures, prone to State Aliasing, are strictly limited to $U=2$ (Hierarchical), $U=4$ (Abilene), and $U=6$ (Grid) to mitigate overfitting to stochastic drops that aggregate states cannot represent. Conversely, high-dimensional Vectorial architectures require more updates, so the parameter $U$ is scaled proportionally to the initial lifetime, which enlarges the observation space. We assign $U=5,8,10$ for $L = 3, 5, 7$ on Hierarchical, $ U=8,12,15$ for $L = 5, 8, 11$ on Abilene, and $U=7,10,13$ for $L = 4, 7, 10$ on Grid, respectively.
	
	\subsection{Evaluation Metrics}
	To assess the routing strategies presented in this work, we adopt two complementary metrics: \emph{reliability}, which captures the network-wide effectiveness in meeting packet deadlines, and \emph{spatial drop rate}, which localises where and how severely this effectiveness breaks down across individual interfaces. Together, they support both the aggregate performance comparisons and the interface-level analyses reported in Section~\ref{sec:numerical_results}.

	\paragraph{Reliability}
	The primary evaluation metric for our routing strategies is reliability, defined as the time‑average ratio of timely throughput to the aggregate arrival rate, namely the number of packets successfully delivered within their deadline to the total number of generated packets. This metric captures the core objective of deadline-constrained routing, reflecting the algorithm's effectiveness in meeting temporal requirements under varying network conditions.
	\begin{equation}
		Reliability = \mathbb{E}\!\left[ \frac{\text{Timely\_Throughput}(t)}{ \text{Arrival\_Rate}(t)}\right] \in [0,1]
	\end{equation}
	
	\paragraph{Spatial Drop Rate}
	To gain deeper insights into network congestion dynamics and the spatial footprint of different routing policies, we evaluate the spatial drop rate. This metric quantifies the time-averaged number of packets dropped---either due to deadline expiration or active queue management---at each individual network interface $(i,j) \in \mathcal{E}$. By analyzing this metric, we can effectively identify localized congestion hotspots and evaluate the load-balancing capabilities of the routing algorithms.
	\begin{equation}
		D_{ij} = \mathbb{E}\!\left[ \text{Dropped\_Packets}_{ij}(t) \right], \quad \forall (i,j) \in \mathcal{E}
	\end{equation}
	
	\section{Numerical Results}
	\label{sec:numerical_results}
	
	This section presents the numerical evaluation of the proposed framework. Section~\ref{ssec:considered_approaches} introduces the compared routing strategies and the experimental baselines; Section~\ref{sec:results_model_based_approaches} evaluates the prior-guided heuristics of Section~\ref{sec:model_based_policies}; Section~\ref{sec:results_rl_based_approaches} assesses the RL-based approaches trained via \ac{mgarl}; and Section~\ref{ssec:best_approach_by_objective} summarises the best-performing policy for each operational objective.
	
	\subsection{Considered Approaches}
	\label{ssec:considered_approaches}
	To evaluate the performance of our latency-aware routing framework, we compare a diverse set of network control strategies, broadly categorized into prior-guided heuristics and data-driven RL approaches. In the following, we outline the full set of compared strategies---starting from the prior-guided heuristics, then the data-driven RL approaches---followed by the theoretical performance ceiling against which their reliability is measured; a closing remark then clarifies the rationale guiding our choice of baselines relative to classical network controllers. All reported metrics (reliability and spatial drop rate) are computed over a 500-episode test phase per approach.
	
	\paragraph{Prior-guided Heuristic Approaches} All five policies of Section~\ref{sec:model_based_policies} are included in the comparison: \ac{mwprc}, \ac{mwpecp}, \ac{mwpecps}, \ac{upgecp}, and \ac{upgecps}.

	\paragraph{RL-Based Approaches} 
	For the data-driven strategies based on \ac{ec}, we focus on \ac{marlecps}, presented in Section~\ref{sec:marl_ec_p_star} in its Vectorial and Scalar forms. Combining each form with either training paradigm --- \ac{mgarl}  or fully Online from scratch.  --- yields four configurations, denoted, as shorthand, \emph{\ac{marlecps} Vectorial \ac{mgarl}}, \emph{\ac{marlecps} Scalar \ac{mgarl}}, \emph{\ac{marlecps} Vectorial Online}, and \emph{\ac{marlecps} Scalar Online}. 
	Comparing \emph{\ac{mgarl}} against \emph{Online} isolates the specific contribution of the \ac{mgarl} training protocol, since the two variants differ only in how the Actor-Critic pair is trained.
	Two further agents share the same \ac{marlecps} architecture but observe the network through the baseline congestion metrics of Secs.~\ref{sec:regular_congestion}--\ref{sec:lifetime_aware_congestion} instead of \ac{ec}: \textit{\ac{marlrc} Scalar Online}, a variant of the routing agent originally proposed in \citet{vitale2025flexible} using the volume-based Regular Congestion metric, and \textit{\ac{marllac} Vectorial Online}, using the deadline-agnostic Lifetime-Aware Congestion metric. Together with the four \ac{marlecps} configurations, these complete the set of six \ac{rl}-based approaches, isolating the contribution of the \ac{ec} congestion metric from that of the training protocol. Throughout the experimental analysis, we abbreviate  these last two schemes as \emph{\ac{madrl} RC Scalar Online} and \emph{\ac{madrl} LAC Vectorial Online}.
	
	\paragraph{Theoretical Upper Bound} As a reference ceiling for the reliability of all compared strategies, we consider an optimistic upper bound, defined as the ratio of the network's Min-Cut Capacity to the aggregate arrival rate:
	\begin{equation}
		\min\left(1, \frac{\text{Min-Cut Capacity}}{\text{Aggregate Arrival Rate}}\right)    
	\end{equation}
	It remains at $1.0$ until the injected traffic exceeds the absolute physical limits of the network, providing a fundamental ceiling for concurrent flow throughput.
	
	\paragraph{Remark (Baseline Scope)} We restrict our baselines to methods operating under the same EL-driven queueing and LELF scheduling studied here, rather than to classical cloud-network controllers, for which a head-to-head comparison would be methodologically misleading rather than informative. UMW~\citep{sinha2017optimal} and UCNC~\citep{zhang2021optimal} optimize loop-free average delay or throughput, while RCNC~\citep{cai2022ultra} enforces lifetimes through an LDP formulation; none yields deadline-aware routing weights compatible with the EL/LELF setting studied here, and re-deriving them into comparable deadline-constrained routers is itself a non-trivial research effort --- undertaken in \citet{vitale2025flexible} --- and beyond our present scope. We therefore benchmark against the matched, published \ac{mwprc} router of \citet{vitale2025flexible} and against from-scratch online agents, all operating under identical EL-driven queueing and LELF scheduling, so that observed differences are attributable to the routing metric and training paradigm rather than to mismatched objectives.

	\subsection{Prior-guided Approaches Results}
	\label{sec:results_model_based_approaches}
	
	This section evaluates the five prior-guided policies of Section~\ref{sec:model_based_policies} across the three network topologies. We first examine reliability and spatial drop patterns topology by topology, then draw cross-topology observations, and conclude by identifying the best-performing prior-guided policy overall.
	
	\begin{figure*}[htb!]
		\centering
		\begin{subfigure}{0.85\textwidth}
			\centering
			\caption{Hierarchical}\label{fig:thru_model_based_hierarchical_sub}
			\includegraphics[width=\linewidth]{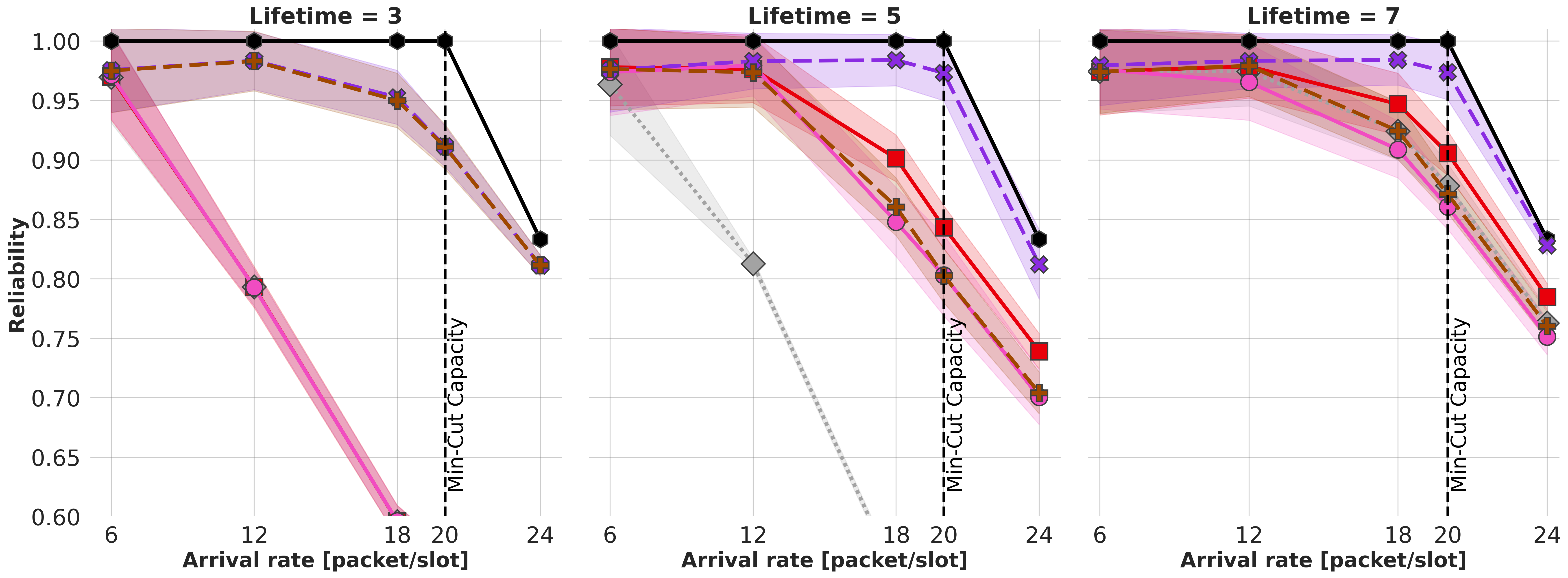}
			\vspace{-0.5cm}
		\end{subfigure}
		\begin{subfigure}{0.85\textwidth}
			\centering
			\caption{Abilene}\label{fig:thru_model_based_abilene_sub}
			\includegraphics[width=\linewidth]{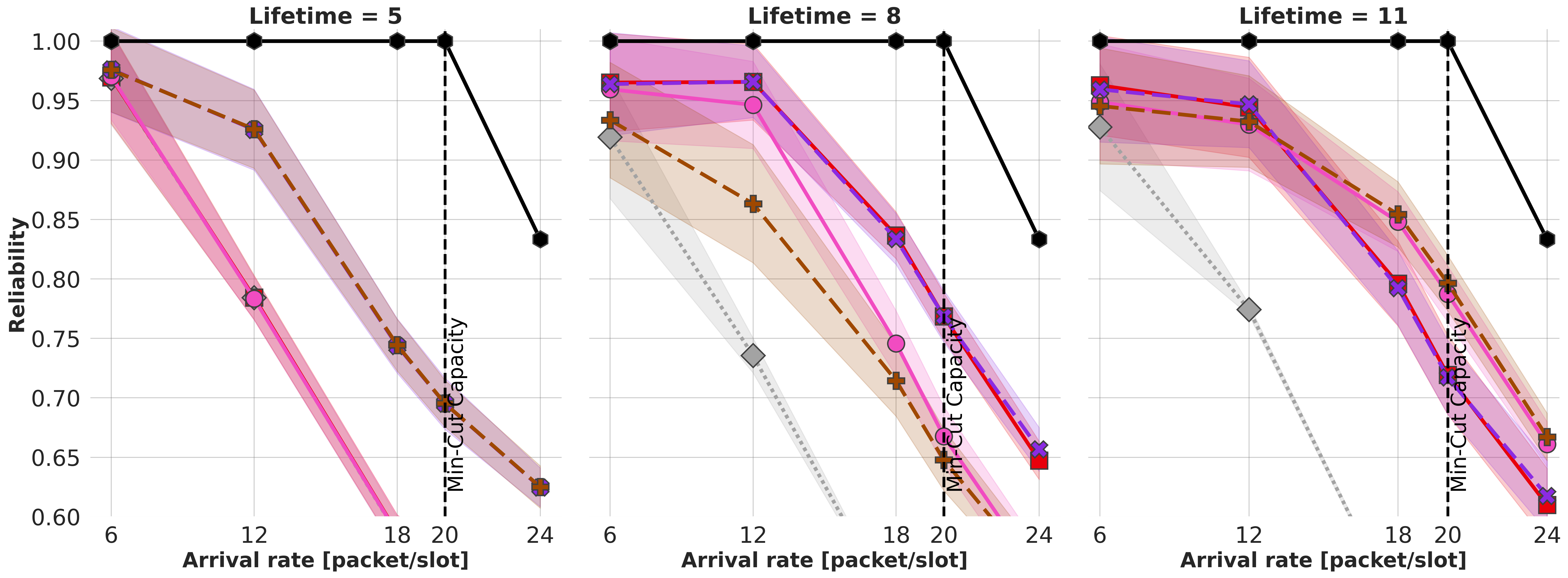}
			\vspace{-0.5cm}
		\end{subfigure}
		\begin{subfigure}{0.85\textwidth}
			\centering
			\caption{Grid}\label{fig:thru_model_based_grid_sub}
			\includegraphics[width=\linewidth]{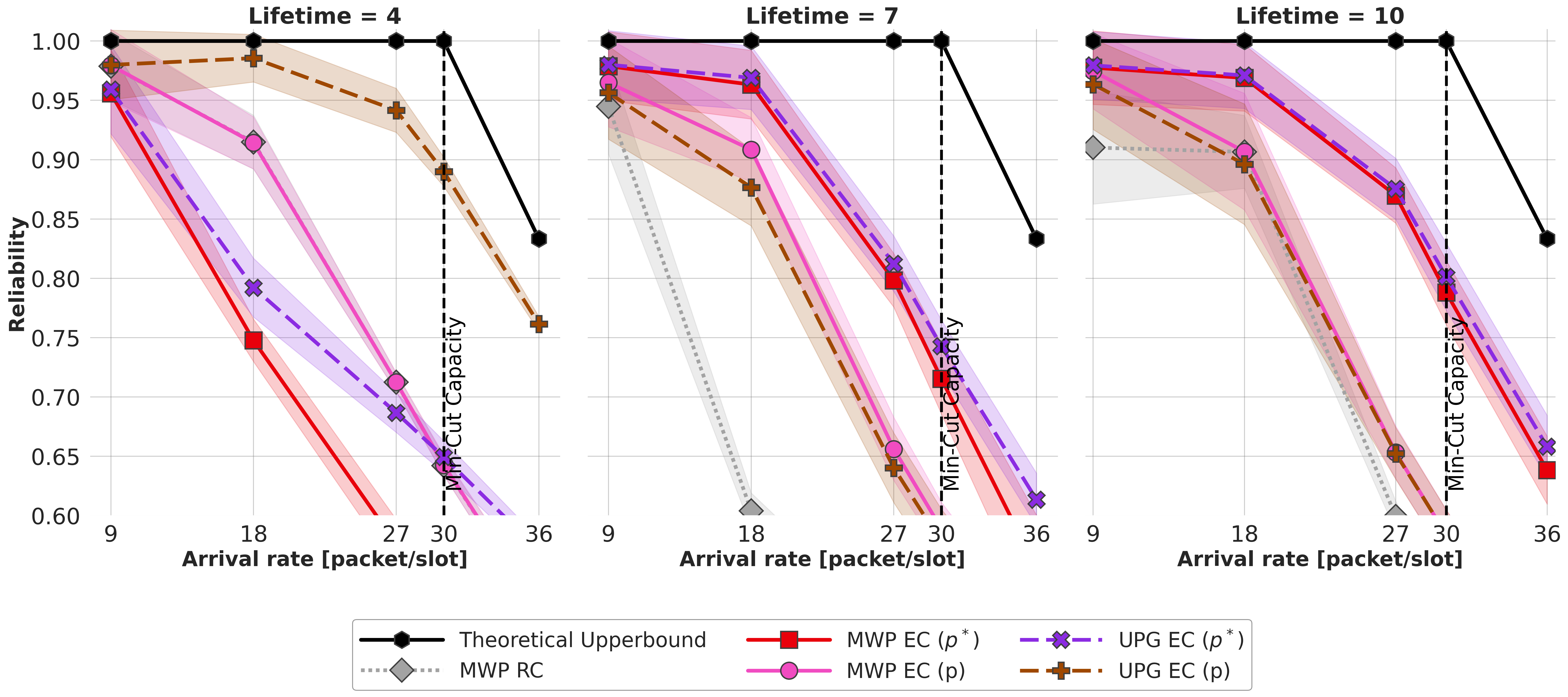}
			\vspace{-0.5cm}
		\end{subfigure}
		\caption{Reliability performance comparison of prior-guided approaches on the considered network topologies. The line's shading indicates the standard deviation (SD).}
		\label{fig:model_based_policies_comparison}
		\vspace{-0.5cm}
	\end{figure*}
	
	Figure~\ref{fig:model_based_policies_comparison} reports the reliability of the five prior-guided policies across the three network topologies, as traffic load and packet lifetime vary.
	
	\begin{figure*}[ht]
		\begin{subfigure}[t]{0.95\textwidth}
			\centering
			\includegraphics[width=\linewidth]{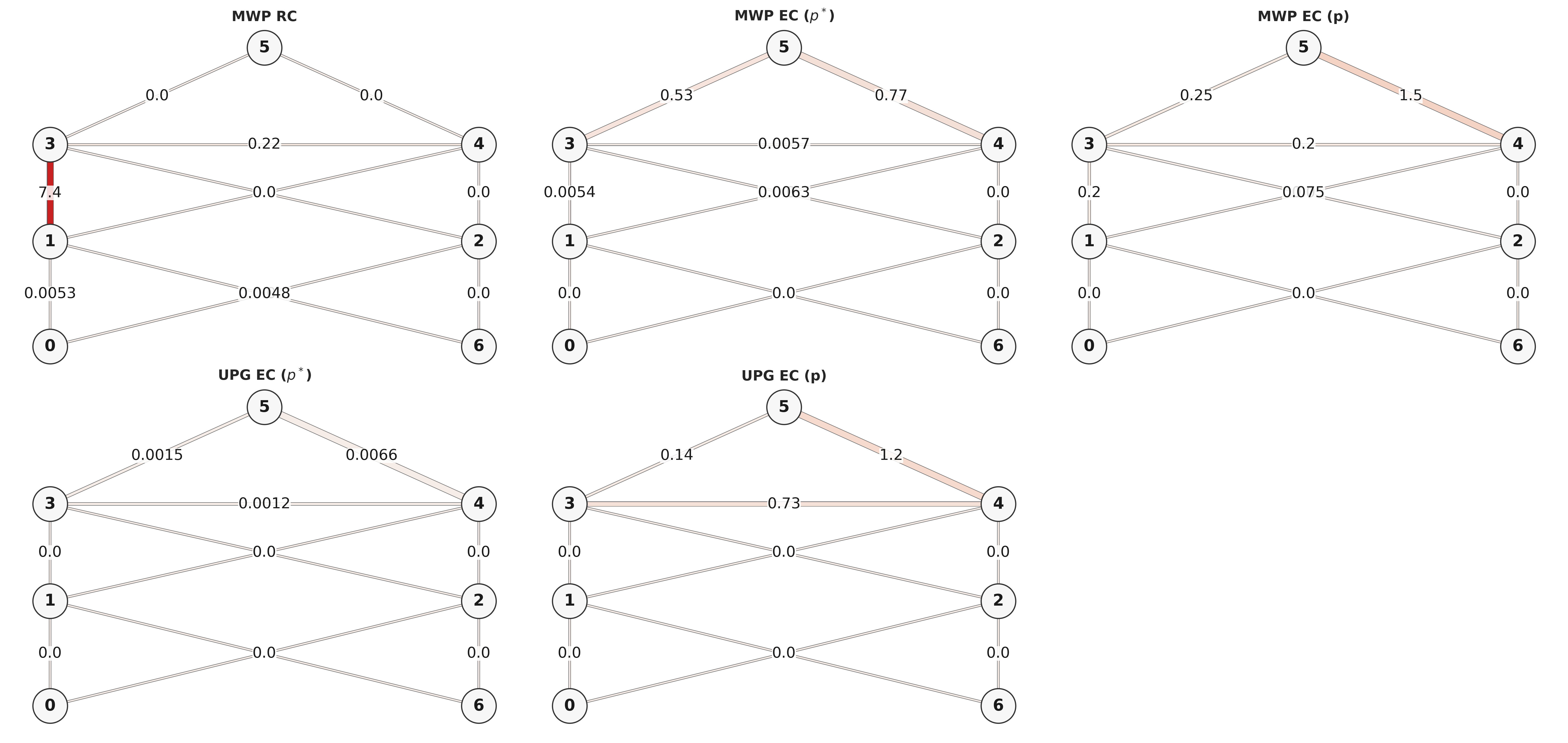}
			\label{fig:expired_hierarchical_spatial_topology}
			\vspace{-0.5cm}
		\end{subfigure}
		\centering
		\begin{subfigure}[c]{0.95\textwidth}
			\centering        
			\includegraphics[width=\linewidth,height=0.85\textheight,keepaspectratio]{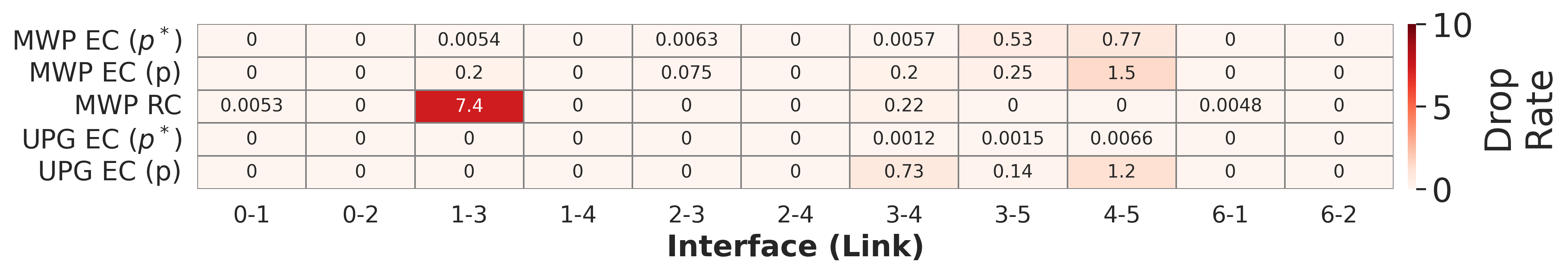}
			\label{fig:expired_hierarchical_spatial_heatmap}
			\vspace{-0.5cm}
		\end{subfigure}
		\caption{\textbf{Spatial distribution of expired packets for prior-guided approaches on Hierarchical topology} with $L=5, b=18$. The top figure reports the dropped packet counts on each edge of the network topology. The bottom figure reports the same information at the interface level in the form of a heatmap. While \ac{mwprc} heavily concentrates drops at a few bottleneck interfaces (e.g., 4-5 and 3-4), EC-based approaches distribute the load, with \ac{upg} variants effectively eliminating localized hotspots.}
		\label{fig:expired_hierarchical_spatial}
		\vspace{-0.5cm}
	\end{figure*}
	
	\paragraph{Hierarchical Topology}Figure~\ref{fig:thru_model_based_hierarchical_sub} reports the reliability performance for the Hierarchical topology across three packet lifetimes ($L \in \{3, 5, 7\}$) and varying traffic loads (up to $b=24$ packets/slot). Three distinct regimes emerge as the lifetime increases (i.e., the deadline relaxes). Under the \textit{tight} deadline ($L=3$), only shortest paths are feasible, the congestion metric becomes irrelevant under greedy assignment, and all \ac{mwp} variants collapse at peak load while \ac{upg}-based policies sustain above 80\% reliability by spreading traffic across equivalent shortest paths. Under the \textit{intermediate} deadline ($L=5$), additional paths become viable and the limitation of volume-based metrics surfaces: \ac{mwprc} drops by over 40\% relative to \ac{ec}-based policies at $b=24$, while \ac{upgecps} remains the most robust by combining urgency-aware filtering with balanced distribution. Under the \textit{relaxed} deadline ($L=7$), queuing slack absorbs sub-optimal routing and the gap between greedy and \ac{upg} narrows, although \ac{mwprc} remains consistently penalized by its inability to filter expiring traffic. The spatial drop distribution of Figure~\ref{fig:expired_hierarchical_spatial} ($L=5,\, b=18$) visually confirms the mechanism: \ac{mwprc} concentrates expirations on links 4-5 and 3-4, whereas EC-based policies, particularly \ac{upg} variants, eliminate these hotspots.
	
	\paragraph{Abilene Topology}
	This topology introduces real-world asymmetries, heterogeneous node centralities, and distinct bottleneck links, making load distribution critical. In Figure~\ref{fig:thru_model_based_abilene_sub}, we report the reliability performance for this topology across lifetimes $L \in \{5, 8, 11\}$, with the theoretical Min-Cut Capacity at $b=20$. Under the \textit{tight} deadline ($L=5$), the absence of queuing margin exposes a $\sim$15\% reliability gap between greedy and balanced assignments at the Min-Cut Capacity. Under the \textit{intermediate} deadline ($L=8$), \ac{mwprc} remains heavily penalized, underperforming by over 30\% compared to \ac{ec}-based routing at $b=18$. Under the \textit{relaxed} deadline ($L=11$), \ac{mwprc} still plummets to over 35\% below the best configuration at $b=20$; yet, Abilene's structural asymmetry highlights a performance gap for the decoupled approximation, where \ac{upgecps} sacrifices over 5-6\% reliability compared to the exact path-dependent evaluation of \ac{upgecp}. The spatial drop analysis of Figure~\ref{fig:expired_abilene_spatial} ($L=11,\, b=18$) visually confirms these trends: \ac{mwprc} concentrates expirations at specific bottleneck interfaces, whereas EC-based policies, particularly when coupled with \ac{upg}, reduce hotspot intensity by proactively routing traffic away from saturated paths.
	
	\paragraph{Grid $3\times3$ Topology}
	This topology represents a regular mesh structure with high symmetry and multiple equivalent paths, making it highly susceptible to central bottlenecks. In Figure~\ref{fig:thru_model_based_grid_sub}, we analyze this topology across lifetimes $L \in \{4, 7, 10\}$, with the theoretical Min-Cut Capacity at $b=30$. Under the \textit{tight} deadline ($L=4$), the decoupled interface-level evaluation of $p^*$ oversimplifies the congestion state, exposing a massive $\sim$36\% reliability gap compared to the exact path-dependent evaluation of \ac{upgecp}, which is strictly necessary to navigate the central bottleneck. Under the \textit{intermediate} deadline ($L=7$), \ac{mwprc} collapses under heavy load, underperforming by over 20\% compared to \ac{ec}-based routing; however, the performance gap between exact and approximated metrics narrows significantly, as the increased lifetime provides sufficient margin for the $p^*$ approximation to correct sub-optimal local routing decisions. Under the \textit{relaxed} deadline ($L=10$), \ac{ecps} and \ac{ecp} policies cluster around 85\% and 65\% reliability, respectively, at $b=27$, whereas \ac{mwprc} falls below 80\% at the saturation threshold ($b=30$), revealing a sizable performance gap compared to the $p^*$ approximation, which efficiently distributes diagonal flows. The spatial drop analysis of Figure~\ref{fig:expired_grid_spatial} ($L=10,\, b=27$) visually confirms these mechanics: \ac{mwprc} forces cross-diagonal traffic directly through the grid's center, generating severe core congestion, whereas EC-based approaches dynamically bypass the central region, improving load uniformity and mitigating cascading congestion.
	
	\paragraph{Cross-Topology Observations}
	A joint analysis of the results across all three topologies reveals the following insights:
	
	\begin{itemize}
		\item \em{Superiority of EC metrics across varying loads.} Regardless of the network structure, as the arrival rate increases and the lifetime tightens, metrics based on \ac{ec} systematically outperform traditional congestion routing. By filtering queued traffic that cannot compete with the routed packets upon its arrival, \ac{ec} prevents expiring packets from inflating the congestion estimates, securing relative reliability gains ranging from 20\% to over 40\% when the network approaches saturation. This proactive filtering directly reduces peak interface congestion, effectively preventing the localized packet drops that affect the \ac{mwprc} baseline.
		
		\item \em{\ac{upg} vs.\ greedy assignment.} The benefit of \ac{upg} over greedy assignment scales with the regime rather than with the topology itself. Under tight deadlines or near saturation, greedy variants funnel traffic into a single low-weight path, triggering rapid hotspot formation and a sharp reliability drop regardless of topology. The \ac{upg} mechanism mitigates this funnelling effect by spreading traffic across paths of comparable weight, with gains that grow with network symmetry and path diversity (most pronounced in Grid, where multiple equivalent routes coexist).
		
		\item \em{Robustness of the $p^*$ approximation.} The lightweight \ac{ecps} model performs consistently across most of the considered settings. It relies on a single reference path to compute interface congestion, thereby reducing computational complexity with respect to EC~$p$. The exact model (EC~$p$) maintains an edge in highly asymmetric environments (Abilene) or under strict deadlines in symmetric grids; the $p^*$ approximation sacrifices only a few percentage points of reliability across the remaining operational regimes.
	\end{itemize}
	\paragraph{Best Prior-guided Policy}
	Across the considered topologies, the UPG EC policies---both in their exact ($p$) and approximated ($p^*$) formulations---consistently deliver the highest reliability under stress, with peak gains of up to $40\%$ over the \ac{mwprc} baseline. The spatial drop maps (Figs.~\ref{fig:expired_hierarchical_spatial}, \ref{fig:expired_abilene_spatial}, \ref{fig:expired_grid_spatial}) confirm the underlying mechanism: \ac{upgecps} preserves the urgency-aware filtering of the exact metric while mitigating the localized hotspots that can affect greedy variants under heavy load. This is precisely why \ac{upgecps} was anticipated in Section~\ref{sec:mga_rl_instantiation} as the reference-policy/demonstrator for \ac{mgarl}.
	
	\begin{figure*}[htb!]
		\centering
		\begin{subfigure}{0.85\textwidth}
			\centering
			\caption{\textbf{Hierarchical}}\label{fig:thru_ddpg_based_finetuned_hierarchical_sub}
			\includegraphics[width=\linewidth]{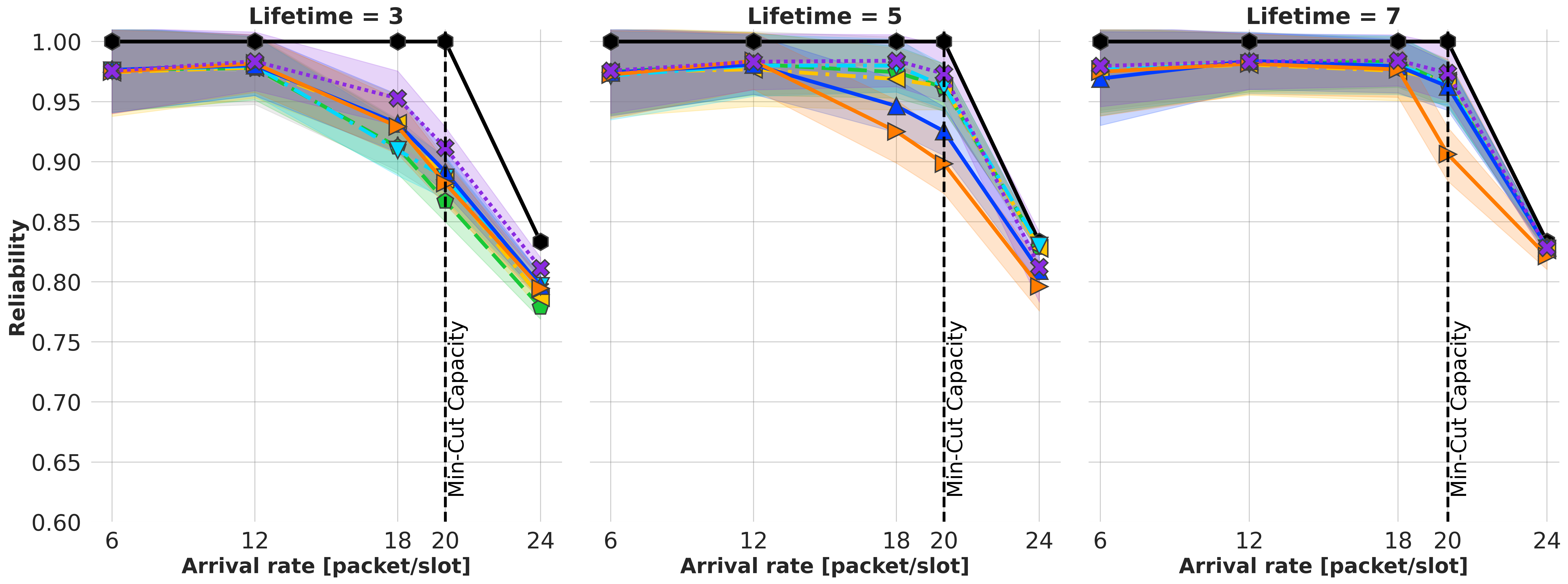}
			\vspace{-0.5cm}
		\end{subfigure}
		
		\begin{subfigure}{0.85\textwidth}
			\centering
			\caption{\textbf{Abilene}}\label{fig:thru_ddpg_based_finetuned_abilene_sub}
			\includegraphics[width=\linewidth]{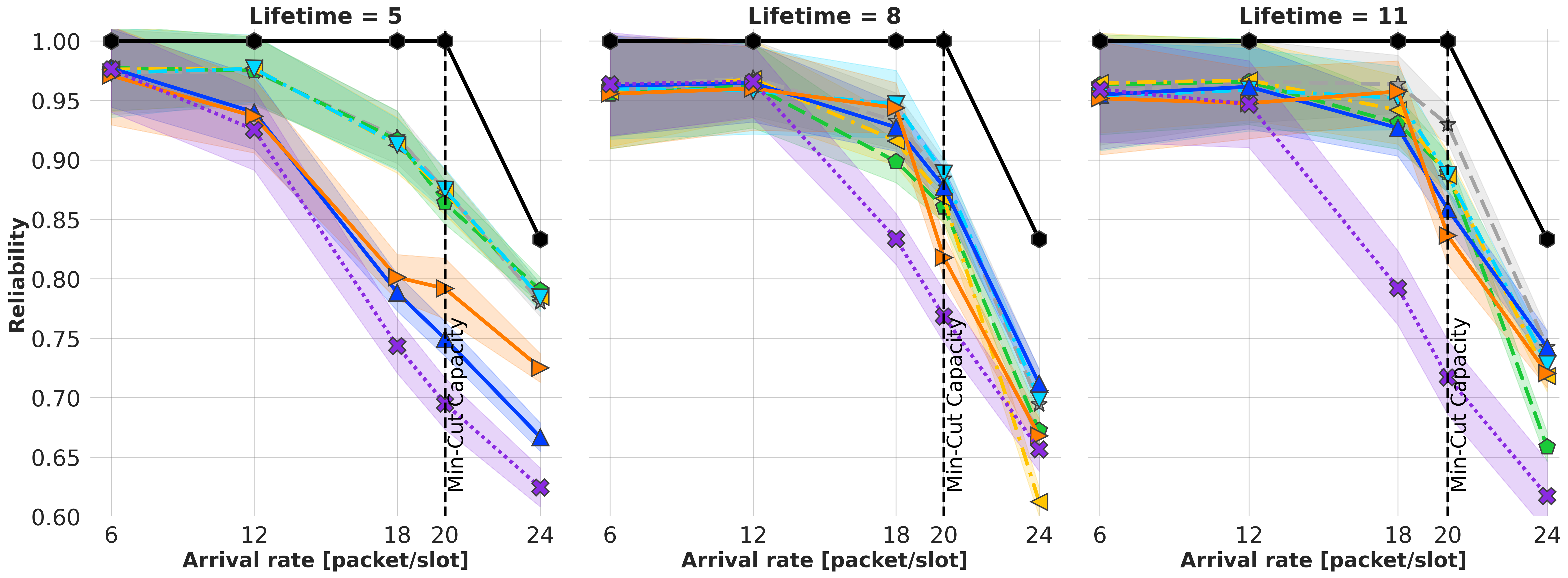}
			\vspace{-0.5cm}
		\end{subfigure}
		
		\begin{subfigure}{0.85\textwidth}
			\centering
			\caption{\textbf{Grid}}\label{fig:thru_ddpg_based_finetuned_grid_sub}
			\includegraphics[width=\linewidth]{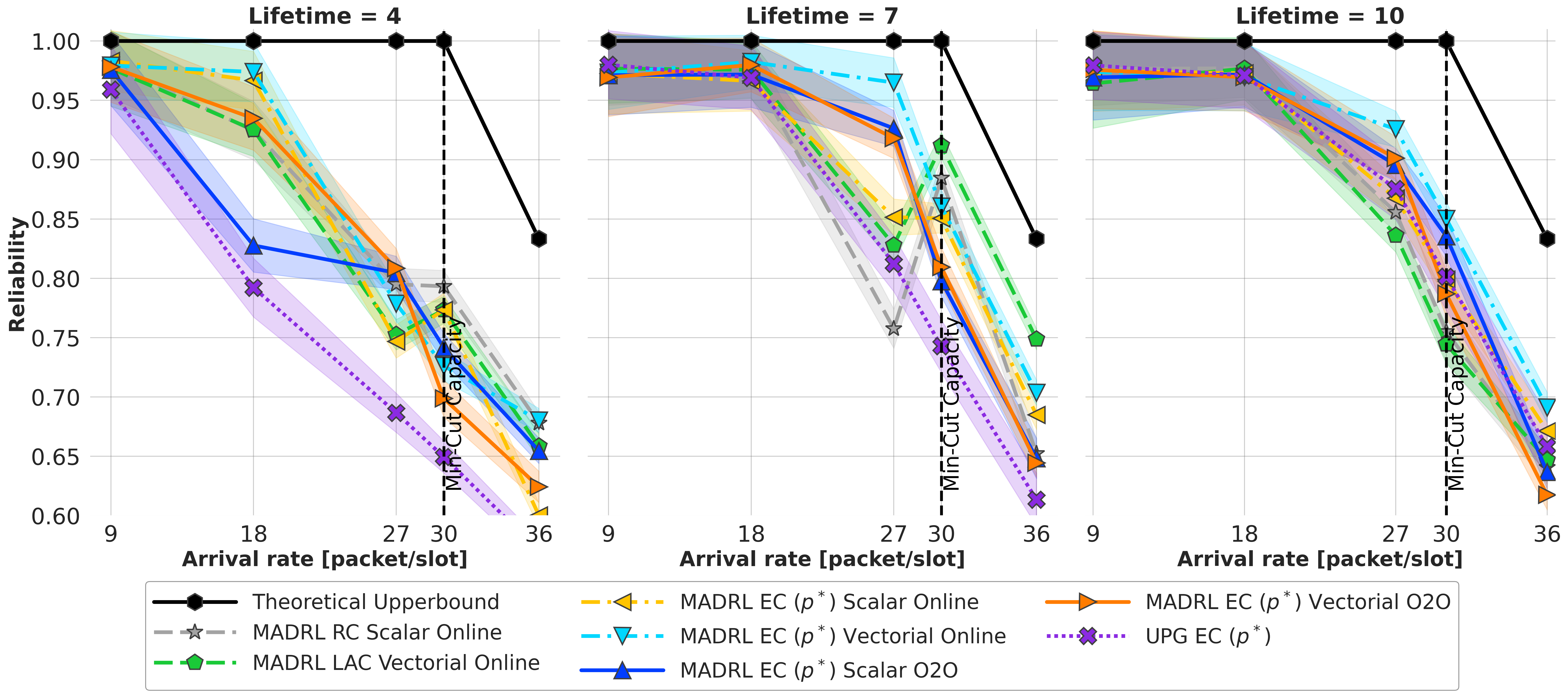}
			\vspace{-0.4cm}
		\end{subfigure}
		
		\caption{Reliability performance comparison of DDPG-based approaches on the considered topologies.  The line's shading indicates the Standard Deviation (SD).
		}
		\label{fig:ddpg_based_finetuned_model_based_policies_comparison}
		\vspace{-0.5cm}
	\end{figure*}
	
	\subsection{RL-based Approaches}
	\label{sec:results_rl_based_approaches}
	
\begin{figure}[htb!]
		\begin{subfigure}{0.5\textwidth}
			\centering
			\caption{Hierarchical}
			\includegraphics[width=\linewidth]{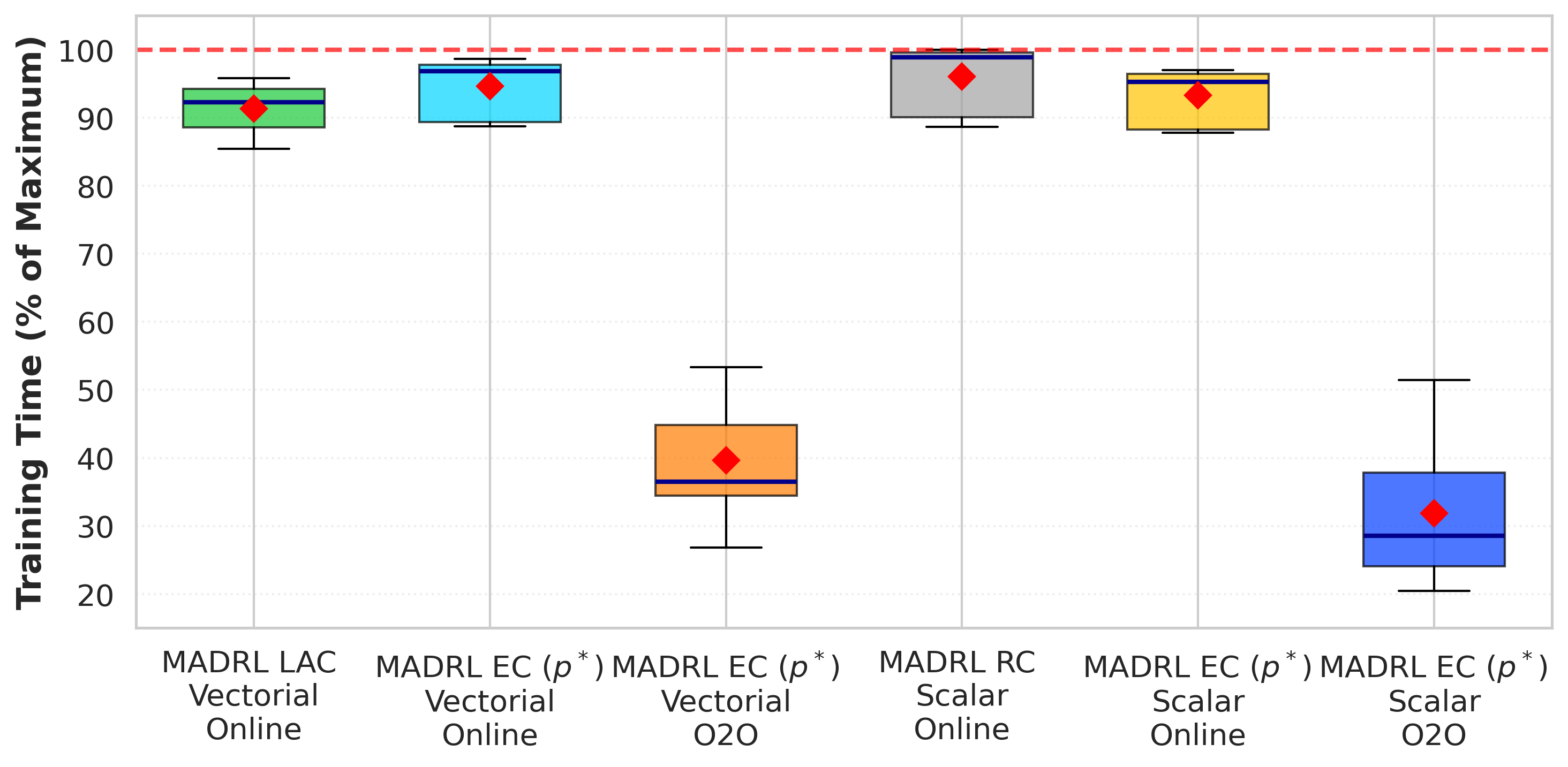}
			\label{fig:hierarchical_o2o_training_time}
			\vspace{-0.5cm}
		\end{subfigure}
		\hfill
		\begin{subfigure}{0.5\textwidth}
			\centering
			\caption{Abilene}
			\includegraphics[width=\linewidth]{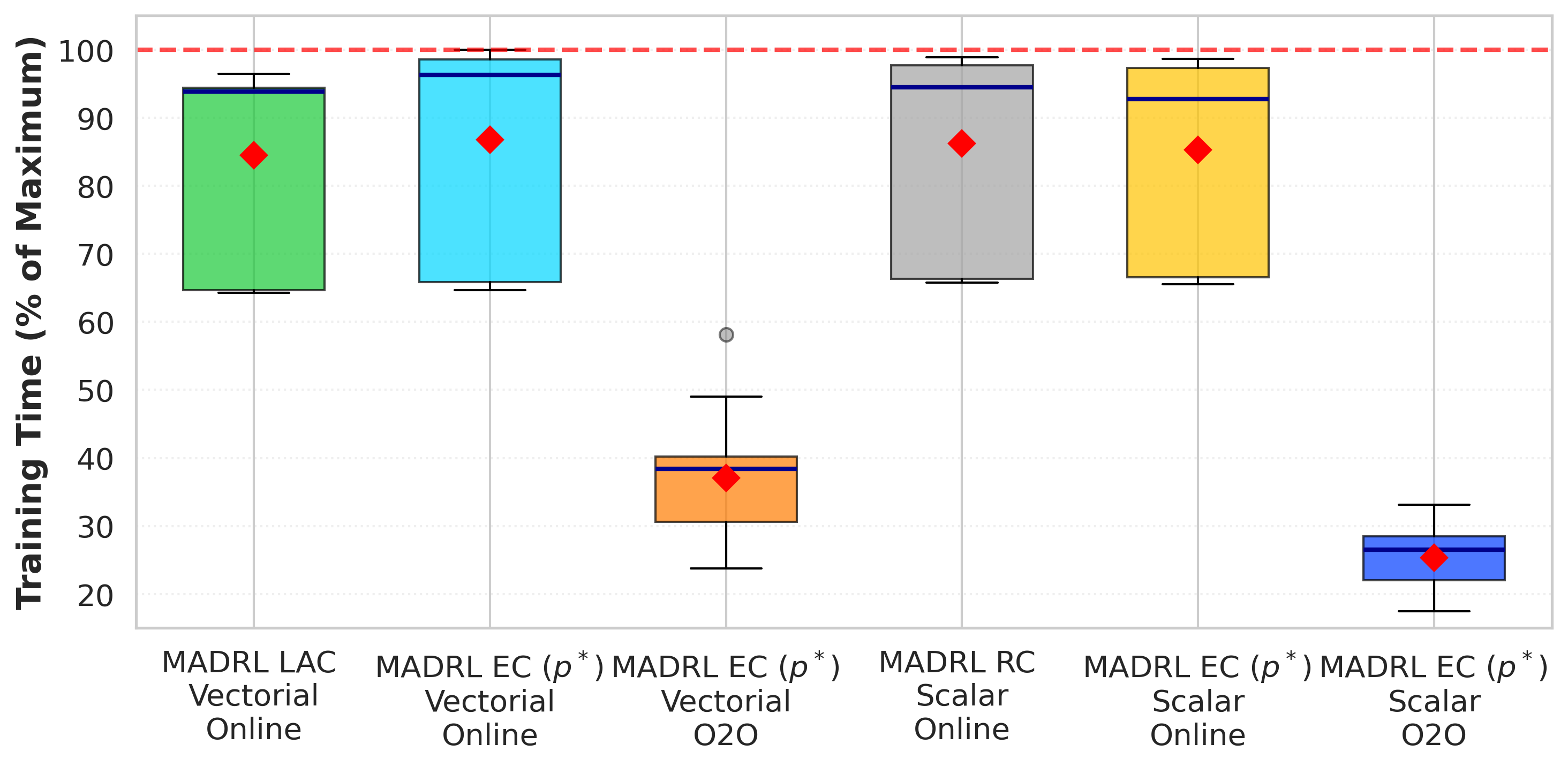}
			\label{fig:abilene_o2o_training_time}
			\vspace{-0.5cm}
		\end{subfigure}
		\hfill
		\begin{subfigure}{0.5\textwidth}
			\centering
			\caption{Grid}
			\includegraphics[width=\linewidth]{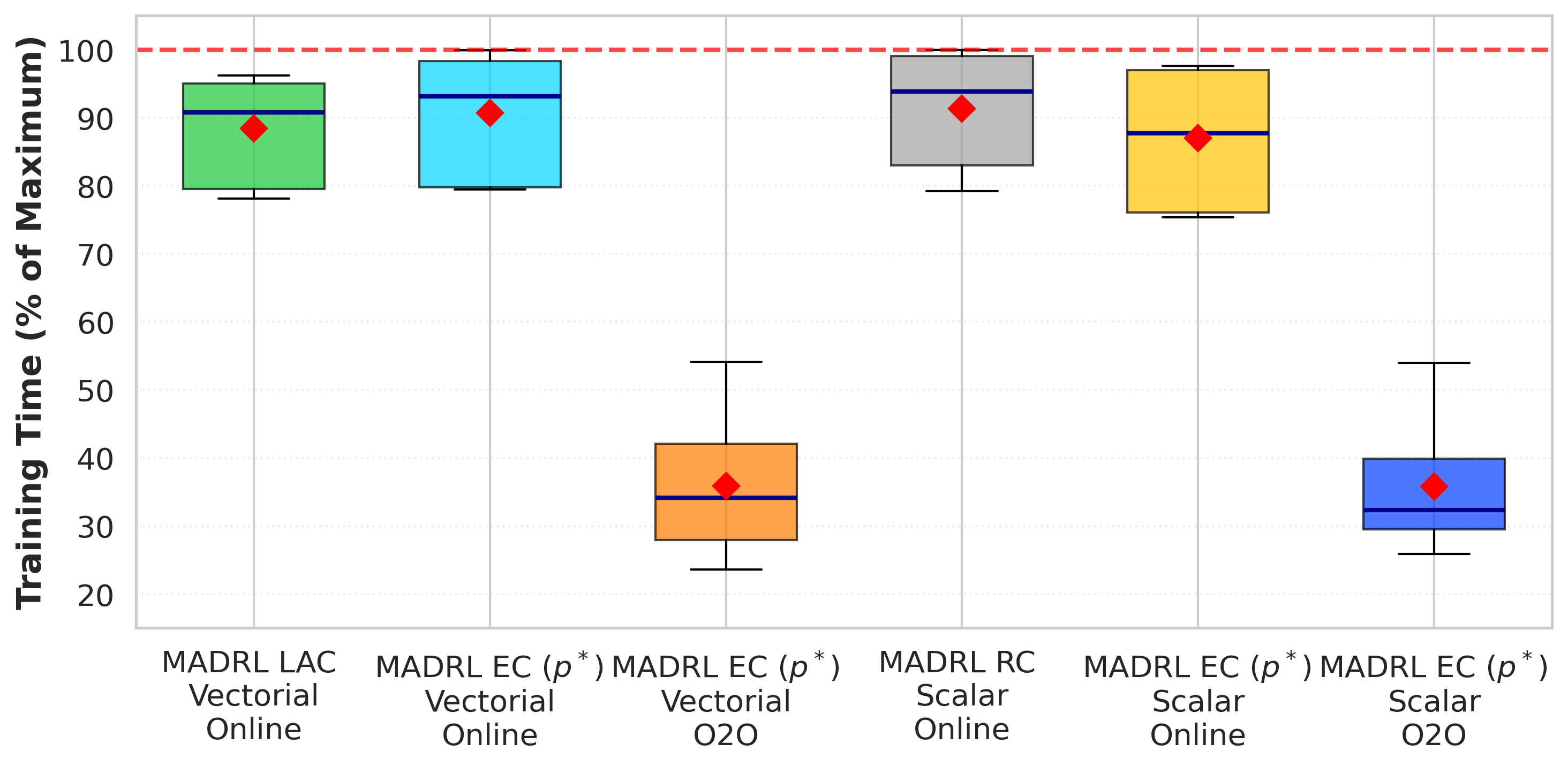}
			\label{fig:grid_o2o_training_time}
			\vspace{-0.3cm}
		\end{subfigure}
		\caption{Training time reduction for considered network topologies; Hierarchical \ref{fig:hierarchical_o2o_training_time}, Abilene \ref{fig:abilene_o2o_training_time} and Grid \ref{fig:grid_o2o_training_time}. The red diamond denotes the mean training time, while the blue line denotes the median. Reductions are computed as a percentage of the longest wall-clock run on the same topology (dashed red line) and identical hardware. The \ac{mgarl} policies include the full pre-training and online fine-tuning phases.}
		\label{fig:020_training_time_advantage}
		\vspace{-0.5cm}
	\end{figure}

	For each network topology---Hierarchical, Abilene, and Grid---we compare the six RL-based approaches defined in Section~\ref{ssec:considered_approaches} against the theoretical upper bound and against the best prior-guided policy, \ac{upgecps} (Section~\ref{sec:results_model_based_approaches}). Figure~\ref{fig:ddpg_based_finetuned_model_based_policies_comparison} reports their reliability, while Figure~\ref{fig:020_training_time_advantage} shows the corresponding training time reductions. We discuss each topology in turn, before drawing cross-topology conclusions.

	\paragraph{Hierarchical Topology} In this highly structured environment, \ac{upgecps} provides an excellent baseline. By utilizing it as a bootstrap, both the Vectorial and Scalar \ac{mgarl} agents smoothly inherit this behavior. The structure of the network topology allows even the compressed Scalar representation to avoid severe aliasing penalties. Consequently, both \ac{mgarl} configurations reach performance comparable to their fully Online counterparts, but in a fraction of the training time (as evidenced by the boxplots in Figure~\ref{fig:hierarchical_o2o_training_time}). This effectively bypasses the costly and unstable exploration phase that heavily penalizes from-scratch learning. Notably, the deliberately high residual floor ($\lambda_{\text{res}} = 0.8$, Sec.~\ref{sec:hyperparameters}) constrains the agent to primarily refine the pre-trained policy rather than substantially deviate from it. Given the baseline's near-optimal performance in this topology, RL exploration would yield only marginal gains at the cost of increased stability risks. Thus, the RL agent operates largely as a policy fine-tuner providing incremental stability rather than high performance improvements, while \ac{upgecps} remains an efficient, self-sufficient control strategy for the Hierarchical topology.
	
	\paragraph{Abilene Topology} The asymmetric nature of the Abilene backbone highlights the adaptability of \ac{mgarl} pre-training across different temporal constraints. Under the minimum lifetime configuration ($L=5$), the strictness of the deadlines combined with the network's structural asymmetry creates a unique dynamic that inverts the usual benefits of the demonstrator. While the pre-trained \ac{mgarl} policies improve upon the \ac{upgecps} baseline, they fall short of the reliability levels achieved by their fully Online counterparts. This indicates that although the prior-guided heuristic provides a solid starting point for the learning agents, in such highly constrained environments, it acts as an overly restrictive anchor. It hinders the exploration needed to discover the better routing strategies that from-scratch learning eventually finds (albeit at a higher computational training cost). Additionally, as deadlines relax, the true value of \ac{mgarl} pre-training emerges. \ac{upgecps} serves as an effective bootstrap, yielding a favorable trade-off between performance and training time. When comparing state representations, the compressed Scalar \ac{mgarl} variant provides fast convergence, at the cost of state aliasing, which empirically leads to sub-optimal performance. Conversely, the Vectorial \ac{mgarl} agent leverages its full state expressiveness to match the reliability of the fully Online counterpart, while reducing the required training time.
	
	\paragraph{Grid $3\times3$ Topology} The Grid topology, characterized by high path diversity and multiple equivalent routes, presents the hardest exploration challenge. Here, \ac{upgecps} acts as a suitable bootstrap. Both Vectorial and Scalar \ac{mgarl} agents achieve strong reliability with significantly higher stability and lower training times compared to the fully Online agents, which frequently collapse or exhibit high variance under heavy load. \ac{upgecps} relies on a simplified \ac{ecps} approximation that performs on average better than the exact counterpart \ac{ecp} (see Figure~\ref{fig:model_based_policies_comparison}), making it effective for bootstrapping RL agents. Notably, under intermediate and relaxed deadlines ($L=7$ and $L=10$), \ac{mgarl} performance is comparable to or even better than that of the alternatives, showing stability across saturation regimes and underscoring the benefits of pre-training in complex, symmetric environments.
	
	\paragraph{Cross-Topology Observations \& RL Implications}
	A joint analysis of the \ac{madrl} experiments highlights three fundamental dynamics governing the accelerated deployment of RL in latency-critical networks:
	
	\begin{itemize}
		\item \emph{The power of sub-optimal bootstrapping.} Across all configurations, using a computationally light, albeit imperfect, prior-guided heuristic like \ac{upgecps} emerges as a beneficial starting point for \ac{rl}. It mitigates the catastrophic performance drops typical of early-stage exploration, promoting a safer initialization and allowing the agent to reach operationally acceptable reliability from the very first online interactions, rather than after thousands of trial-and-error episodes.
		
		\item \emph{\ac{mgarl} sample efficiency over online learning.} By leveraging a limited offline pre-training phase of $200$ epochs (drawn from a static dataset collected over $500$ episodes, i.e., $N_{\mathrm{off}}=|\mathcal{D}_{\text{off}}|\approx 25{,}000$ transitions) followed by $2{,}000$ online fine-tuning episodes, \ac{mgarl} agents reach reliability levels comparable to---and on the Grid topology even exceeding---the fully Online. This is achieved while consuming only $1/7$ of the online interaction budget ($2{,}000$ vs.\ $14{,}000$ episodes). Once the offline phase is included in the total cost, the resulting wall-clock reduction reaches $55\%$--$65\%$ (Fig.~\ref{fig:020_training_time_advantage}), supporting the practical viability of the paradigm under tight deployment budgets.
		
		\item \emph{Vectorial expressiveness vs.\ Scalar compression.} While \ac{mgarl} supports rapid deployment regardless of the observation granularity, the choice of state representation governs the final policy ceiling. The Scalar form converges faster thanks to its compressed state but is structurally penalized by state aliasing, which hides differences between high-urgency and low-urgency queue compositions. The Vectorial form, by preserving the full urgency distribution, enables the agent to fine-tune the bootstrapped policy more accurately, consistently delivering the highest reliability among the \ac{rl}-based configurations; its larger state space is effectively absorbed by the \ac{mgarl} warm-start.
	\end{itemize}

	\subsection{Best Approach by Operational Objective}
	
	\label{ssec:best_approach_by_objective}
	
	No single policy dominates across all objectives: which one to adopt depends on the specific goal at hand.  Based on the simulation results presented above, Table~\ref{tab:winner_by_objective} summarizes our recommendation for six representative objectives, spanning both prior-guided and RL-based approaches. Among the prior-guided policies, the exact \ac{ecp} formulation is preferable when the objective is to maximize raw reliability under the most adverse conditions (tight lifetimes and structurally asymmetric topologies), whereas its approximation, EC~$p^*$, offers the most attractive reliability-complexity trade-off, and is therefore both our recommended prior-guided policy overall and the natural choice as \ac{mgarl} demonstrator. Among the RL-based configurations, the two design axes serve distinct objectives: the choice of state representation (Vectorial vs.\ Scalar) governs the reliability ceiling the agent can reach, while the choice of training paradigm (\ac{mgarl} vs.\ fully Online) governs the online interaction cost required to reach it. Consequently, Scalar \ac{mgarl} is preferable when fast convergence with a compact state is the priority; Vectorial \ac{mgarl} offers the best balance of reliability and deployment cost; and Vectorial Online remains the reference choice when training cost is not a constraint and maximal reliability is sought.

	\begin{table}[h]
		\centering
		\caption{\small{Best policy depending on the operational objective.}}
		\label{tab:winner_by_objective}
		\begin{tabular}{ >{\raggedright\arraybackslash}m{0.6\columnwidth} >{\centering\arraybackslash}m{0.3\columnwidth} }
			\toprule
			\textbf{Objective} & \textbf{Best Policy} \\
			\midrule
			\midrule
			Peak reliability under tight deadlines and saturation
			& \ac{upgecp} \\
			\midrule
			Best reliability–complexity trade-off across regimes
			& \ac{upgecps} \\
			\midrule
			Lightweight, stable demonstrator to bootstrap \ac{mgarl}
			& \ac{upgecps} \\
			\midrule
			Fastest convergence with a compact state representation
			& \ac{marlecps} Scalar \ac{mgarl}\\
			\midrule
			Deployment-oriented RL: high reliability at low online cost
			& \ac{marlecps} Vectorial \ac{mgarl}\\
			\midrule
			Highest RL reliability when training cost is unconstrained
			& \ac{marlecps} Vectorial Online \\
			\bottomrule
		\end{tabular}
	\end{table}
	
	\section{Conclusions and Future Directions}
	\label{sec:conclusions}
	In this work, we presented a comprehensive methodology designed to reduce the complexities of network control for latency-sensitive applications, with a specific focus on meeting strict delivery deadlines. Relying on the \ac{dcmt} \citep{vitale2025flexible} problem formulation, the introduction of spatially-aware \acf{ec} metrics and the \acf{upg} strategy provides a computationally efficient basis for deriving routing policies aimed at timely packet delivery. Specifically, by proactively filtering out non-viable packets, the \ac{ec}-based policies achieved reliability gains ranging from 20\% to over 40\% compared to traditional volume-based routing. Furthermore, the \ac{upg} strategy effectively balanced spatial loads, mitigating localized congestion hotspots across structurally diverse environments.
	
	A critical insight derived from our evaluation is the robustness of the decoupled interface-level approximation ($p^*$) across structurally diverse regimes. Analytically anchoring the congestion evaluation to a single reference path reduces the per-step evaluation cost from $\bigO(|\mathcal{P}|)$ to $\bigO(|\mathcal{E}|)$ while preserving the filtering benefits of the exact \ac{ecp} model; empirically, as deadlines relax it matches or exceeds the exact model's reliability on the Hierarchical and Grid topologies, while trailing it by a few percentage points on the structurally asymmetric Abilene backbone. Validating this complexity advantage on substantially larger instances remains an open direction.
	
	Building upon this prior-guided foundation, the baseline \ac{madrl} framework was extended by incorporating these predictive metrics into the agents' observation space. To bridge the gap between theoretical learning and practical deployment, we unified the considered policy-learning objectives under the \acf{gpr} and embedded the training pipeline within the resulting \acf{mgarl} protocol, grounded in the \ac{rlfd} paradigm: the computationally light \ac{upgecps}  acts as a programmatic demonstrator whose trajectories drive both the offline pre-training and the online fine-tuning phases through a residual \ac{bc} regulariser. To ensure that this demonstration-driven warm-start survives the offline-to-online transition, we introduce the Anchored Transfer Stabilization Protocol. Together, these mechanisms allow the framework to bypass the performance collapse and sample inefficiency typical of from-scratch online learning, matching---and on the Grid topology exceeding---the reliability of fully Online agents while reducing by a factor of seven the online interaction budget.

	Within the \ac{mgarl} framework, our analysis revealed a fundamental expressiveness-complexity trade-off governed by the state-space representation. The compressed Scalar observation space enables faster training. However, due to severe state aliasing, the Scalar \ac{mgarl} agent struggles to accurately evaluate complex congestion patterns, occasionally degrading its routing performance below the prior-guided heuristic baseline it was bootstrapped from. Conversely, the Vectorial representation preserves the full urgency distribution. This expressiveness allows the Vectorial \ac{mgarl} agent to safely navigate the online fine-tuning phase, matching or exceeding the baseline's performance and achieving the highest reliability among the scalable RL configurations, with sporadic degradations due to the combination of a large state space with limited training budget.
	
	Ultimately, this work contributes an extensible architecture for deadline-aware routing in NextG environments. While this architecture provides a robust foundation, several promising directions for future research remain open. To further assess the framework's scalability and adaptability, future studies will evaluate its performance on significantly larger network instances operating under highly dynamic conditions, such as fluctuating link capacities. Additionally, exploring alternative neural architectures presents a compelling avenue; for instance, integrating Graph Neural Networks (GNNs) into the \ac{mgarl} protocol could enable the agents to natively capture and exploit the topological properties of evolving networks. 
	A systematic ablation isolating the individual contributions of the Anchored Transfer components---the residual BC floor $\lambda_{\mathrm{res}}$, the SymLog reward transform, and the frozen Z-score normalization---is left to future work. A further open question concerns the sensitivity of \ac{mgarl} to the Stage-1/Stage-2 transition point $T_1$: a premature transition may leave the Actor and Critic insufficiently converged on $\mathcal{D}_{\mathrm{off}}$, exacerbating the Actor-Critic Misalignment discussed in Section~\ref{sec:mgarl_design_rationale}, whereas an excessively long pre-training phase may over-anchor the Actor to $\mu^{\mathrm{MB}}$, compounding the restrictive-anchor effect empirically observed under the Abilene $L=5$ configuration (Section~\ref{sec:results_rl_based_approaches}). A systematic study of $T_1$---jointly with the residual floor $\lambda_{\mathrm{res}}$ and the Early Stopping criterion that determines it in practice---is left to future work.
	Finally, a critical next step involves validation beyond simulation. Deploying the trained \ac{madrl} policies on programmable data plane hardware, such as P4-enabled equipment, would allow measuring true inference latencies and assessing the practical viability of the \ac{ec} metric under operational constraints.
	\vspace{-0.25cm}
	\section*{Declaration of competing interest}
	\vspace{-0.25cm}
	\small{The authors declare that they have no known competing financial
		interests or personal relationships that could have appeared to
		influence the work reported in this paper.}
	\vspace{-0.25cm}
	\section*{Acknowledgements}
	\vspace{-0.25cm}
	\small{This work was partially supported by the European Union under the Italian National Recovery and Resilience Plan (NRRP) of NextGenerationEU, partnership on "Telecommunications of the Future" (PE00000001 - program "RESTART"), by the PRIN project "Resilient delivery of real-time interactive services over NextG compute-dense mobile networks" (E53D2300055000), and by funds from the US National Science Foundation as specified in the RINGS program (CNS-2148315).}
	
	\bibliographystyle{elsarticle-harv}
	\bibliography{bibliography}
	
	\appendix
	\setcounter{figure}{0}
	\setcounter{table}{0}
	\setcounter{equation}{0}
	
	\section{Spatial Drop Rate Distribution}
	\label{sec:appendinx_spatial_drop}
	This appendix complements the macroscopic reliability analysis of Section~\ref{sec:numerical_results} with a microscopic view of where packet expirations concentrate across network interfaces. For each topology, we focus on settings with an arrival rate equal to $90\%$ of the Min-Cut capacity and a lifetime sufficient to activate all candidate paths, so that differences across policies stem from routing decisions rather than infeasibility. Two complementary representations are used throughout: a network topology overlay showing per-edge (undirected) drop counts, and a heatmap aggregating drops at the interface level.
	\begin{figure*}[b!]
		\centering
		\begin{subfigure}[t]{0.9\textwidth}
			\centering
			\includegraphics[width=\linewidth]{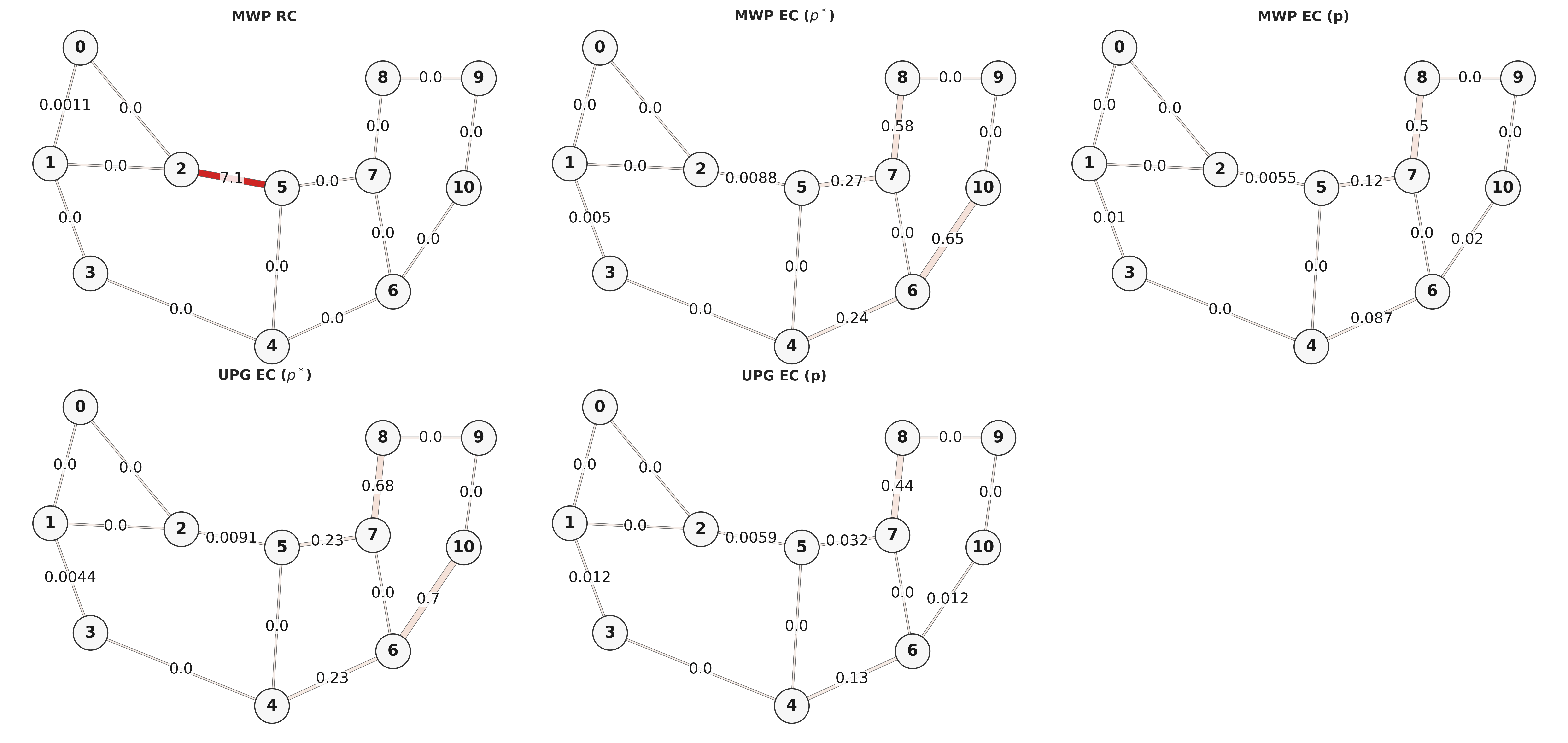}
			\label{fig:expired_abilene_spatial_topology}
			\vspace{-0.8cm}
		\end{subfigure}
		\centering
		\begin{subfigure}[c]{0.9\textwidth}
			\centering
			\includegraphics[width=\linewidth,height=0.42\textheight,keepaspectratio]{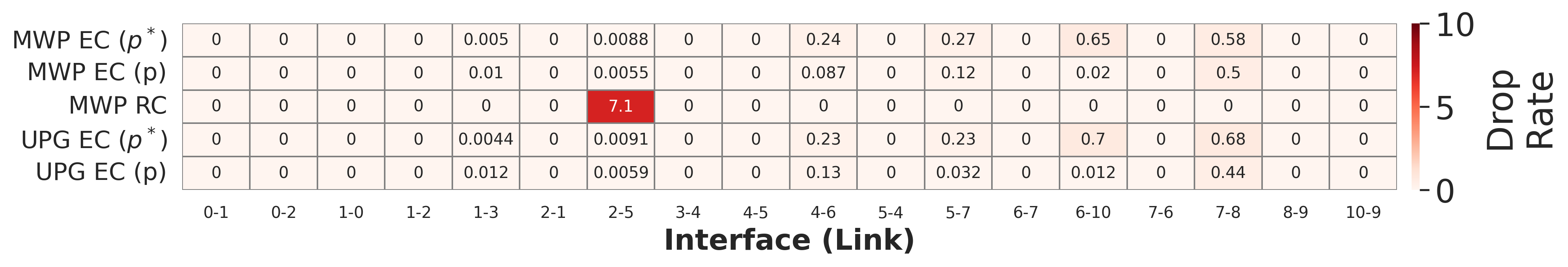}
			\vspace{-0.5cm}
			\label{fig:expired_abilene_spatial_heatmap}
		\end{subfigure}
		\caption{\small\textbf{Spatial distribution of expired packets for prior-guided approaches on Abilene topology with $L=11, b=18$.}
		}
		\label{fig:expired_abilene_spatial}
		\vspace{-0.3cm}
	\end{figure*}
	\begin{figure*}[htb!]
		\centering
		\begin{subfigure}[t]{0.9\textwidth}
			\centering
			\includegraphics[width=\linewidth]{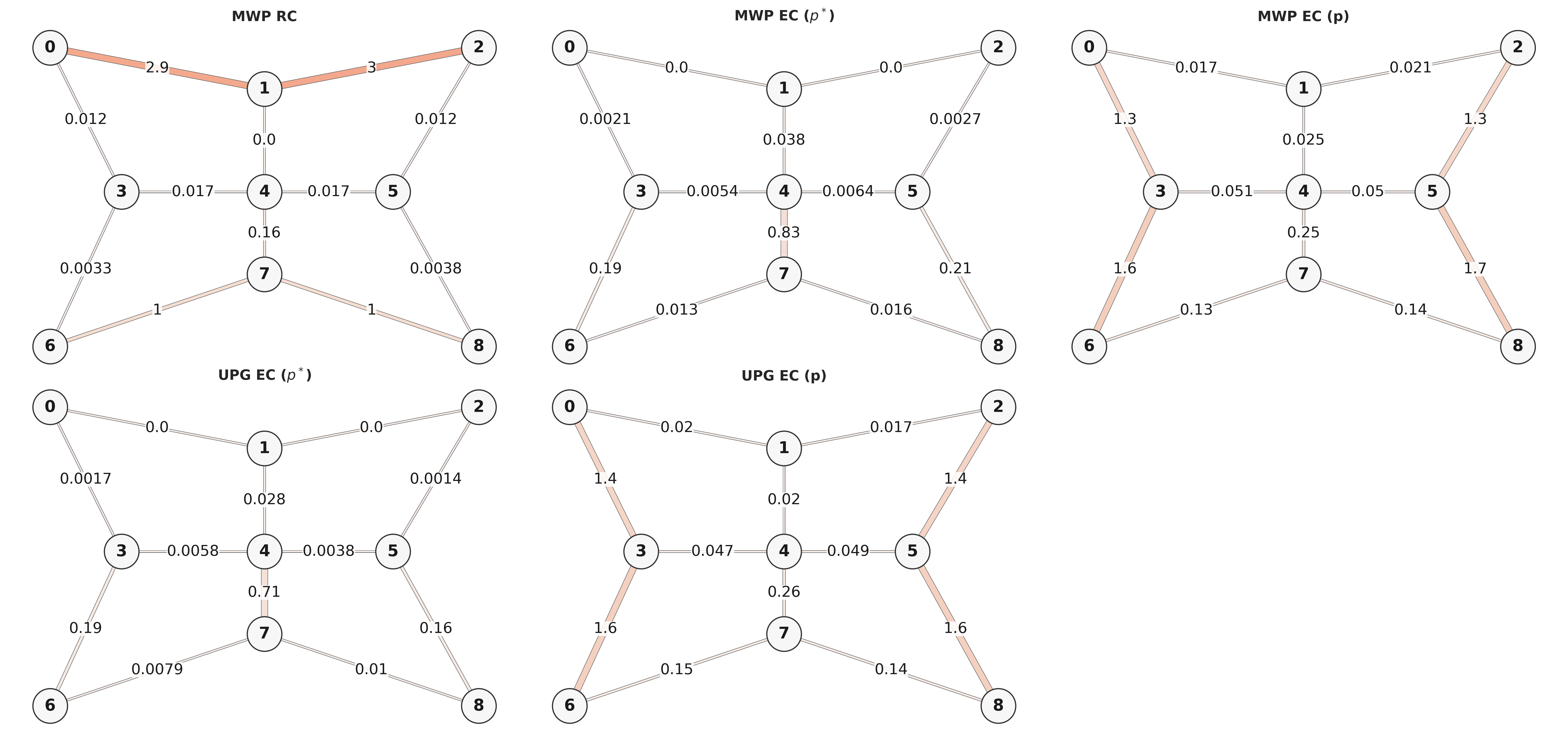}
			\vspace{-0.5cm}
		\end{subfigure}
		\centering
		\begin{subfigure}[b]{0.9\textwidth}
			\centering
			\includegraphics[width=\linewidth,height=0.95\textheight,keepaspectratio]{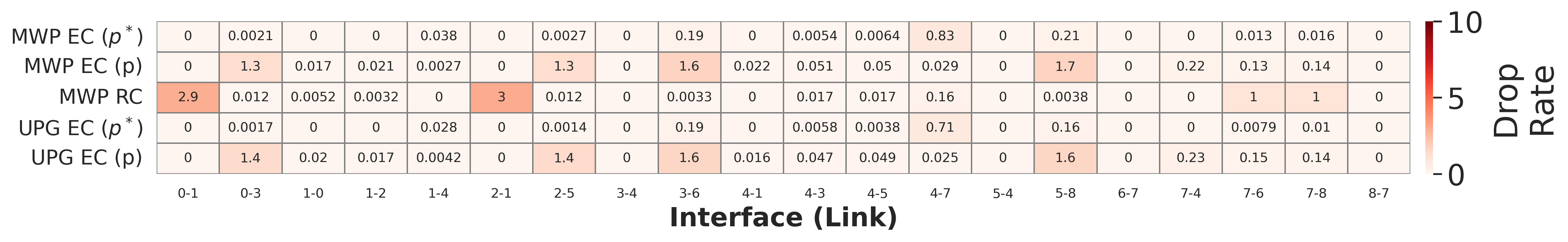}
			\vspace{-0.5cm}
		\end{subfigure}
		\caption{\textbf{Spatial distribution of expired packets for prior-guided approaches on Grid $3\times3$ network with $L=10, b=27$.}
		}
		\label{fig:expired_grid_spatial}
		\vspace{-0.5cm}
	\end{figure*}
	\begin{figure*}[htb!]
		\centering
		\begin{subfigure}[t]{0.9\textwidth}
			\centering
			\includegraphics[width=\linewidth]{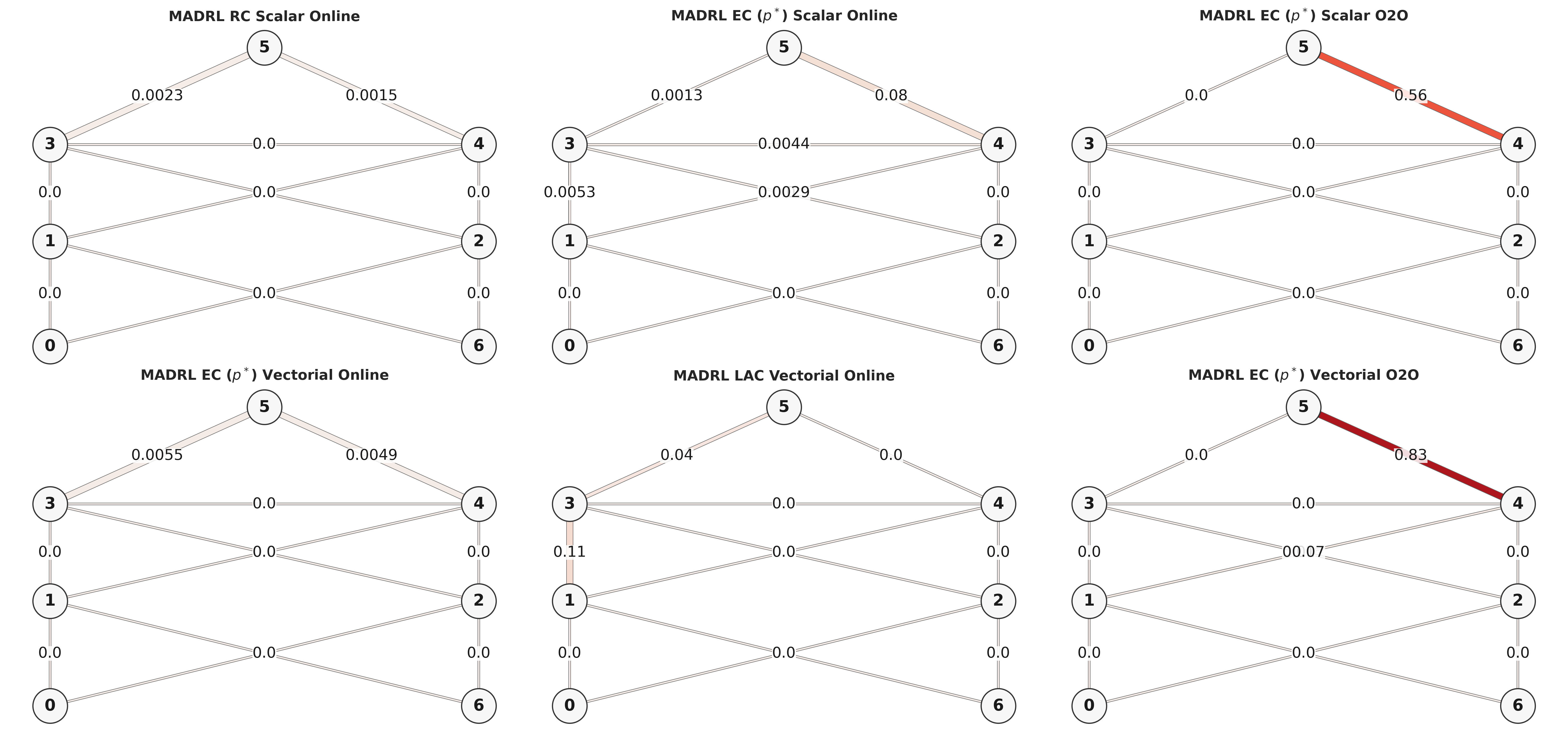}
			\label{fig:rl_expired_hierarchical_spatial_topology}
			\vspace{-0.5cm}
		\end{subfigure}
		\centering
		\begin{subfigure}[c]{0.9\textwidth}
			\centering
			\includegraphics[width=\linewidth,height=0.85\textheight,keepaspectratio]{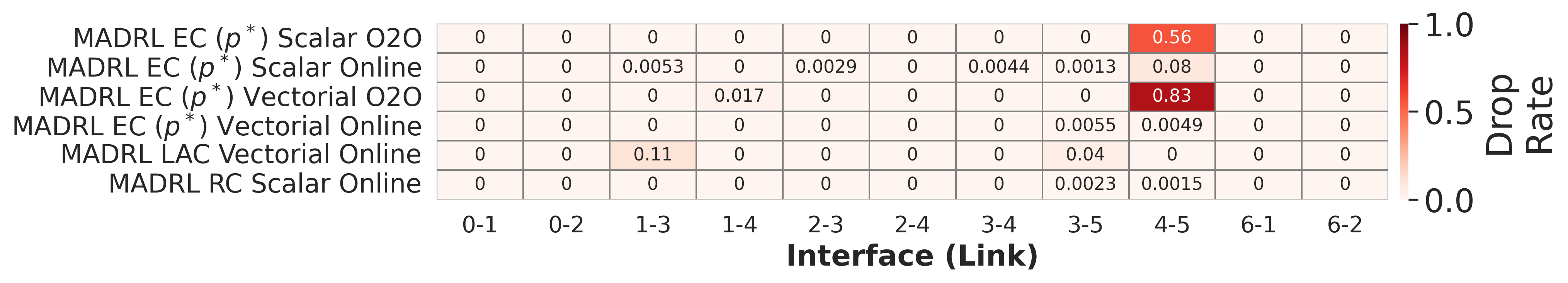}
			\label{fig:rl_expired_hierarchical_spatial_heatmap}
			\vspace{-0.5cm}
		\end{subfigure}
		\caption{\textbf{Spatial distribution of expired packets for DDPG-based approaches on Hierarchical topology with $L=5, b=18$.}
		}
		\label{fig:rl_expired_hierarchical_spatial}
		\vspace{-0.3cm}
	\end{figure*}
	\subsection{Prior-guided  Policies}
	Figures~\ref{fig:expired_hierarchical_spatial}, \ref{fig:expired_abilene_spatial}, and~\ref{fig:expired_grid_spatial} report the spatial drop distributions for the Hierarchical, Abilene, and Grid topologies, respectively. Three recurrent patterns emerge across topologies. First, the traditional \ac{mwprc} baseline systematically concentrates expirations on the topological bottlenecks dictated by shortest-path routing (link 4-5 in Hierarchical, link 2-5 in Abilene, the central crossbar in Grid), since its volume-based metric cannot discriminate between viable and expiring traffic. Second, \ac{ec}-based policies redistribute the load proactively by filtering out non-viable packets, lowering peak per-interface drops by more than an order of magnitude in the most stressed scenarios. Third, the \ac{upg} assignment compounds this benefit by avoiding the single-path saturation that affects greedy variants, leading to a near-uniform residual drop profile. These observations directly motivate the choice of \ac{upgecps} as the bootstrap demonstrator for the \ac{mgarl} framework: it provides the stable, well-distributed initial behavior that prevents the early-stage value-overestimation cascade typical of from-scratch \ac{rl}.
	
	\begin{figure*}[htb!]
		\centering
		\begin{subfigure}[t]{0.9\textwidth}
			\centering
			\includegraphics[width=\linewidth]{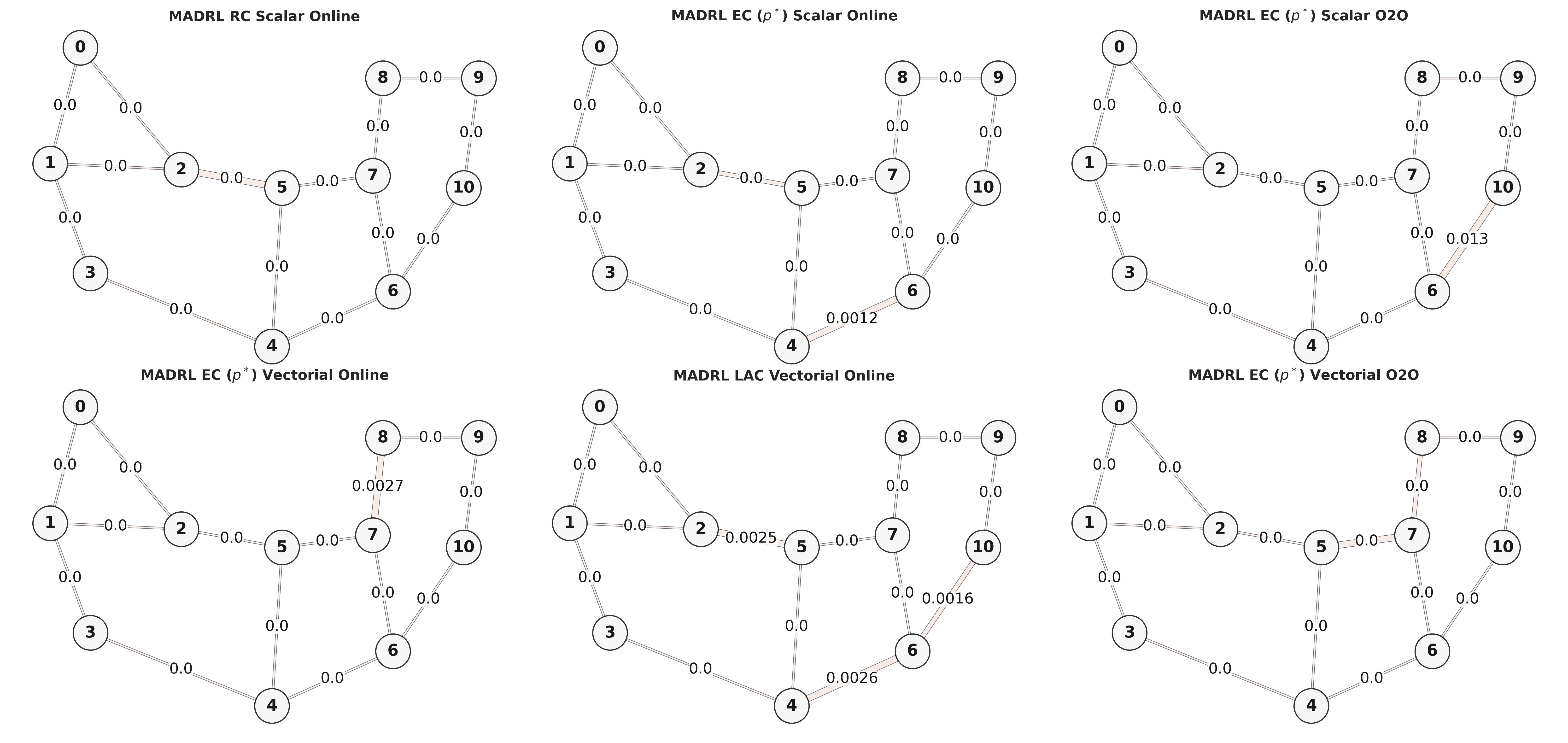}
			\label{fig:rl_expired_abilene_spatial_topology}
			\vspace{-0.5cm}
		\end{subfigure}
		\centering
		\begin{subfigure}[c]{0.9\textwidth}
			\centering
			\includegraphics[width=\linewidth,height=0.42\textheight,keepaspectratio]{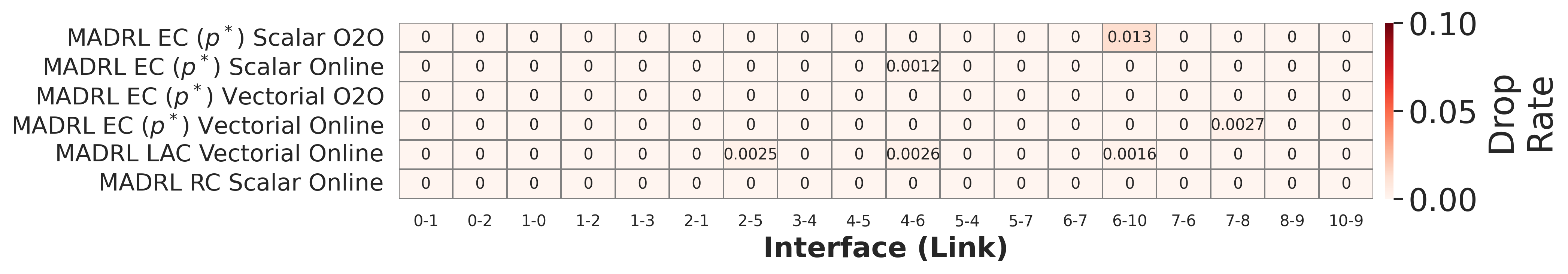}
			\label{fig:rl_expired_abilene_spatial_heatmap}
			\vspace{-0.5cm}
		\end{subfigure}
		\caption{\textbf{Spatial distribution of expired packets for DDPG-based approaches on Abilene topology with $L=11, b=18$.}
		}
		\label{fig:rl_expired_abilene_spatial}
		\vspace{-0.5cm}
	\end{figure*}

	\begin{figure*}[htb!]
		\centering
		\begin{subfigure}[t]{0.9\textwidth}
			\centering
			\includegraphics[width=\linewidth]{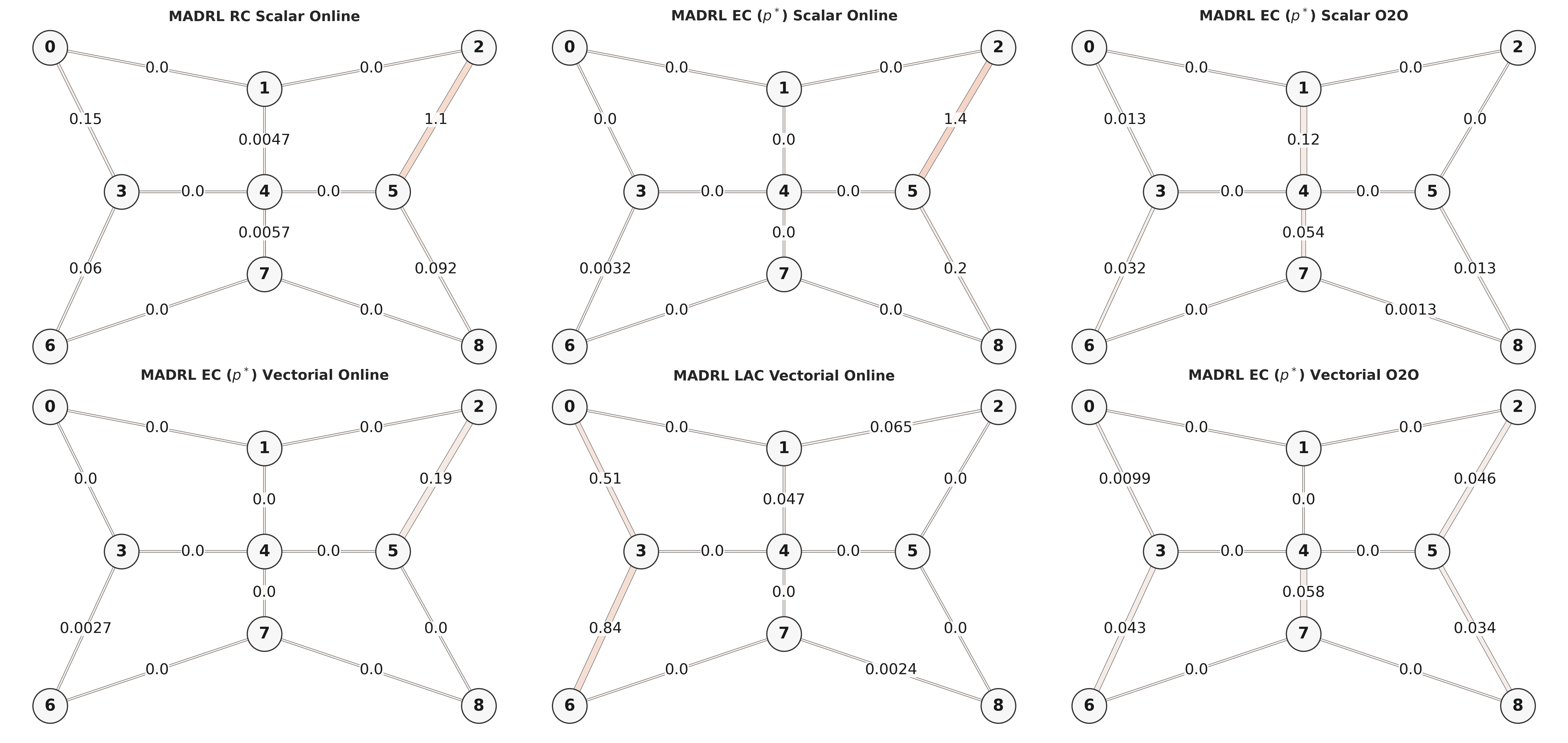}
			\vspace{-0.5cm}
		\end{subfigure}
		\centering
		\begin{subfigure}[b]{0.9\textwidth}
			\centering
			\includegraphics[width=\linewidth,height=0.95\textheight,keepaspectratio]{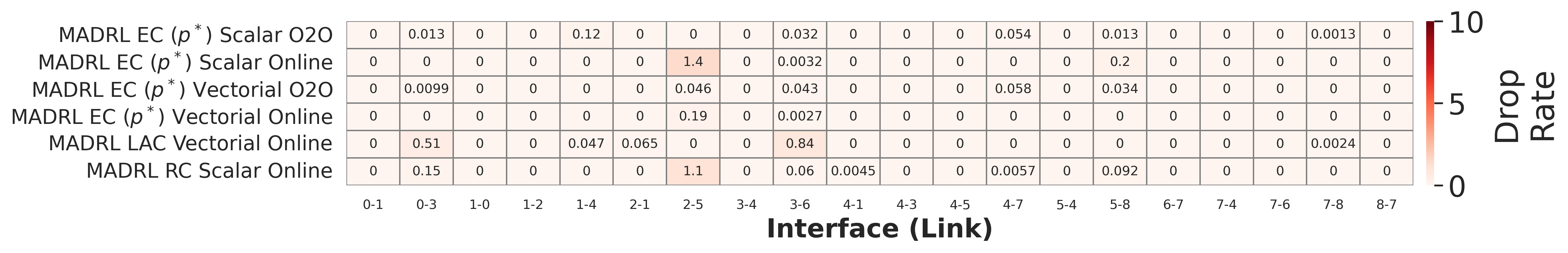}
			\vspace{-0.5cm}
		\end{subfigure}
		\caption{\textbf{Spatial distribution of expired packets for DDPG-based approaches on Grid $3\times3$ topology with  $L=10, b=27$.}
		}
		\label{fig:rl_expired_grid_spatial}
		\vspace{-0.5cm}
	\end{figure*}
	
	\subsection{RL-based Policies}
	Figures~\ref{fig:rl_expired_hierarchical_spatial}, \ref{fig:rl_expired_abilene_spatial}, and~\ref{fig:rl_expired_grid_spatial} extend the spatial analysis to the \ac{madrl} variants. Two design dimensions become visually apparent. The choice of the congestion metric (\ac{rc}/\ac{lac} vs.~EC~$p^*$) governs whether the agent learns to discriminate urgency: \ac{rc}- and \ac{lac}-based agents leave residual hotspots on the structural bottlenecks (e.g., interfaces 1-3 in Hierarchical, 6-10 in Abilene, the central crossbar in Grid), whereas EC-guided agents flatten the drop profile across the network. The choice of training paradigm (Online vs.~\ac{mgarl}) then determines whether the agent reaches this optimum reliably: fully Online agents exhibit residual scattered drops attributable to exploration noise, while \ac{mgarl} Vectorial agents consistently achieve the cleanest profile by inheriting the spatially balanced behavior of the \ac{upgecps} bootstrap and refining it under the expressive Vectorial observation. The combination \textit{EC~$p^*$ $+$ Vectorial $+$ \ac{mgarl}} emerges as the configuration that simultaneously minimizes the peak drop rate and the spatial variance, empirically supporting its selection as the recommended framework instantiation.
	
	\begin{table*}[!b]
		\section{Notation Table\label{sec:appendix_table}}
		
		\centering
		\caption{\small Table of Notations}
		\label{tab:notations_part1}
		\begin{tabular}{@{\extracolsep{\fill}} l p{0.80\textwidth} }
			\toprule
			\textbf{Symbol} & \textbf{Description} \\
			\midrule
			\multicolumn{2}{l}{\textit{Network Topology and Service Model}} \\
			$\mathcal{G}(\mathcal{V}, \mathcal{E})$ & Directed graph representing the network topology (nodes $\mathcal{V}$, links $\mathcal{E}$). \\
			$\rho_{i}^{+}, \rho_{i}^{-}$ & Sets of outgoing and incoming neighbors of node $i \in \mathcal{V}$. \\
			$C_{ij}(t)$ & Capacity of link $(i, j)$ at time $t$ (maximum packets per time slot). \\
			$\mathcal{C}$ & Set of commodities (latency-sensitive services). \\
			$s^c, d^c, L^c$ & Source node, destination node, and initial lifetime (TTL) for commodity $c$. \\
			$b^c(t), \bar{b}^c$ & Instantaneous and mean packet arrival rate for commodity $c$ at the source node. \\
			$\mathcal{P}^c, \mathcal{P}$ & Set of feasible candidate paths for commodity $c$ and the union of all paths in the network. \\
			$\mathcal{P}_{ij}, \mathcal{P}_{ij}^c$ & Subset of paths (global or specific to commodity $c$) traversing link $(i, j)$. \\
			\midrule
			\multicolumn{2}{l}{\textit{Queueing and Deadline-Aware Model}} \\
			$l$ vs. $\ell$ & Absolute lifetime (time-to-live) $l$ vs. Effective Lifetime (EL) $\ell$. \\
			$EL(p, l, i)$ & Function determining the effective lifetime based on remaining hops to destination. \\
			$q_{ij}^{(c,l)}(t)$ & Number of packets of commodity $c$ with lifetime $l$ currently in the queue at interface $(i,j)$. \\
			$\mathbf{q}_{ij}(t)$ & Aggregate queuing state vector of interface $(i,j)$ for all commodities and lifetimes. \\
			$f_{ij}^{(c, l)}(t)$ & Flow variables: number of packets of commodity $c$ with lifetime $l$ transmitted over link $(i,j)$. \\
			$g_{ij}^{(c, l)}(t)$ & Intentional dropping variables for packets with lifetime $l$ at interface $(i,j)$. \\
			$b_{ij}^{(c,l)}(t)$ & Packets exogenously arriving at node $i$ assigned to path $p \in \mathcal{P}_{ij}^c$. \\
			$f_{\rightarrow ij}^{(c, l)}(t)$ & Packets arriving at node $i$ from neighbors $\rho_i^-$ assigned to paths $p \in \mathcal{P}_{ij}^c$. \\
			$f_{\rightarrow d^c}^{(c, l)}(t)$ & Packets arriving at destination node $d^c$ from neighbors $\rho_{d^c}^-$ assigned to paths $p \in \mathcal{P}^c$. \\
			\midrule
			\multicolumn{2}{l}{\textit{Congestion Metrics and Effective Congestion (EC)}} \\
			$T^{p}_{ij}$ & Number of hops from the source of path $p$ to reach interface $(i,j)$. \\
			$\text{\sf L}^{p}$ & Initial effective lifetime of a packet assigned to path $p$ at its source. \\
			$p^*$ & Reference path ($p^*$) for the shared EC $(p^*)$-model. \\
			$Q_{ij}(t)$ & Total scalar packet count at interface $(i,j)$. \\
			$\mathbf{Q}_{ij}(t)$ & Vector of packets enqueued at interface $(i,j)$ indexed by their residual effective lifetime $\ell$. \\
			$\mathbf{\bar{Q}}_{ij}^{(p)}(t), \bar{Q}_{ij}^{(p)}(t)$ & Vector and Scalar filtered competing traffic for path $p$ at interface $(i,j)$ according to \ac{ecp} metric. \\
			$\mathbf{\hat{Q}}_{ij}^{(\mathcal{P}_{ij})}(t), \hat{Q}_{ij}^{(\mathcal{P}_{ij})}(t)$ & Vector and Scalar filtered competing traffic at interface $(i,j)$ according to \ac{ecps} metric. \\
			\midrule
			\multicolumn{2}{l}{\textit{\ac{gpr} and \ac{mgarl} Framework}}\\
			$\mu^{\mathrm{MB}}$ & Analytical prior-guided reference policy (demonstrator), queryable at any $s \in \mathcal{S}$. \\
			$\hat{r}_{on}(t),\, \hat{r}_{\mathrm{off}}(t)$ & Empirical reward estimates from the live ($\mathcal{B}(t)$) and pre-collected ($\mathcal{B}_{\mathrm{off}}$) mini-batches. \\
			$\hat{\mathcal{M}}_{MSE}$ & Empirical policy deviation between $\mu_\theta$ and $\mu^{\mathrm{MB}}$. \\
			$\alpha,\, \beta$ & \ac{gpr} weights of the live and pre-collected reward contributions. \\
			$\mathcal{D}(t)$ & Live replay buffer containing interaction tuples up to time $t$. \\
			$\mathcal{B}(t)$ & Mini-batch drawn from the live replay buffer $\mathcal{D}(t)$ at time $t$. \\
			$\mathcal{D}_{\text{off}}$ & Static dataset of transitions $(s, a, r, s')$ collected while executing the expert prior-guided heuristic policy. \\
			$\mathcal{B}_{\text{off}}$ & Mini-batch drawn from the static dataset of transitions $\mathcal{D}_{\text{off}}$. \\
			$N_{buf},\, N,\, M$ & Live replay buffer capacity; live and pre-collected mini-batch sizes. \\
			$\theta, \phi$ & Trainable parameters for the Actor ($\theta$) and Critic ($\phi$) neural networks. \\
			$\lambda, \lambda_{0}, \lambda_{res}$ & Imitation Factor: Weight of the Behavior Cloning (BC) loss (with decay parameters). \\
			$\omega$ & Q-Normalization Factor: Batch-wise mean absolute Q-value to scale the RL loss component. \\
			$\text{SymLog}(x)$ & Symmetric Logarithmic transformation used to compress reward spikes. \\
			$\mu_{\mathcal{D}}, \sigma_{\mathcal{D}}$ & Frozen Z-score statistics (mean and std dev) extracted from the offline dataset. \\
			$K, E, W$ & Number of offline Epochs ($K$), online Episodes ($E$), and Warm-up episodes ($W$). \\
			$D_{\mathrm{dec}}$ & buffer size at which $\lambda$ reaches its floor $\lambda_{\mathrm{res}}$.\\
			$\rho$ & Fraction of offline samples mixed into online training batches (Seeded Experience Replay). \\
			\bottomrule
		\end{tabular}
	\end{table*}
	
\end{document}